\documentclass[a4paper,fleqn]{cas-sc}

\usepackage[T1]{fontenc}
\usepackage[polutonikogreek,italian,english]{babel}
\usepackage[page,toc,titletoc,title]{appendix}
\usepackage[numbers,sort&compress]{natbib}
\usepackage{graphicx}
\usepackage{graphics}
\usepackage{braket}
\usepackage{bbold}
\usepackage{amssymb}
\usepackage{amsmath}
\usepackage{nicefrac}
\usepackage{dcolumn}
\usepackage{bm}
\usepackage{slashed}
\usepackage{datetime}
\usepackage{mciteplus}
\usepackage{multirow}
\usepackage{bbold}
\usepackage{color}
\usepackage{xcolor}
\usepackage{float}
\usepackage[utf8]{inputenc}
\usepackage[normalem]{ulem}
\usepackage{acro}

\DeclareAcronym{AI}{
  short=AI,
  long=Artificial Intelligence,
}
\DeclareAcronym{BSM}{
  short=BSM,
  long=Beyond-the-Standard-Model,
}
\DeclareAcronym{QCD}{
  short=QCD,
  long=Quantum ChromoDynamics,
}
\DeclareAcronym{SM}{
  short=SM,
  long=Standard Model,
}
\DeclareAcronym{CP}{
  short=CP,
  long=Charge-Parity,
}
\DeclareAcronym{SLAC}{
  short=SLAC,
  long=Stanford Linear Accelerator Center,
}
\DeclareAcronym{BNL}{
  short=BNL,
  long=Brookhaven National Laboratory,
}
\DeclareAcronym{FCNCs}{
  short=FCNCs,
  long=Flavor-Changing Neutral Currents,
}
\DeclareAcronym{GIM}{
  short=GIM,
  long=Glashow--Iliopoulos--Maiani,
}
\DeclareAcronym{CEM}{
  short=CEM,
  long= Color Evaporation Model,
}
\DeclareAcronym{CSM}{
  short=CSM,
  long= Color Singlet Mechanism,
}
\DeclareAcronym{CO}{
  short=CO,
  long=Color Octet,
}
\DeclareAcronym{NRQCD}{
  short=NRQCD,
  long=NonRelativistic QCD,
}
\DeclareAcronym{LDME}{
  short=LDME,
  long=Long-Distance Matrix Element,
}
\DeclareAcronym{LO}{
  short=LO,
  long=Leading Order,
}
\DeclareAcronym{NLO}{
  short=NLO,
  long=Next-to-Leading Order,
}
\DeclareAcronym{NNLO}{
  short=NNLO,
  long=Next-to-NLO,
}
\DeclareAcronym{MHOUs}{
  short=MHOUs,
  long=Missing Higher-Order Uncertainties,
}
\DeclareAcronym{DIS}{
  short=DIS,
  long=Deep Inelastic Scattering,
}
\DeclareAcronym{DGLAP}{
  short=DGLAP,
  long=Dokshitzer--Gribov--Lipatov--Altarelli--Parisi,
}
\DeclareAcronym{PDFs}{
  short=PDFs,
  long=Parton Distribution Functions,
}
\DeclareAcronym{FFs}{
  short=FFs,
  long=Fragmentation Functions,
}
\DeclareAcronym{MPI}{
  short=MPI,
  long=Multi-Parton Interaction,
}
\DeclareAcronym{DPS}{
  short=DPS,
  long=Double-Parton Scattering,
}
\DeclareAcronym{SCET}{
  short=SCET,
  long=Soft and Collinear Effective Theory,
}
\DeclareAcronym{TM}{
  short=TM,
  long=Transverse-Momentum,
}
\DeclareAcronym{TMD}{
  short=TMD,
  long=Transverse-Momentum-Dependent,
}
\DeclareAcronym{FFNS}{
  short=FFNS,
  long=Fixed-Flavor Number Scheme,
}
\DeclareAcronym{VFNS}{
  short=VFNS,
  long=Variable-Flavor Number Scheme,
}
\DeclareAcronym{GM-VFNS}{
  short=GM-VFNS,
  long=General-Mass Variable-Flavor Number Scheme,
}
\DeclareAcronym{HFMP}{
  short=HFMP,
  long=Heavy-Flavor Matching Point,
}
\DeclareAcronym{ABF}{
  short=ABF,
  long=Altarelli--Ball--Forte,
}
\DeclareAcronym{BFKL}{
  short=BFKL,
  long=Balitsky--Fadin--Kuraev--Lipatov,
}
\DeclareAcronym{LL}{
  short=LL,
  long=Leading Logarithmic,
}
\DeclareAcronym{NLL}{
  short=NLL,
  long=Next-to-Leading Logarithmic,
}
\DeclareAcronym{NNLL}{
  short=NNLL,
  long=Next-to-NLL,
}
\DeclareAcronym{LVM}{
  short=LVM,
  long=Light Vector Meson,
}
\DeclareAcronym{UGD}{
  short=UGD,
  long=Unintegrated Gluon Distribution,
}
\DeclareAcronym{LHC}{
  short=LHC,
  long=Large Hadron Collider,
}
\DeclareAcronym{EIC}{
  short=EIC,
  long=Electron-Ion Collider,
}
\DeclareAcronym{HFAG}{
  short=HFAG,
  long=Heavy Flavor Averaging Group,
}
\DeclareAcronym{BLM}{
  short=BLM,
  long=Brodsky--Lepage--Mackenzie,
}

\def\tsc#1{\csdef{#1}{\textsc{\lowercase{#1}}\xspace}}
\tsc{WGM}
\tsc{QE}

\newcommand{\drv}{{\rm d}}

\newcommand{\LQCD}{\Lambda_{\rm QCD}}
\newcommand{\MSb}{\overline{\rm MS}}

\newcommand{\NLO}{{\rm NLO}}

\newcommand{\LL}{{\rm LL/LO}}

\newcommand{\NLLp}{{\rm NLL/NLO^+}}
\newcommand{\NLLpp}{{\rm NLL/NLO^{(+)}}}
\newcommand{\HENLOp}{{\rm HE}\mbox{-}{\rm NLO^+}}

\newcommand{\CnLL}{{\cal C}_n^\LL}

\newcommand{\CnNLLp}{{\cal C}_n^\NLLp}

\newcommand{\CnHENLOp}{{\cal C}_n^{{\rm HE}\text{-}{\rm NLO}^+}}
\newcommand{\DY}{\Delta Y}

\newcommand{\E}{{\cal E}}
\newcommand{\F}{{\cal F}}
\newcommand{\J}{{\cal J}}

\newcommand{\Hb}{{\cal H}_b}
\newcommand{\Jpsi}{J/\psi}

\newcommand{\BCs}{B_c(^1S_0)}
\newcommand{\Bss}{B_c(^3S_1)}
\newcommand{\B}{{\cal B}}

\newcommand{{\Jethad}}{\textsc{Jethad}}
\newcommand{{\Hell}}{\textsc{Hell}}
\newcommand{{\RadISH}}{\textsc{RadISH}}

\begin{document}
\let\WriteBookmarks\relax
\def\floatpagepagefraction{1}
\def\textpagefraction{.001}

\shorttitle{High-energy QCD dynamics from bottom flavor fragmentation at the Hi-Lumi LHC}    

\shortauthors{Celiberto, Francesco Giovanni}  

\title []{\Huge High-energy QCD dynamics from bottom \\ flavor fragmentation at the Hi-Lumi LHC}  

\author[1]{Francesco Giovanni Celiberto}[orcid=0000-0003-3299-2203]


\ead{francesco.celiberto@uah.es}


\affiliation[1]{organization={Universidad de Alcal\'a (UAH), Departamento de F\'isica y Matem\'aticas},
            addressline={Campus Universitario}, 
            city={Alcal\'a de Henares},
            postcode={E-28805}, 
            state={Madrid},
            country={Spain}}




\begin{abstract}
We study the inclusive production of hadrons with bottom flavor at the LHC and its luminosity upgrade.
We describe the collinear fragmentation of singly $b$-flavored hadrons, $B$ mesons and $\Lambda_b$ baryons, via the {\tt KKSS07} determination of fragmentation functions, while for charmed $B$ mesons, $\BCs$ and $\Bss$ particles, we employ the novel {\tt ZCFW22} set, built on the basis of state-of-the-art nonrelativistic QCD inputs.
We use the {\Jethad} multimodular working environment to analyze rapidity and transverse-momentum distributions for observables sensitive to the associated emission of two hadrons or a hadron-plus-jet system.
Our reference formalism is the $\NLLp$ hybrid collinear and high-energy factorization, where the standard collinear description is improved by the inclusions of energy logarithms resummed up to the next-to-leading approximation and beyond.
We provide corroborating evidence that $b$-flavor emissions act as fair stabilizers of the high-energy resummation, thus serving as valuable tools for precision studies of high-energy QCD.
As a bonus, we highlight that the predicted production-rate hierarchy between noncharmed $b$-hadrons and charmed $\BCs$ mesons is in line with recent LHCb estimates.
This serves as simultaneous benchmark both for the hybrid factorization and for the single-parton fragmentation mechanism based on initial-scale inputs calculated via the nonrelativistic QCD effective theory.
\end{abstract}



\begin{keywords}
 Bottom flavor \sep 
 Hi-Lumi LHC \sep 
 Precision studies \sep 
 QCD resummation \sep 
 Natural stability
\end{keywords}

\maketitle

\newcounter{appcnt}


\tableofcontents
\clearpage

\setlength{\parskip}{3pt}%

\section{Opening remarks}
\label{sec:introduction}

Exploring the heavy-flavor domain stands as a frontier research field whereby addressing unresolved quests of particle physics. 
Here, possible experimental evidences of the associated production of heavy quarks and \ac{BSM} particles are milestones in the search for long-awaited signals of New Physics. 
Still they can play a crucial role in precision studies of strong interactions, the values of charm- and bottom-quark masses making perturbative \ac{QCD} applicable.

Special emphasis deserves the hadroproduction of the bottom quark, which represents the heaviest quark species that can fragment and generate hadrons. 
A rigorous description of $(b \bar{b})$ pair production by the hands of collinear factorization and within the \ac{NLO} perturbative accuracy was formalized at the end of the eighties~\cite{Nason:1987xz,Nason:1989zy,Beenakker:1988bq,Frixione:1994nb}, while detailed analyses on differential distributions at \ac{NNLO} have been undertaken only recently~\cite{Catani:2020kkl,Mazzitelli:2023znt}.
A key actor here is the bottom mass, $m_b$.
It genuinely marks the transition point between two distinct factorization schemes.

For $b$-flavored particles emitted with low transverse momentum, so that $m_b \gtrsim |\vec q|$, a suitable description is found in the so-called Fixed Flavor Number Scheme (FFNS, see Ref.~\cite{Alekhin:2009ni} for insight).
It exclusively incorporates collinear \ac{PDFs} and/or \ac{FFs} of light partons (light-flavored quarks and the gluon).
To be precise, here we refer to a scheme with a number of active flavors $n_f = n_l$, with $n_l$ being the number of light flavors.\footnote{We remark that also a scheme with $n_f = n_l + 1$ would still be a FFNS, provided that the (first) heavy-quark species is considered light.}
Additionally, in the FFNS, heavy quarks manifest solely in the final state and their masses must not be set to zero.
This is a required condition not to face perturbative-convergence issues due to the rise of power corrections of the form $|\vec q|/m_b$.

Conversely, when the transverse momentum is large, $ m_b \ll |\vec q|$, contributions scaling with $\ln (|\vec q|/m_b)$ progressively grow and demand for an all-order logarithmic resummation of collinear logarithms of the factorization scale over the mass of the heavy quark(s)~\cite{Mele:1990cw,Cacciari:1993mq}.
In that case, the Zero-Mass Variable-Flavor Number Scheme (ZM-VFNS, or simply VFNS) emerges as the most suitable framework whereby combining fixed-order radiative corrections with the resummation~\cite{Buza:1996wv,Cacciari:1993mq,Binnewies:1997xq,Bierenbaum:2009mv,Helenius:2023wkn,Ghira:2023bxr}. Within the VFNS, all quark flavors are treated as massless particles and participate to the \ac{DGLAP} evolution.
The flavor number increases by one every time a heavy-quark threshold is crossed.

An elegant matching between the FFNS scheme and the VFNS one is realized by means of the so-called \ac{GM-VFNS}. 
It embodies calculations involving both massive quarks at lower scales and massless quarks at higher scales, while the heavy-quark masses serve as control parameters for the transition between the two descriptions.
Several implementations of the GM-VFNS have been proposed so far~\cite{Kramer:2000hn,Forte:2010ta,Blumlein:2018jfm,Aivazis:1993pi,Thorne:1997ga}.
For an exhaustive exploration of these studies, we direct the reader to Refs.~\cite{Gao:2017yyd,Buza:1996wv,Bierenbaum:2009mv}.

While the use of a GM-VFNS undoubtedly improves the description of both the kinematic limits, it still suffers from some ambiguities.
On one hand, it systematically misses power corrections rising from initial-state quarks, which might be quenched, however, by the smallness of heavy-flavor PDFs.
On the other hand, it does not behave as desired in the intermediate region, $ m_b \sim |\vec q|$, since the difference between the pure ZM-VFNS contribution and its massless limit does not automatically vanish and it must be suppressed by hand (see Ref.~\cite{Forte:2010ta} for a detailed discussion).

A simplest solution comes instead from the use of a ZM-VFNS with a suitable choice of the \ac{HFMP}, namely the transition scale starting from which the massive flavor is treated as if it were massless.
As extensively shown in Ref.~\cite{Bertone:2017djs}, numerically accurate predictions can be achieved with HFMP(s) ranging from five to 10 times the mass of the heavy quark(s).
The main benefits of this choice are: the possibility of keeping the size of missing power corrections steadily below $4\div1\%$, a fully preserved resummation accuracy, and an improved continuity in predictions for observables.

Notably, bottom-flavored emissions are useful not only to study direct-production channels, but also to access the top sector.
For this reason, mechanisms governing $b$-quark fragmentation are believed to play an important role within the top-physics domain.
The production of $b$-particles at hadron colliders in NNLO QCD and its application to $(t \bar{t})$ events with leptonic decays were extensively investigated in Refs.~\cite{Czakon:2021ohs,Czakon:2022pyz}.
The same approach finds application in precision studies of electroweak interactions, particularly in the analysis of photon radiation originating from massive charged fermions, like via $(b \bar{b})$ decays from a Higgs boson~\cite{Buonocore:2017lry}. 
Refs.~\cite{Maltoni:2012pa,Bagnaschi:2018dnh,Lim:2016wjo} investigate the role of the bottom quark in the associated production of lepton pairs. 
The semi-inclusive emission of a Higgs boson decaying into $b$-quarks accompanied by a weak vector within NNLO QCD was considered in Ref.~\cite{Gauld:2019yng}.
Ref.~\cite{Barze:2012tt} examines electroweak emissions from heavy fermions in $W^\pm$ boson production processes with Monte Carlo techniques.

The total cross section for the production of a Higgs boson or a $Z$ boson in bottom fusion and via the FONLL~\cite{Cacciari:1998it,Cacciari:2012ny} matching procedure\footnote{See Ref.~\cite{Forte:2010ta} for a FONLL application to \ac{DIS} structure functions.} was computed in Refs.~\cite{Forte:2015hba,Forte:2016sja} and~\cite{Forte:2018ovl}, respectively.
The associated production of an electroweak boson with $b$-quarks was investigated in Refs.~\cite{Frederix:2011qg,Wiesemann:2014ioa}.
Effects of collinear logarithms and finite bottom-mass powers in $(b \bar{b} H)$ tags were addressed in Refs.~\cite{Bonvini:2015pxa,Bonvini:2016fgf}.
Refs.~\cite{Ridolfi:2019bch,Maltoni:2022bpy} contain a detailed study on the inclusion of the resummation of soft logarithms at NNLO in the perturbative component of the bottom fragmentation function.
Matching issues arising from the the noncommutativity of the light-mass limit and soft-emission one were recently highlighted~\cite{Gaggero:2022hmv}.
Analyses on heavy-quark fragmentation in lepton collisions at NNLO plus logarithmic corrections were presented in Refs.~\cite{Bonino:2023vuz,Bonino:2023icn}.
QCD corrections to the hadroproduction of $(t \bar{t} b \bar{b})$ systems were considered in Refs.~\cite{Bevilacqua:2009zn,Cascioli:2013era,Garzelli:2014aba,Bevilacqua:2017cru,Bevilacqua:2022twl}.

Studies on bottom-flavored jets and hadrons at high energies can be found in Refs.~\cite{Chachamis:2013bwa,Chachamis:2015ona,Karpishkov:2016eaj,Karpishkov:2017kph}.
Ref.~\cite{Maciula:2010yw} investigates kinematic correlations of lepton pairs from semi-leptonic decays of $B$ mesons.
Multiple parton interactions in double $B^+$ productions at the LHC were considered in Ref~\cite{Maciula:2018mig}.
The VFNS collinear fragmentation to $B$-mesons was first studied at NLO in Ref.~\cite{Binnewies:1998vm}. 
The obtained FFs came out from a fitting procedure on LEP1 data. A subsequent fit to data from LEP1 and SLAC-SLC was made~\cite{Kniehl:2008zza}. 
The resulted FF parametrization was then employed in the computation of NLO cross section for the inclusive hadroproduction of $B$~mesons within the GM-VFNS. 
There, $b$-quark finite-mass effects were also assessed~\cite{Kniehl:2011bk}.

Predictions for the semi-inclusive emission of singly $b$-flavored particles, encompassing $B$ mesons and $\Lambda_b$ baryons which we collectively denote as $\Hb$ hadrons, were recently compared with LHCb and CMS data~\cite{Kramer:2018vde}.
That analysis relied upon the assumption that a single set of functions could aptly describe the VFNS fragmentation all species of $\Hb$ particles. 
Consequently, FFs for any singly bottomed hadron species might be derived from the general ones by just multiplying it by an energy-independent branching fragmentation fraction.

However, analyses conducted by the \ac{HFAG} highlighted a departure from the universality assumption regarding the branching fraction, particularly in the case of $\Lambda_b$ tags, as evidenced by data from LEP and Tevatron~\cite{Amhis:2016xyh}. 
On the contrary, the universality seems to be valid of $B$-meson emissions. 
A recent examination of semi-inclusive $\Lambda_b$ rates at LHCb and CMS emphasized the need to go beyond the branching-fraction description when the observed transverse momentum enlarges~\cite{Kramer:2018rgb}. 
Therefore, for the time being the use of the FF determination of Ref.~\cite{Kniehl:2008zza,Kramer:2018vde} is safe provided that only $B$ mesons or both $B$ mesons and $\Lambda_b$ baryons are inclusively considered, but it is not valid for $\Lambda_b$ baryons only.
Forthcoming data with diminished experimental uncertainties will contribute to a better understanding of this aspect.

Particular relevance in accessing the heavy-flavor sector deserves the case of mesons with both the charm and the bottom quark in their leading Fock level: $|c \bar b\rangle$ for positive-charged particles, $|\bar c b\rangle$ for negative ones.
Because of the presence of two heavy quarks, these charmed $B$ mesons can be thought of as generalized quarkonium states.
However, contrarily to charmonia and bottomonia, they cannot annihilate into gluons.
Thus, they are quite stable and exhibit narrow decay widths (for novel determinations of the $B_c$ lifetime, see Refs.~\cite{Alonso:2016oyd,Aebischer:2021eio,Aebischer:2021ilm}).
Being top quarks extremely short-lived and not able to hadronize, charmed $B$ mesons actually represent the final frontier of meson spectroscopy (see, \emph{e.g.} Ref.~\cite{Ortega:2020uvc}). 
The CDF Collaboration at Tevatron observed the $B_c \equiv \BCs$ particle for the first time in 1998~\cite{CDF:1998ihx}.
The $B_c^* \equiv \Bss$ resonance was detected by ATLAS only in 2014~\cite{ATLAS:2014lga}.

The simultaneous presence of a charm and a bottom quark in their leading Fock level makes charmed $B$ mesons excellent testing grounds to shed light on the core nature of strong and weak interactions. 
Analyses on direct emissions as well as on indirect productions from electroweak decays stand as gold-plated channels whereby testing $B_c^{(*)}$ formation mechanisms as well as unraveling peculiar features of their decay products.
As an example, they are useful to investigate rare Higgs decays~\cite{Karyasov:2016hfm,Liao:2018nab,Jiang:2015pah}.

Data analyzed by LHCb have shown that $B_c$ meson production rates are three orders of magnitudes lower than $\Hb$ states, at most~\cite{LHCb:2014iah,LHCb:2016qpe}.
This explains why charmed $B$-meson channels are generally excluded from fits of $b$-hadron fragmentation parameters, thus making our knowledge of their formation mechanism quite scarce.
For this reason, any attempt at catching the core fragmentation dynamics of $B_c^{(*)}$ particles still needs to rely upon model-dependent studies and/or effective approaches.

An intriguing opportunity to access the production mechanism of charmed $B$ mesons comes from an effective theory, known as \ac{NRQCD}~\cite{Caswell:1985ui,Thacker:1990bm,Bodwin:1994jh}.
Originally developed to explain issues in the description of charmonium and bottomonium production, the NRQCD framework builds on the assumption that, to the observed bound state, all the possible Fock states contribute through a linear combination. 
All these contributions are ordered in a double expansion in powers of the strong coupling, $\alpha_s$, and the nonrelativistic relative velocity between two constituent heavy quarks, $v$.

The question whether a NRQCD-inspired description is valid for the perturbative component of the collinear fragmentation of a single parton to the detected $B_c^{(*)}$ meson, or it should be rather adopted for the \emph{short-distance} production of a charm-plus-bottom system directly in the hard subprocess, strictly depends on the transverse momenta at play.
Indeed, while the short-distance mechanism suitably describes the production of the two constituent heavy quarks at low transverse momentum, namely when they feature a relative transverse separation of order $1/|\vec q|$,
when $|\vec q|$ grows the fragmentation of a single parton, followed by its inclusive hadronization into the physical state, starts to prevail.
Phenomenological efforts to unravel the transition region between the two regimes were originally made in the context of charmonia~\cite{Cacciari:1994dr,Roy:1994ie,Cacciari:1995yt,Cacciari:1996dg,Lansberg:2019adr}.
The main outcome was that the (gluon) fragmentation contribution starts to dominate when $|\vec q| \gtrsim 10 \div 15$~GeV.
An analogous threshold was then set also for charmed $B$-mesons~\cite{Kolodziej:1995nv}, whereas more recent analyses~\cite{Artoisenet:2007xi} pointed out how this lower bound might be larger.

Although NRQCD can model the initial-scale input of the $B_c^{(*)}$ fragmentation, the other key ingredient for a proper collinear description is the DGLAP evolution.
Thus, by starting from NLO NRQCD calculations of heavy quarks~\cite{Zheng:2019gnb} and gluon~\cite{Zheng:2021sdo} fragmentation channels, first sets of VFNS-compatible, DGLAP-evolving FFs for $\BCs$ and $\Bss$ mesons were derived in Ref.~\cite{Celiberto:2022keu}.
They were named {\tt ZCFW22} NLO functions.
A similar strategy was applied to obtain the {\tt ZCW19$^+$} NLO FFs for vector quarkonia~\cite{Celiberto:2022dyf,Celiberto:2023fzz}.

In this article we will study the inclusive emission, at the LHC ad its luminosity upgrade, of a $\Hb$ hadron or a $B_c^{(*)}$ meson accompanied by another $b$-particle or by a light-flavored jet, as a testing ground for the manifestation of imprints of high-energy QCD dynamics.
We will make use of the hybrid collinear and high-energy factorization framework~\cite{Colferai:2010wu,Celiberto:2020wpk,Celiberto:2020tmb,Bolognino:2021mrc,Celiberto:2022rfj,Celiberto:2022dyf}, as implemented in the {\Jethad} code~\cite{Celiberto:2020wpk,Celiberto:2022rfj,Celiberto:2023fzz} to analyze rapidity-interval and double transverse-momentum differential rates for observables sensitive to the QCD resummation of high-energy logarithms within the next-to-leading approximation and beyond.

We will hunt for stabilizing effects of our distributions under radiative corrections and energy-scale variations, discovering that they are present and allow for a consistent description of these observables at LHC energies and at the natural scales provided by process kinematics.
We will come out with a bonus result, namely that the aforementioned production-rate hierarchy between charmed $\BCs$ mesons and noncharmed $b$-hadrons is in fair agreement with recent LHCb findings~\cite{LHCb:2014iah,LHCb:2016qpe}.
This will serve as simultaneous benchmark both for our resummed calculations and for the single-parton fragmentation mechanism from NRQCD.

For completeness, we mention Refs.~\cite{Deak:2009xt,Deak:2018obv,vanHameren:2022mtk}, where another hybrid-factorization approach, designed for single forward detections, was presented.
Then, Refs.~\cite{Bonvini:2018ixe,Silvetti:2022hyc} investigate small-$x$ resummed inclusive or differential Higgs and heavy-quark LHC rates from the {\Hell} method~\cite{Bonvini:2016wki,Bonvini:2017ogt,Bonvini:2018iwt}.
The latter builds on the basis of the \ac{ABF} approach~\cite{Ball:1995vc,Ball:1997vf,Altarelli:2001ji,Altarelli:2003hk,Altarelli:2005ni,Altarelli:2008aj}, whose purpose is connecting collinear factorization and small-$x$ resummation. It makes use of high-energy factorization theorems~\cite{Catani:1990xk,Catani:1990eg,Collins:1991ty,Catani:1994sq,Ball:2007ra,Caola:2010kv}.

The structure of this article reads as follows. 
Section~\ref{sec:b-flavor} gives introductory remarks and technical details on $\NLLpp$ hybrid collinear and high-energy factorization applied to $b$-flavor production.
Section~\ref{sec:phenomenology} contains phenomenological analyses on our high-energy distributions.
Section~\ref{sec:conclusions} summarizes results and provides outlooks.

\section{Bottom flavor production within $\NLLp$ hybrid factorization}
\label{sec:b-flavor}

The first part of this Section (\ref{ssec:HE_resummation}) consists in a brief summary of recent phenomenological progresses concerning the QCD semi-hard regime. 
Then (\ref{ssec:nll_cross_section}) the $\NLLp$ hybrid collinear and high-energy is explained and adapted to the study of inclusive double $b$-hadron and $b$-hadron plus light-jet final states at the Hi-Lumi LHC.
The third part (\ref{ssec:collinear_inputs}) provides us with details on collinear ingredient.
Finally in (\ref{ssec:natural_stabilization}) the \emph{natural stability}~\cite{Celiberto:2022grc} of $b$-flavor sensitive distributions is discussed.

\subsection{High-energy resummation at work}
\label{ssec:HE_resummation}

The theoretical description of hadron collisions at high energies relies upon the well-defined \emph{collinear factorization}, where long-distance and short-distance dynamics are decoupled. This remarkable features permits to factorize the nonperturbartive hadronic content from the perturbatively calculable information on parton scatterings.
Nevertheless, there exist particular kinematic limits where the purely collinear description is spoiled by the emergence of large logarithmic corrections. Their impact on cross sections can be enough sizable to offset the narrowness of the QCD running coupling, $\alpha_s$, thus harming the perturbative series convergence. In such cases the collinear factorization must be improved via the inclusion to all orders of one or more logarithmic \emph{resummations}.

The kinematic sector matter of our analysis is the \textit{semi-hard} regime (see Refs.~\cite{Celiberto:2017ius,Bolognino:2021bjd} for novel applications), where the energy scale hierarchy $\sqrt{s} \gg \{Q_i\} \gg \LQCD$ is strongly valid. Here, $\sqrt{s}$ stands for the center-of-mass energy, $\{Q_i\}$ represents one or a set of reaction-related hard scales, and $\LQCD$ is the QCD hadronization scale. Whereas the second inequality simply states that perturbative QCD can be employed in our calculations, the first one heralds the fact that we have accessed the QCD \emph{Regge limit}. Here, large $\ln (s/Q_i^2)$ type logarithms enter the perturbative series with a power increasing with the order of $\alpha_s$. Thus, any pure fixed-order perturbative techniques are not anymore applicable. As anticipated, they need to be enhanced by considering the all-order resummation of terms proportional to energy logarithms. 

The \ac{BFKL} approach~\cite{Fadin:1975cb,Kuraev:1977fs,Balitsky:1978ic} is the well-established tool to resum terms proportional to $(\alpha_s \ln s )^n$, namely the ones constituting the \ac{LL} approximation, and also factors proportional to $\alpha_s(\alpha_s \ln s)^n$, namely the ones specific of the \ac{NLL} approximation. Following the BFKL factorization method, any high-energy amplitude can be rewritten as a transverse-momentum convolution of a universal Green's function with two singly off-shell \emph{emission functions} (also known as \emph{impact factors}) describing the subprocess that leads, in the corresponding fragmentation region, to the emission of a forward or backward final-state particle from the stemming hadron. The Green’s function evolves according to the BFKL integral equation, whose kernel has been calculated within the NLO fixed-order accuracy. Recent studies have been done to calculate its NNLO correction~\cite{Caola:2021izf,Falcioni:2021dgr,DelDuca:2021vjq,Byrne:2022wzk,Byrne:2023nqx} (see also Ref.~\cite{Fadin:2023roz}).

While the BFKL kernel is known at NLO, only a few emission functions have been computed within the same accuracy, such as: $(i)$ the ones for collinear incoming partons~\cite{Fadin:1999de,Fadin:1999df},
which are the common basis for $(ii)$ forward light hadron~\cite{Ivanov:2012iv} and jet~\cite{Bartels:2001ge,Ivanov:2012ms,Colferai:2015zfa} emissions, and $(iii)$ the emission function for forward Higgs in the large top-mass range~\cite{Hentschinski:2020tbi,Celiberto:2022fgx}. Conversely, emission functions for forward Drell--Yan~\cite{Motyka:2014lya,Brzeminski:2016lwh,Motyka:2016lta,Celiberto:2018muu}, heavy-quark pair~\cite{Celiberto:2017nyx,Bolognino:2019ccd,Bolognino:2019yls}, and $\Jpsi$ states at small transverse momentum in the short-distance case~\cite{Boussarie:2017oae} are currently known at LO only.
Those impact factors served as key ingredients for phenomenological advancements in our knowledge of high-energy QCD through the study of semi-inclusive final states characterized by the detection of a forward particle always accompanied by a backward one. Here is an incomplete list: production of two jets at the LHC with large transverse momenta and strongly separated in rapidity (Mueller--Navelet reaction~\cite{Mueller:1986ey}, for relevant applications see Refs.~\cite{Colferai:2010wu,Ducloue:2013hia,Ducloue:2013bva,Caporale:2014gpa,Colferai:2015zfa,Celiberto:2015yba,Celiberto:2015mpa,Celiberto:2016ygs,Caporale:2018qnm,deLeon:2021ecb,Celiberto:2022gji}), promptly complemented by the investigation of the double light hadron channel~\cite{Celiberto:2016hae,Celiberto:2017ptm}, multi-jet states~\cite{Caporale:2015int,Caporale:2016soq,Caporale:2016xku,Celiberto:2016vhn,Caporale:2016zkc}, hadron plus jet~\cite{Bolognino:2018oth,Bolognino:2019cac,Bolognino:2019yqj,Celiberto:2020wpk,Celiberto:2020rxb,Celiberto:2022kxx}, Higgs plus jet~\cite{Celiberto:2020tmb}, heavy-light two-jet~\cite{Bolognino:2021mrc}, forward Drell–Yan plus jet~\cite{Golec-Biernat:2018kem}, and hadron with heavy flavor~\cite{Boussarie:2017oae,Bolognino:2019ouc,Bolognino:2019yls,Celiberto:2021dzy,Celiberto:2021fdp,Celiberto:2022dyf,Celiberto:2022grc,Bolognino:2022paj,Celiberto:2022keu,Celiberto:2022zdg} hadroproductions.

Notably, single forward production processes gives us a direct channel to the gluon content of the proton at small-$x$ by means of the \ac{UGD}, whose evolution is driven by the BFKL kernel. Exploratory, model-dependent studies of the UGD have been carried out via the exclusive $\rho$- and $\phi$-meson electroproduction at HERA~\cite{Anikin:2011sa,Besse:2013muy,Bolognino:2018rhb,Bolognino:2018mlw,Bolognino:2019bko,Bolognino:2019pba,Celiberto:2019slj,Luszczak:2022fkf} and the EIC~\cite{Bolognino:2021niq,Bolognino:2021gjm,Bolognino:2022uty,Celiberto:2022fam,Bolognino:2022ndh} and the exclusive photoproduction of vector quarkonia~\cite{Bautista:2016xnp,Garcia:2019tne,Hentschinski:2020yfm}.
Moreover, the information on the gluon content at small-$x$ embodied in the UGD was the starting point to improve the collinear-factorization picture in the context of prime extractions of small-$x$ resummed PDFs~\cite{Ball:2017otu,Abdolmaleki:2018jln,Bonvini:2019wxf}, as well as to determine small-$x$ enhanced unpolarized and polarized gluon TMD distributions~\cite{Bacchetta:2020vty,Bacchetta:2024fci,Celiberto:2021zww,Bacchetta:2021oht,Bacchetta:2021lvw,Bacchetta:2021twk,Bacchetta:2022esb,Bacchetta:2022crh,Bacchetta:2022nyv,Celiberto:2022omz,Bacchetta:2023zir}. Work done in Ref.~\cite{Hentschinski:2021lsh} sheds light on the connection existing between TMD and BFKL dynamics, whereas Refs.~\cite{Boroun:2023goy,Boroun:2023ldq} investigate the link between the UGD and color-dipole cross sections.

A crucial problem rising from the BFKL description of Mueller--Navelet rapidity distributions and azimuthal-angle correlations is the weight of the NLL part, which is of the same order, but of opposite sign than the pure LL case. This is at the origin of strong instabilities observed in the resummed series that become manifest when factorization and renormalization scales are varied around their natural values.
Therefore, Muller--Navelet observables assume unphysical values when the jet rapidity separation becomes sufficiently large.
In the same way, azimuthal correlations exhibit an unphysical behavior both at small and at large rapidity distances.
Different solutions have been tried to cure this problem.
In particular, the \ac{BLM} prescription~\cite{Brodsky:1996sg,Brodsky:2002ka} as specifically built for semi-hard reactions~\cite{Caporale:2015uva} slightly dampens those instabilities on azimuthal correlations and thus averagely ameliorates the agreement with data. 
However, BLM is almost ineffectual on light di-hadron or hadron plus jet semi-hard distributions. 
The main reason is that the optimal values for renormalization scales prescribed by BLM are much higher than the process natural scales~\cite{Celiberto:2017ius,Bolognino:2018oth,Celiberto:2020wpk}. 
Thus, total cross sections suffer from a large loss of statistics.

Fair signals of a stabilization of the high-energy resummation under NLL corrections and energy-scale variations recently emerged from semi-hard reactions whose final states show the signature of Higgs bosons~\cite{Celiberto:2020tmb,Mohammed:2022gbk,Celiberto:2023rtu,Celiberto:2023uuk,Celiberto:2023eba,Celiberto:2023nym,Celiberto:2023rqp}.
Then, a fair stabilizing trend came out from the study of $\Lambda_c$ hyperons at the LHC~\cite{Celiberto:2021dzy}, and also from analogous observables sensitive to $B$ mesons and $\Lambda_b$ baryons~\cite{Celiberto:2021fdp}.
In those works it was proven that the stabilizing effect is encoded in the peculiar pattern of VFNS collinear FFs portraying the production of these singly heavy-flavored particles at high transverse momentum. Similar results were observed when vector quarkonia~\cite{Celiberto:2022dyf,Celiberto:2023fzz}, charmed $B$ mesons~\cite{Celiberto:2022keu}, or heavy-light tetraquarks~\cite{Celiberto:2023rzw} are emitted in semi-hard configurations.
All these results gave us corroborating evidence that the found \emph{natural stabilization} of BFKL emerges as an \emph{intrinsic} property genuinely correlated with final states sensitive to heavy flavor~\cite{Celiberto:2022grc}.

\subsection{NLL-resummed differential cross section}
\label{ssec:nll_cross_section}

\begin{figure*}[!t]
\centering

\includegraphics[width=0.475\textwidth]{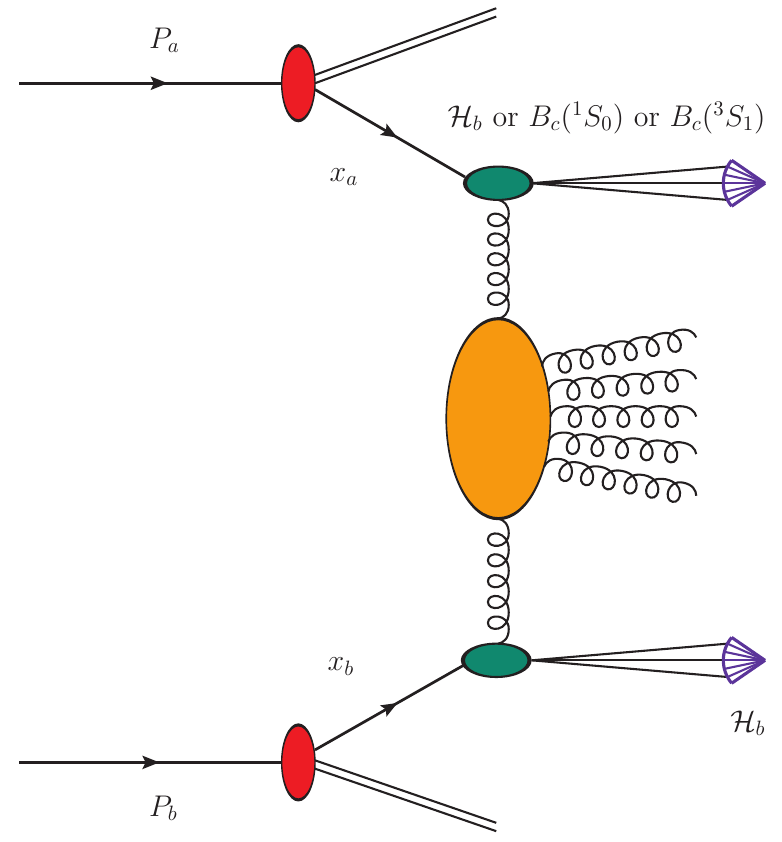}
   \hspace{0.50cm}
\includegraphics[width=0.475\textwidth]{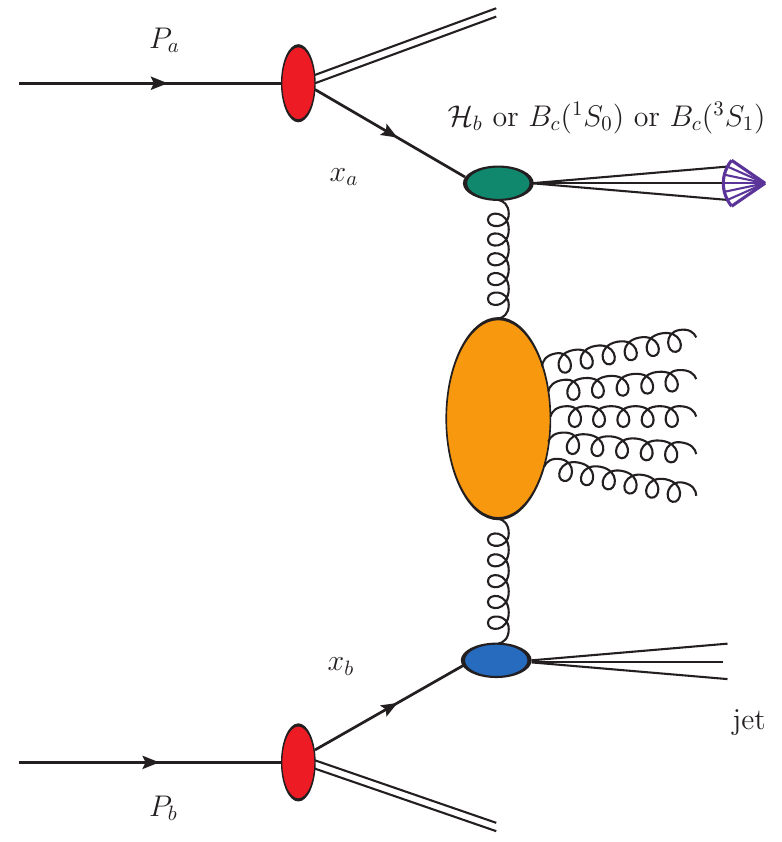}

\caption{Hybrid collinear and high-energy factorization for double-hadron detection (left) and hadron-plus-jet inclusive hadroproduction (right). Collinear parton densities are represented by red ovals. The green (blue) blob depicts the off-shell hard factor embodied in the hadron (jet) emission function, while indigo arrows depicts the emission of $b$-flavored hadrons. The Green's function is described by the orange blob. The diagram was created with {\tt JaxoDraw 2.0}~\cite{Binosi:2008ig}.}
\label{fig:process}
\end{figure*}

We will study the two kinds of processes depicted in Fig.~\ref{fig:process}
\begin{equation}
\label{process}
\setlength{\jot}{10pt} 
\begin{split}
    {\rm p}(P_a) \;+\; {\rm p}(P_b) &\;\rightarrow\; \B_1(q_1, y_1, \phi_1) \;+\; {\cal X} \;+\; \B_2(q_2, y_2, \phi_2) \; ,
    \\
    {\rm p}(P_a) \;+\; {\rm p}(P_b) &\;\rightarrow\; \B(q_1, y_1, \phi_1) \;+\; {\cal X} \;+\; {\rm jet}(q_2, y_2, \phi_2) \; ,
\end{split}
\end{equation}
where ${\rm p}(P_{a,b})$ is an initial-state proton possessing four-momentum $P_{a,b}$, from which a parton with longitudinal-momentum fraction $x_{a,b}$ is struck.
Then, $\B_{(1,2)}(q_{1,2}, y_{1,2})$ stands for a singly bottom-flavored hadron $\Hb \equiv \{B + \Lambda_b\}$, or a charmed-bottomed meson $\BCs \equiv B_c$, or its resonance $\Bss \equiv B_c^*$, emitted with four-momentum $q_{1,2}$, rapidity $y_{1,2}$ and azimuthal angle $\phi_{1,2}$, and a light-favored jet is possibly tagged with four-momentum $q_2$, rapidity $y_2$ and azimuthal angle $\phi_2$. Furthermore, ${\cal X}$ represents an inclusively radiated gluon system. The large final state transverse momenta, $|\vec q_{1,2}|$, together with the high rapidity interval, $\DY \equiv y_1 - y_2$, are required ingredients to make the final state be both semi-hard and diffractive. 
In addition, transverse momenta of the observed particles need to be large enough to make the VFNS fragmentation approach be valid.  

Stemming protons' four-momenta can be written as Sudakov vectors with the conditions $P_a^2= P_b^2=0$ and $2 (P_a\cdot P_b) = s$, so that final-state four-momenta are cast via the following decomposition
\begin{equation}\label{sudakov}
q_{1,2} = x_{1,2} P_{a,b} - \frac{q_{1,2\perp}^{\,2}}{x_{1,2} s}P_{b,a} + q_{1,2\perp} \ , \quad
q_{1,2\perp}^2 = - \, \vec q_{1,2}^{\,2}\;.
\end{equation}
The outgoing-object longitudinal-momentum fractions are labeled as $x_{1,2}$.
They are connected with the respective rapidities and to $\DY$ via
\begin{equation}\label{xyDY}
y_{1,2} = \pm \, \frac{1}{2}\ln\frac{s \, x_{1,2}^2}
{\vec q_{1,2}^2} \;, \qquad \drv y_{1,2} = \pm \, \frac{\drv x_{1,2}}{x_{1,2}} \;, \qquad \DY = \ln\frac{s \, x_1 x_2}{|\vec q_1||\vec q_2|} \;.
\end{equation}

According to the pure collinear factorization, one would write the LO cross section for our processes (Eq.~\eqref{process}) as a one-dimensional convolution among the partonic-subprocess factor, proton PDFs, and $b$-hadron FFs.
Considering the double hadron case as depicted in the left diagram of Fig.~\ref{fig:process}, we have
\begin{equation}
\label{sigma_collinear_BB} 
\begin{split}
 \frac{\drv \sigma_{[p + p \,\to\, \B + \B]}^{\rm LO}}{\drv x_1 \drv x_2 \drv^2 \vec q_1 \drv^2 \vec q_2}
 &=\sum_{i,j=-5}^{5} \int_0^1 \drv x_i \int_0^1 \drv x_j\ 
 f_i(x_i, \mu_F) f_j(x_j, \mu_F)
\\ 
 &\times \, \int_{x_1}^1 \frac{\drv \xi_1}{\xi_1} \int_{x_2}^1 \frac{\drv \xi_2}{\xi_2}\
 D^{\B_1}_i\left(\frac{x_1}{\xi_1}, \mu_F\right) D^{\B_2}_j\left(\frac{x_2}{\xi_2}, \mu_F\right)
 \frac{\drv {\hat\sigma}_{ij}}
 {\drv x_i \drv x_j \drv \xi_1 \drv \xi_2 \drv^2 \vec q_1 \drv^2 \vec q_2}\;,
\end{split}
\end{equation}
with the $i$ and $j$ indices running over parton species, $f_{i,j}(x_{1,2}, \mu_F)$ being the proton PDFs, $D^{\B_{1,2}}_{i,j}(x_{1,2}/\xi_{1,2} \equiv z_{1,2}, \mu_F)$ denoting $b$-hadron FFs, $x_{1,2}$ and $\xi_{1,2}$ respectively standing for incoming- and fragmenting-partons' longitudinal fractions, and $\drv \hat\sigma_{ij}$ representing the partonic cross section.
Analogously, we write the following collinear-convolution formula for the $b$-hadron plus light-jet case depicted in the right diagram of Fig.~\ref{fig:process}
\begin{equation}
\label{sigma_collinear_BJ}
\begin{split}
 \frac{\drv \sigma_{[p + p \,\to\, \B + {\rm jet}]}^{\rm LO}}{\drv x_1 \drv x_2 \drv^2 \vec q_1 \drv^2 \vec q_2}
 &=\sum_{i,j=-5}^{5} \int_0^1 \drv x_i \int_0^1 \drv x_j\ 
 f_i(x_i, \mu_F) f_j(x_j, \mu_F)
\int_{x_1}^1 \frac{\drv \xi}{\xi}D^{\B}_i\left(\frac{x_1}{\xi}, \mu_F\right) 
\frac{\drv {\hat\sigma}_{ij}}
{\drv x_1\drv x_2\drv \xi\,\drv ^2\vec q_1\drv ^2\vec q_2}\;.
\end{split}
\end{equation}

At variance with the collinear framework, to write the expression for the high-energy resummed cross section in the hybrid factorization, one first makes use of BFKL and then completes the picture by adding collinear PDFs and FFs.
We suitably rewrite the differential cross section as a Fourier expansion in terms of azimuthal coefficients
\begin{equation}
 \label{dsigma_Fourier}
 \frac{\drv \sigma^\NLLp}{\drv y_1 \drv y_2 \drv \vec q_1 \drv \vec q_2 \drv \phi_1 \drv \phi_2} =
 \frac{1}{2\pi^2} \left[ \frac{1}{2} \, {\cal C}_0^\NLLp + \sum_{n=1}^{+\infty}
 {\cal C}_n^\NLLp \, \cos \left(n (\pi - \Phi)\right) \right] \, ,
\end{equation}
with $\Phi \equiv \phi_1 - \phi_2$.
The ${\cal C}_n^\NLLp$ coefficients are calculated within the BFKL formalism at a full NLL accuracy. Making use of the $\MSb$ renormalization scheme~\cite{PhysRevD.18.3998}, we obtain~\cite{Caporale:2012ih}
\begin{equation}
\label{Cn_NLLq_MSb}
\begin{split}
 \CnNLLp &= \int_0^{2\pi} \drv \phi_1 \int_0^{2\pi} \drv \phi_2\,
 \frac{\drv \sigma^\NLLp}{\drv y_1 \drv y_2\, \drv |\vec q_1| \, \drv |\vec q_2| \drv \phi_1 \drv \phi_2} \, \cos \left(n (\pi - \Phi)\right) \;
\\
 &= \; \frac{e^{\DY}}{s} 
 \int_{-\infty}^{+\infty} \drv \nu \, e^{{\DY} \bar \alpha_s(\mu_R)\chi^\NLO(n,\nu)}
\\
 &\times \; \alpha_s^2(\mu_R) \, 
 \biggl\{
 \E_1^\NLO(n,\nu,|\vec q_1|, x_1) \,[\E_2^\NLO(n,\nu,|\vec q_2|,x_2)]^*\,
 + \,
 \left.
 \bar \alpha_s^2(\mu_R)
 \, \DY
 \frac{\beta_0}{4 N_c}\chi(n,\nu)\F(\nu)
 \right\} \;,
\end{split}
\end{equation}
where $\bar \alpha_s(\mu_R) \equiv \alpha_s(\mu_R) N_c/\pi$ with $N_c$ the color number, $\beta_0 = 11N_c/3 - 2 n_f/3$ is the first coefficient of the QCD $\beta$-function with $n_f$ the flavor number.
A two-loop running-coupling choice with $\alpha_s\left(M_Z\right)=0.118$ and with dynamic thresholds for the flavor number $n_f$ is made.
The kernel entering the exponent of Eq.~\eqref{Cn_NLLq_MSb} encompasses the resummation of LL and NLL terms. Its analytic expression is
\begin{eqnarray}
 \label{chi}
 \chi^\NLO(n,\nu) = \chi(n,\nu) + \bar\alpha_s \hat \chi(n,\nu) \;,
\end{eqnarray}
where
\begin{eqnarray}
 \label{kernel_LO}
 \chi\left(n,\nu\right) = -2\gamma_{\rm E} - 2 \, {\rm Re} \left\{ \psi\left(\frac{n}{2} + \frac{1}{2} + i \nu \right) \right\} \, 
\end{eqnarray}
stand for the LO kernel eigenvalues, $\gamma_{\rm E}$ is the Euler-Mascheroni constant, and $\psi(z) \equiv \Gamma^\prime
(z)/\Gamma(z)$ the Gamma-function logarithmic derivative. 
The $\hat\chi(n,\nu)$ function in Eq.~\eqref{chi} represents the NLO kernel correction in the Mellin space
\begin{equation}
\begin{split}
\label{chi_NLO}
\hat \chi\left(n,\nu\right) &= \bar\chi(n,\nu)+\frac{\beta_0}{8 N_c}\chi(n,\nu)
\left(\frac{10}{3}+\ln\frac{\mu_R^4}{m_{1 \perp}^2 m_{2 \perp}^2}-\chi(n,\nu)\right) \;,
\end{split}
\end{equation}
where $m_{1,2 \perp}$ are the transverse masses of the two outgoing states. 
For a $b$-hadron one has $m_{\B \perp} = \sqrt{m_\B^2 + |\vec q_\B|^2}$, with $m_\B \equiv m_{\Hb} \equiv m_{\Lambda_c} = 5.62$~GeV or $m_\B \equiv m_{B_c^{(*)}} = 6.275$~GeV.
Conversely, the transverse mass of a light-flavored jets can be safely set to transverse momentum, say $m_{\J \perp} \equiv |\vec q_\J|$.
The NLO $\bar\chi(n,\nu)$ function was computed in Ref.~\cite{Kotikov:2000pm}
\begin{equation}
 \label{kernel_NLO}
 \bar \chi(n,\nu)\,=\, - \frac{1}{4}\left\{\frac{\pi^2 - 4}{3}\chi(n,\nu) - 6\zeta(3) - \frac{\drv^2 \chi}{\drv\nu^2} + \,2\,\Xi(n,\nu) + \,2\,\Xi(n,-\nu)
 \right.
\end{equation}
\[
 \left.
 +\; \frac{\pi^2\sinh(\pi\nu)}{2\,\nu\, \cosh^2(\pi\nu)}
 \left[
 \left(3+\left(1+\frac{n_f}{N_c^3}\right)\frac{11+12\nu^2}{16(1+\nu^2)}\right)
 \delta_{n0}
 -\left(1+\frac{n_f}{N_c^3}\right)\frac{1+4\nu^2}{32(1+\nu^2)}\delta_{n2}
\right]\right\} \, .
\]
The $\Xi(n,\nu)$ function reads
\begin{equation}
\label{kernel_NLO_phi}
 \Xi(n,\nu)\,=\,-\int_0^1 \drv x\,\frac{x^{-1/2+i\nu+n/2}}{1+x}\left\{\frac{1}{2}\left(\psi^\prime\left(\frac{n+1}{2}\right)-\zeta(2)\right)+\mbox{Li}_2(x)+\mbox{Li}_2(-x)\right.
\end{equation}
\[
\left.
 +\; \ln x\left[\psi(n+1)-\psi(1)+\ln(1+x)+\sum_{k=1}^\infty\frac{(-x)^k}{k+n}\right]+\sum_{k=1}^\infty\frac{x^k}{(k+n)^2}\left[1-(-1)^k\right]\right\}
\]
\[
 =\; \sum_{k=0}^\infty\frac{(-1)^{k+1}}{k+(n+1)/2+i\nu}\left\{\psi^\prime(k+n+1)-\psi^\prime(k+1)\right.
\]
\[
 \left.
 +\; (-1)^{k+1}\left[\rho(k+n+1)+\rho(k+1)\right]-\frac{\psi(k+n+1)-\psi(k+1)}{k+(n+1)/2+i\nu}\right\} \; ,
\]
with
\begin{equation}
\label{kernel_NLO_phi_beta_psi}
 \rho(z)=\frac{1}{4}\left[\psi^\prime\left(\frac{z+1}{2}\right)
 -\psi^\prime\left(\frac{z}{2}\right)\right] \; ,
\end{equation}
and
\begin{equation}
\label{dilog}
\mbox{Li}_2(z) = - \int_0^z \drv \zeta \,\frac{\ln(1-\zeta)}{\zeta} \; .
\end{equation}

The two quantities
\begin{equation}
\label{IFs}
\E_{1,2}^\NLO(n,\nu,|\vec q_{1,2}|,x_{1,2}) =
\E_{1,2}(n,\nu,|\vec q_{1,2}|,x_{1,2}) +
\alpha_s(\mu_R) \, \hat \E_{1,2}(n,\nu,|\vec q_{1,2}|,x_{1,2})
\end{equation}
stand for the emission functions of the forward $b$-hadron and the light jet.
At LO one has
\begin{equation}
\label{LOBIF}
\begin{split}
\E_\B(n,\nu,|\vec q_\B|,x_\B) 
&= 2 \sqrt{\frac{C_F}{C_A}}
|\vec q_\B|^{2i\nu-1}\,\int_{x_\B}^1\frac{\drv \xi}{\xi}
\left(\frac{\xi}{x_\B} \right)
^{2 i\nu-1} 
 \left[\frac{C_A}{C_F}f_g(\xi)D_g^\B\left(\frac{x_\B}{\xi}\right)
 +\sum_{i=q,\bar q}f_i(\xi)D_i^\B\left(\frac{x_\B}{\xi}\right)\right] 
\end{split}
\end{equation}
and
\begin{equation}
 \label{LOJIF}
 \E_\J(n,\nu,|\vec q_\J|,x_\J) =  2 \sqrt{\frac{C_F}{C_A}}
 |\vec q_\J|^{2i\nu-1}\,\left(\frac{C_A}{C_F}f_g(x_\J)
 +\sum_{j=q,\bar q}f_j(x_\J)\right) \;,
\end{equation}
with $C_F \equiv (N_c^2-1)/(2N_c)$ and $C_A \equiv N_c$ being the Casimir factors for a gluon emission from a quark and a gluon, respectively.
The $\F(\nu)$ function at the end of the last row of Eq.~\eqref{Cn_NLLq_MSb} embodies the logarithmic derivative of LO emission functions
\begin{equation}
 \F(\nu) = \ln\left(|\vec q_1| |\vec q_2|\right) + \frac{i}{2} \, \frac{\drv}{\drv \nu} \ln \frac{\E_1}{\E_2^*} \;.
\label{fnu}
\end{equation}
What is left in Eq.~(\ref{Cn_NLLq_MSb}) are the NLO corrections to $\E_{\B,\J}$ functions, $\hat \E_{1,2}$.
The NLO correction to the $\B$ hadron at large transverse momentum was computed in Ref.~\cite{Ivanov:2012iv} and is provided in the Appendix~\hyperlink{app:NLOBIF}{A}.
Our choice for the jet NLO emission function builds on analyses described in Refs.~\cite{Ivanov:2012iv,Ivanov:2012ms}. It relies upon a jet algorithm obtained in the ``small-cone'' limit
and with a cone-type distance~\cite{Furman:1981kf,Aversa:1988vb,Colferai:2015zfa}, where the radius is set to $r=0.5$ (see the Appendix~\hyperlink{app:NLOJIF}{B}).

A proper ``resummation versus fixed-order'' study should rely upon comparing NLL predictions from our high-energy approach with NLO calculations in pure collinear factorization.
Unfortunately, a numerical technology aimed at computing NLO distributions for two-particle reactions in hadron collisions is still missing.
Thus, for the sake of comparison with reference fixed-order results, one can truncate the expansion of the azimuthal coefficients of Eq.~\eqref{Cn_NLLq_MSb} up to the ${\cal O}(\alpha_s^3)$ order. 
In this way, one obtains an effective high-energy fixed-order ($\HENLOp$) expression that collects the leading-power asymptotic signal present in a pure NLO calculation, whereas terms proportional by inverse powers of the partonic center-of-mass energy are neglected.
The $\MSb$ expressions for $\HENLOp$ azimuthal coefficients reads
\begin{equation}
\label{Cn_HENLO_MSb}
 \CnHENLOp =
 \frac{e^{\DY}}{s}
 \int_{-\infty}^{+\infty} \drv \nu \,
 \alpha_s^2(\mu_R) \,
 \left[ 1 + \bar \alpha_s(\mu_R) \DY \chi(n,\nu) \right] \,
 \E_1^\NLO(n,\nu,|\bm{\kappa}_1|, x_1) \,[\E_2^\NLO(n,\nu,|\bm{\kappa}_2|,x_2)]^*
 \;,
\end{equation}
with the exponentiated kernel expanded and truncated at ${\cal O}(\alpha_s)$.

We will also compare our NLL results with the pure LL background.
The latter is obtained by discarding NLO corrections of both the kernel and the emission functions
\begin{equation}
\label{Cn_LL_MSb}
  \CnLL = \frac{e^{\DY}}{s} 
 \int_{-\infty}^{+\infty} \drv \nu \, e^{{\DY} \bar \alpha_s(\mu_R)\chi(n,\nu)} \, \alpha_s^2(\mu_R) \, \E_1(n,\nu,|\vec q_1|, x_1) \, [\E_2(n,\nu,|\vec q_2|,x_2)]^* \,.
\end{equation}

Eqs.~(\ref{Cn_NLLq_MSb}) to~(\ref{Cn_LL_MSb}) shed light on the structure of our hybrid factorization. According to BFKL, the cross section is high-energy factorized as a convolution between the Green's function and two singly off-shell emission functions. 
The latters embody collinear inputs, namely collinear convolutions between incoming protons' PDFs and outgoing hadrons' FFs.
The $\NLLp$ label tells us that the resummation of energy logarithms is performed within a full NLL accuracy and within NLO perturbative order.
The `$+$' superscript in Eqs.~(\ref{Cn_NLLq_MSb}) and~(\ref{Cn_HENLO_MSb}) highlights that some \ac{NNLL} extra terms come from the cross product between NLO corrections of the two emission functions.
Finally, factorization ($\mu_F$) and renormalization ($\mu_R$) scales are set to the \emph{natural} scales suggested by kinematics. In particular, we fix $\mu_F = \mu_R = \mu_N \equiv m_{1 \perp} + m_{2 \perp}$.

Our guiding line here is to consider, as a \emph{natural} scale, any energy scale with the same order of magnitude of the typical energy of the emitted particle.
Since, in our case, two objects are emitted, there are in principle to scales (one for each fragmentation region).
Natural values for them are $m_{1 \perp}$ and $m_{2 \perp}$, respectively.
For the sake of comparing our observables with future, possible predictions coming from other approaches, it is convenient to make a simple choice, namely to have just one scale.
As a bonus, choosing the sum of $m_{1 \perp}$ and $m_{2 \perp}$ turns out to match the settings of many other numeric codes aimed at precision QCD phenomenology (see, \emph{e.g.}, Refs.~\cite{Alioli:2010xd,Campbell:2012am,Hamilton:2012rf}).

\subsection{Collinear inputs}
\label{ssec:collinear_inputs}

As for collinear PDFs, we make use of the {\tt NNPDF40\_nlo\_as\_01180} NLO set~\cite{NNPDF:2021uiq,NNPDF:2021njg} as implemented in {\tt LHAPDF v6.5.4}~\cite{Buckley:2014ana}.
It was obtained on the basis of global fits by means of the well-defined \emph{replica} methodology originally proposed in Ref.~\cite{Forte:2002fg} in the context of a neural-network approach (we refer to Ref.~\cite{Ball:2021dab} for an in-depth analysis on ambiguities about \emph{correlations} among different PDF sets).

Parton fragmentation to singly-bottomed $\Hb \equiv \{B + \Lambda_b\}$ hadrons is depicted via the {\tt KKSS07} NLO FF determination, which was originally obtained by fitting data on inclusive productions of noncharmed $B$ mesons in lepton annihilation~\cite{Kniehl:2008zza}.
Here, the threshold for bottom- and antibottom-quark evolution is set to 4.5~GeV and its initial-scale parametrization reads as a power-like function with three parameters~\cite{Kartvelishvili:1985ac}.
Light partons and the charm quarks are perturbatively generated with DGLAP, while their FFs are zero below the threshold for the bottom.
The current {\tt KKSS07} set~\cite{Kramer:2018vde} for $\Hb$ particles was constructed from the original noncharmed $B$-meson ones by switching off the branching fraction for the $b \to B^\pm$ subprocess, which was assumed to be $f_u = f_d = 0.397$~\cite{Kniehl:2008zza}. Such a choice is justified by the assumption that a unique function can be employed in the description of noncharmed $B$-meson fragmentation or of an inclusive sum of $B$ and $\Lambda_b$ channels, but not for the $\Lambda_b$-baryon case alone.

Our strategy for charmed $B$-meson fragmentation is more involved. It was presented for the first time in our recent study focused on accessing the high-energy spectrum of QCD via the hadroproduction of these doubly heavy-flavored states~\cite{Celiberto:2022keu}. 
It takes inspiration from a first determination of VFNS, DGLAP-evolving FFs for vector quarkonia~\cite{Celiberto:2022dyf,Celiberto:2023fzz}.
The starting point is the NLO calculation, within the NRQCD effective theory, of initial-scale inputs for charm, bottom quark~\cite{Zheng:2019gnb} and gluon~\cite{Zheng:2021sdo} fragmentation channels to $\BCs \equiv B_c$ and $\Bss \equiv B_c^*$ mesons (see Refs.~\cite{Feng:2021qjm,Feng:2018ulg} for similar calculations).
To generate DGLAP-evolved sets for our charmed $B$ mesons we need to plug the NRQCD inputs into a numeric evolution code.
Among them, we cite: {\tt QCD-PEGASUS}~\cite{Vogt:2004ns}, {\tt HOPPET}~\cite{Salam:2008qg}, {\tt QCDNUM}~\cite{Botje:2010ay}, {\tt APFEL(++)}~\cite{Bertone:2013vaa,Carrazza:2014gfa,Bertone:2017gds}, and {\tt EKO}~\cite{Candido:2022tld,Hekhorn:2023gul}.
Contrariwise to PDFs, which obey a space-like evolution equation, FFs must be time-like evolved~\cite{Curci:1980uw,Furmanski:1980cm}. 
In Ref.~\cite{Celiberto:2022keu} we employed {\tt APFEL++} to generate novel and phenomenology-ready {\tt LHAPDF} FF grids, which we named {\tt ZCFW22} NLO determinations.

For the sake of comparison, we employ the {\tt NNFF10\_KAsum\_nlo} set~\cite{Bertone:2017tyb} (see Ref.~\cite{Bertone:2018ecm} for advancements) to describe charged-kaon fragmentation at NLO. 

\subsection{Natural stabilization of the NLL series}
\label{ssec:natural_stabilization}

\begin{figure*}[!t]
\centering

   \includegraphics[scale=0.475,clip]{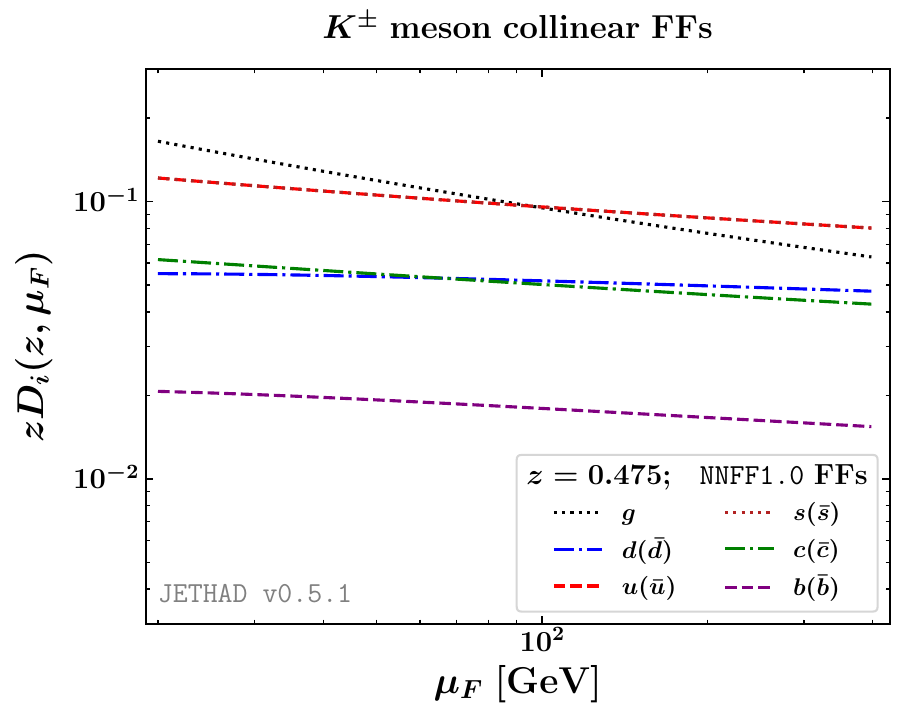}
   \hspace{0.50cm}
   \includegraphics[scale=0.475,clip]{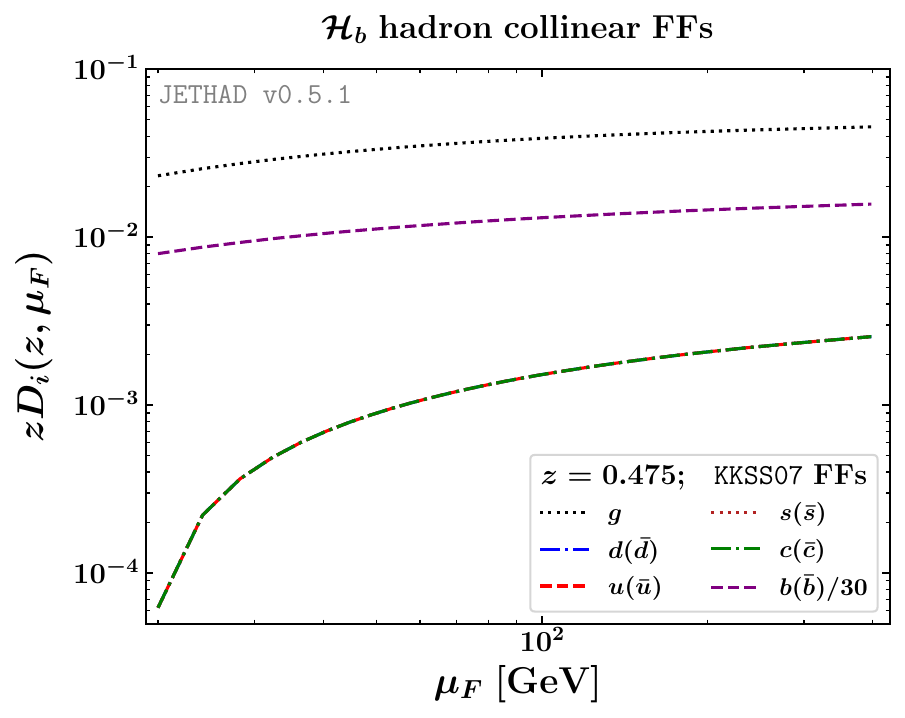}

   \includegraphics[scale=0.475,clip]{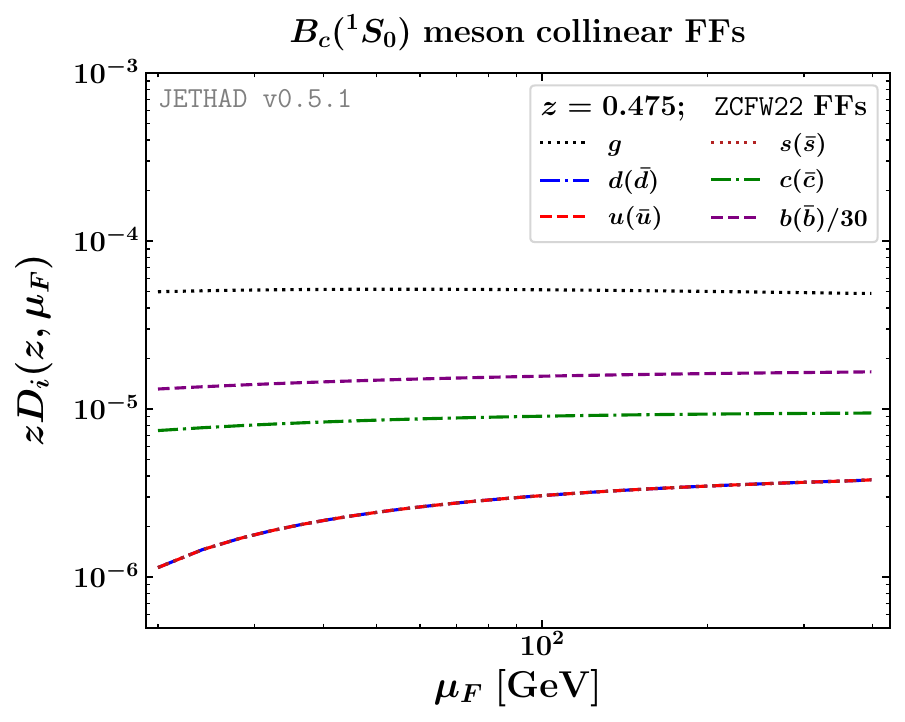}
   \hspace{0.50cm}
   \includegraphics[scale=0.475,clip]{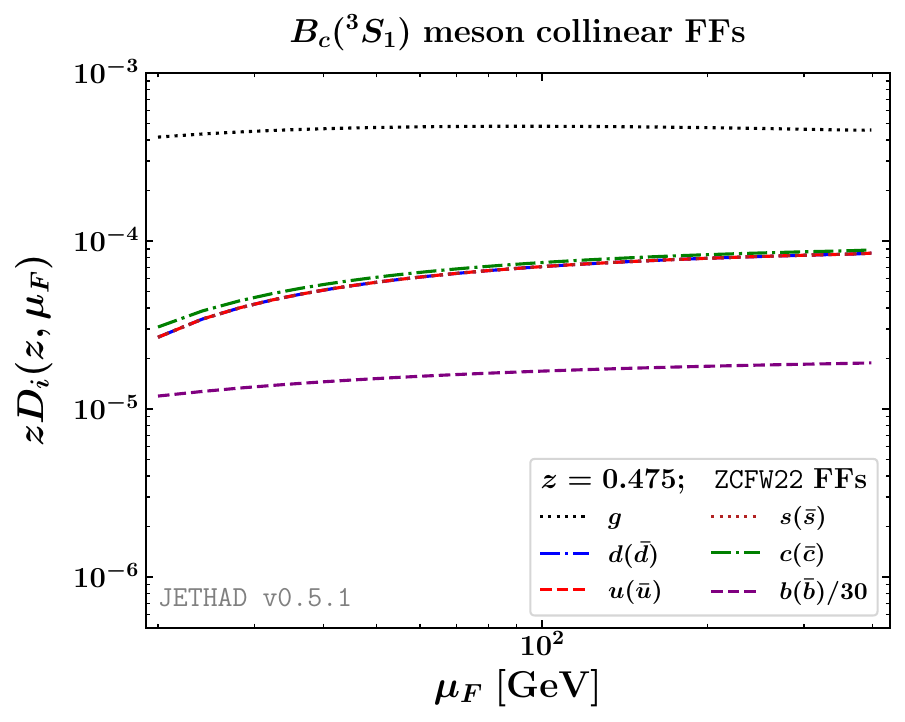}

\caption{Factorization-scale dependence of {\tt NNFF1.0}~\cite{Bertone:2017tyb}, {\tt KKSS07}~\cite{Kniehl:2008zza,Kramer:2018vde}, {\tt and ZCFW22}~\cite{Celiberto:2022keu} NLO FFs respectively describing: kaon (upper left panel), $\Hb$ hadron (upper right panel), and charmed $B$ meson (lower panels) detections, for $\langle z \rangle \simeq 0.475$.}
\label{fig:NLO_FFs}
\end{figure*}

Here we provide details on the stabilizing mechanism emerging from the VFNS collinear fragmentation of $b$-hadrons.
The linking dynamics between heavy-flavor FFs and the observed stability of semi-inclusive distributions calculated within the hybrid collinear and high-energy factorization at $\NLLpp$ was first unveiled through a study of semi-hard emissions of heavy-light hadrons, namely: $D$ particles~\cite{Kniehl:2004fy,Kniehl:2005de,Kniehl:2006mw,Kneesch:2007ey,Corcella:2007tg,Anderle:2017cgl,Salajegheh:2019nea,Salajegheh:2019srg,Soleymaninia:2017xhc}, $\Lambda_c$ baryons~\cite{Kniehl:2005de,Kniehl:2020szu,Delpasand:2020vlb}, and $\Hb$ states~\cite{Binnewies:1998vm,Kniehl:2007erq,Kniehl:2008zza,Kniehl:2011bk,Kniehl:2012mn,Kramer:2018vde,Kramer:2018rgb,Salajegheh:2019ach,Kniehl:2021qep}.
This striking property came out for the first time from an analysis of semi-inclusive detections of $\Lambda_c$~\cite{Celiberto:2021dzy,Celiberto:2022rfj}, $D^{* \pm}$~\cite{Celiberto:2022zdg,Bolognino:2022paj}, and $\Hb$~\cite{Celiberto:2021fdp,Celiberto:2022rfj} hadrons in forward directions of rapidity at the LHC.
Such a remarkable result was then corroborated by similar investigations on vector quarkonia~\cite{Celiberto:2022dyf,Celiberto:2022kza} and $B_c^{(*)}$ mesons~\cite{Celiberto:2022keu}, whose formation mechanism was modeled on the basis of single-parton NRQCD fragmentation~\cite{Braaten:1993mp,Zheng:2019dfk,Braaten:1993rw,Chang:1992bb,Braaten:1993jn,Ma:1994zt,Zheng:2019gnb,Zheng:2021sdo,Feng:2021qjm,Feng:2018ulg}.
A further evidence came recently out in the context of heavy-light tetraquarks~\cite{Celiberto:2023rzw,Celiberto:2024mrq} described via a Suzuki--Nejad--Amiri--Ji initial-scale fragmentation input~\cite{Suzuki:1985up,Nejad:2021mmp,Suzuki:1977km,Amiri:1986zv} (see Refs.~\cite{Lepage:1980fj,Brodsky:1985cr} for further details).
A less pronounced stabilization pattern was also observed in the case of strange-quark flavored, cascade $\Xi^-/\bar\Xi^+$ baryon tags~\cite{Celiberto:2022kxx}.

Panels in Fig.~\ref{fig:NLO_FFs} contain the factorization-scale dependence of {\tt KKSS07}~\cite{Kniehl:2008zza,Kramer:2018vde} and {\tt ZCFW22}~\cite{Celiberto:2022keu} NLO functions respectively describing the collinear fragmentation of $\Hb$ hadrons (right upper plot) and $B_c^{(*)}$ mesons (lower plots).
For comparison, the left upper plot carries analogous information of light-quark flavored species, namely charged kaons, described by the {\tt NNFF1.0}~\cite{Bertone:2017tyb} NLO collinear-FF determination.
For the sake of simplicity, we show values of our FF sets just for one value of the final-state hadron longitudinal-momentum fraction, $z$.
In particular, we choose a value of $z$ that roughly matches the mean value at which FFs are typically probed standard semi-hard configurations (see Section~\ref{ssec:kinematics}).
Thus we have $z = 0.475 \simeq \langle z \rangle$.
As expected, the $b$ channel of {\tt KKSS07} and {\tt ZCFW22} functions heavily prevails over the other flavors.
Particular attention deserves, however, the behavior the gluon FF, which clearly decreases with $\mu_F$ for kaons but smoothly increases for all the three $b$-flavored species, up to reach a plateau.
More quantitative information on the FFs of Fig.~\ref{fig:NLO_FFs} would come from an analysis on the associated uncertainties.
However, apart the kaon {\tt NNFF1.0} determination, the other ones do not carry such information.
{\tt KKSS07} functions describing noncharmed $b$-hadron fragmentation were extracted without any uncertainty estimate, whereas the inclusion of uncertainties for the two {\tt ZCFW22} sets, possibly connected with perturbative scale-variation effects, is planned as a future extension of this work.

As extensively shown in Refs.~\cite{Celiberto:2022grc,Celiberto:2021dzy,Celiberto:2021fdp,Celiberto:2022dyf,Celiberto:2022keu,Celiberto:2023rzw}, the gluon collinear FF plays a crucial role. \emph{De facto}, it has a strong stabilization power on our $\NLLp$ observables. 
Its behavior with $\mu_F$ fairly controls the stability of the high-energy logarithmic series of our distributions.
In the semi-hard regime proton PDFs are typically probed at $10^{-4} \lesssim x \lesssim 10^{-2}$, where the gluon PDF heavily dominates over quark ones.
Given that the gluon FF  diagonally multiplies with the gluon PDF in the LO hadron emission function~(see Eq.~(\ref{LOBIF})), its impact on the cross section is magnified.
This effect is not quenched at NLO, where $\{qg\}$ and $\{gq\}$ nondiagonal channels are opened~\cite{Celiberto:2021dzy} (see the Appendix~\hyperlink{app:NLOBIF}{A}).
On one side, $\alpha_s$ decreases with $\mu_R$ both in the Green's function and in the emission functions (see Section~\ref{ssec:nll_cross_section}). 
At the same time, however, the gluon PDF increases as $\mu_F$ enlarges. When that density multiplies a gluon FF which also increases with $\mu_F$, as it happens for $b$-hadrons, the two effects balance each other. 
This leads to the stabilizing trend of heavy-flavor distributions under energy-scale variations.
If instead the gluon FF falls down with $\mu_F$, as it happens for light-flavored species, such charged kaons, the stabilizing pattern is lost. 
This harms any chance of making precision tests of semi-hard observables at the natural energies given by kinematics~\cite{Celiberto:2020wpk}.
The emergence of the \emph{natural stability}~\cite{Celiberto:2022grc} in presence of both singly and doubly heavy flavored hadron species provides us with clear evidence that this striking feature is an \emph{intrinsic} property of heavy-flavor detections, which becomes manifest whenever a heavy hadron emitted, independently of the choice made for the initial-scale input of the corresponding collinear fragmentation.

\section{Hi-Lumi LHC distributions}
\label{sec:phenomenology}

\begin{figure*}[!t]
\centering

   \hspace{0.00cm}
   \includegraphics[scale=0.36,clip]{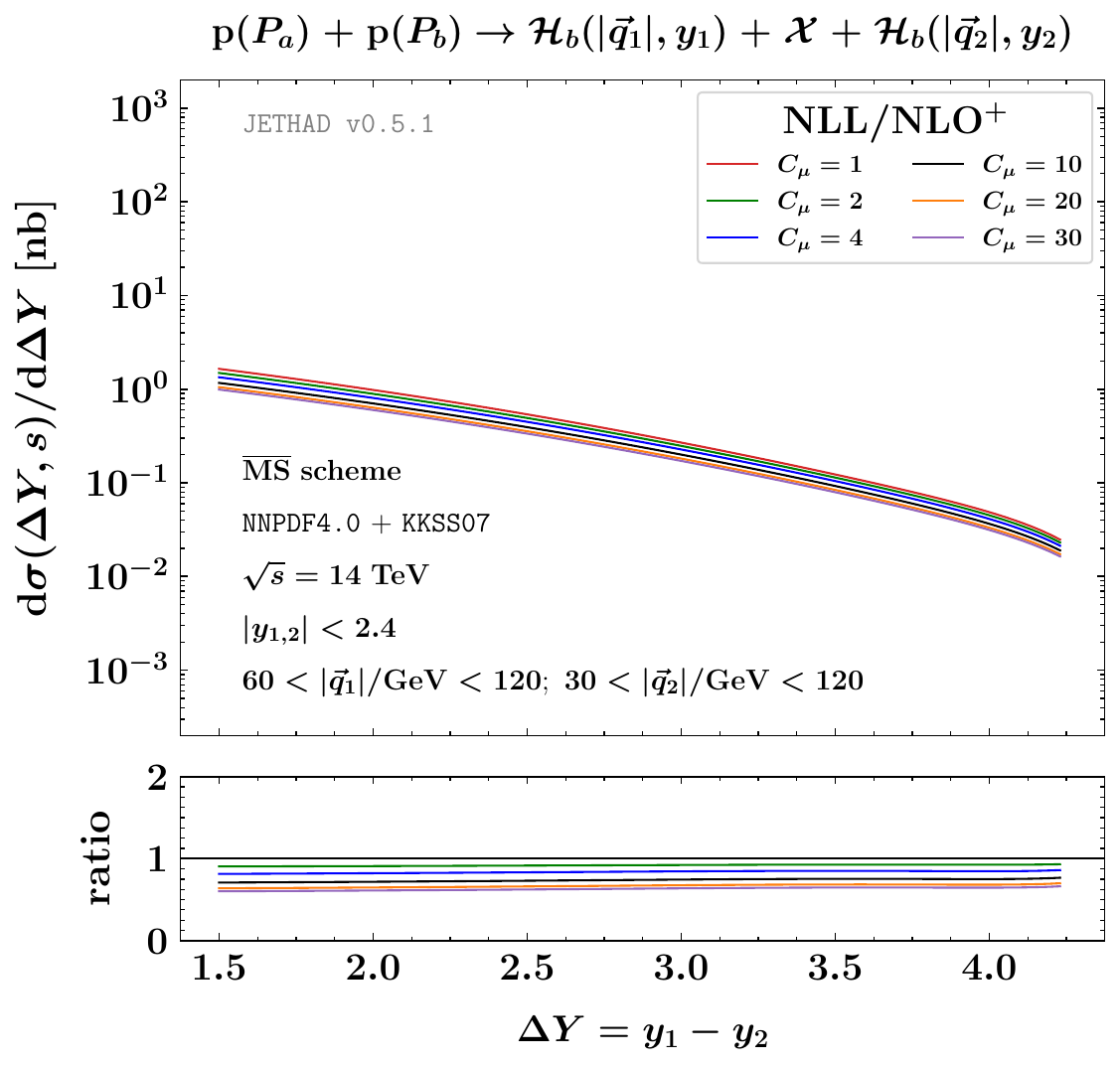}
   \hspace{0.50cm}
   \includegraphics[scale=0.36,clip]{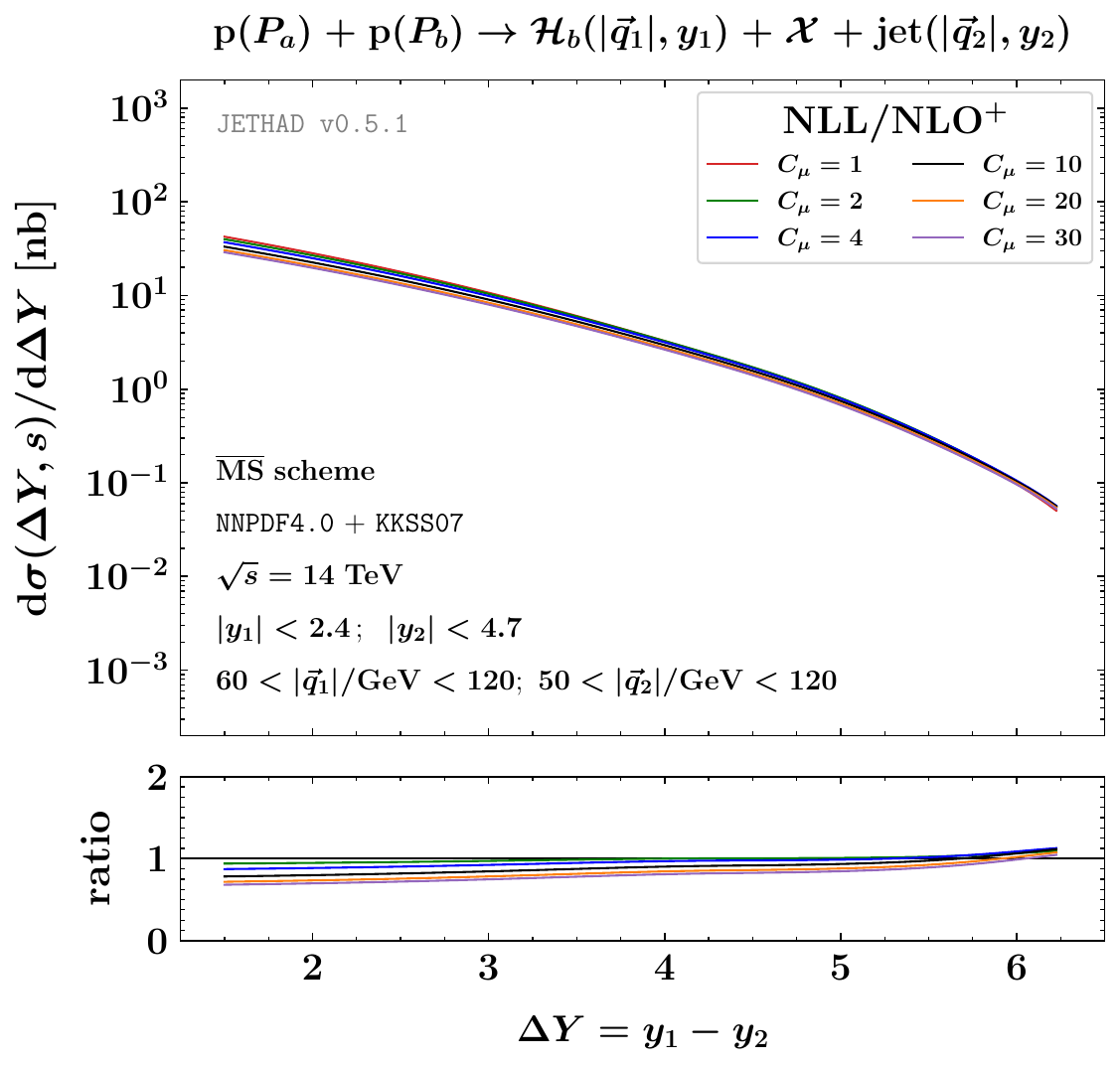}

   \includegraphics[scale=0.36,clip]{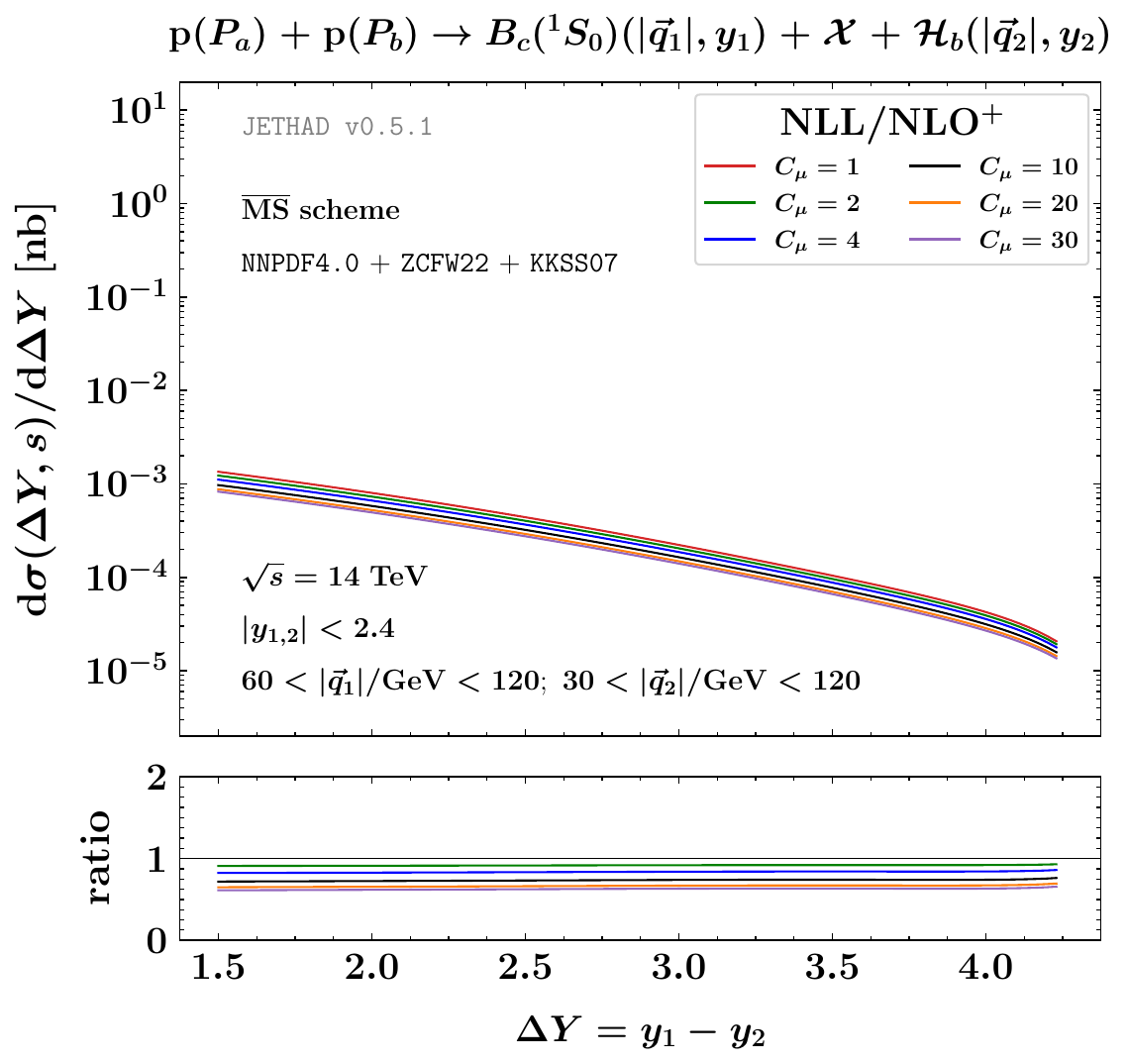}
   \hspace{0.50cm}
   \includegraphics[scale=0.36,clip]{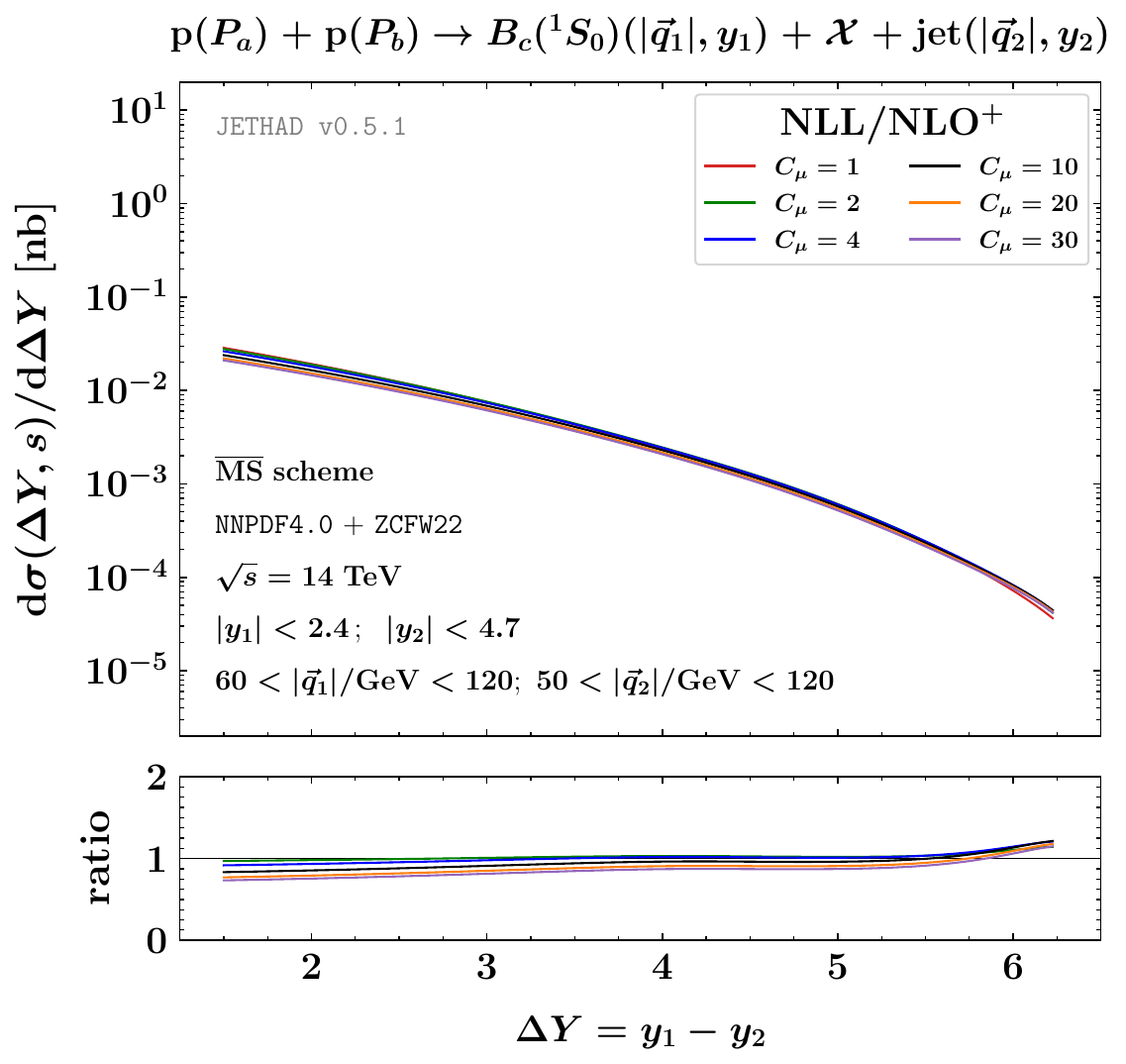}

   \includegraphics[scale=0.36,clip]{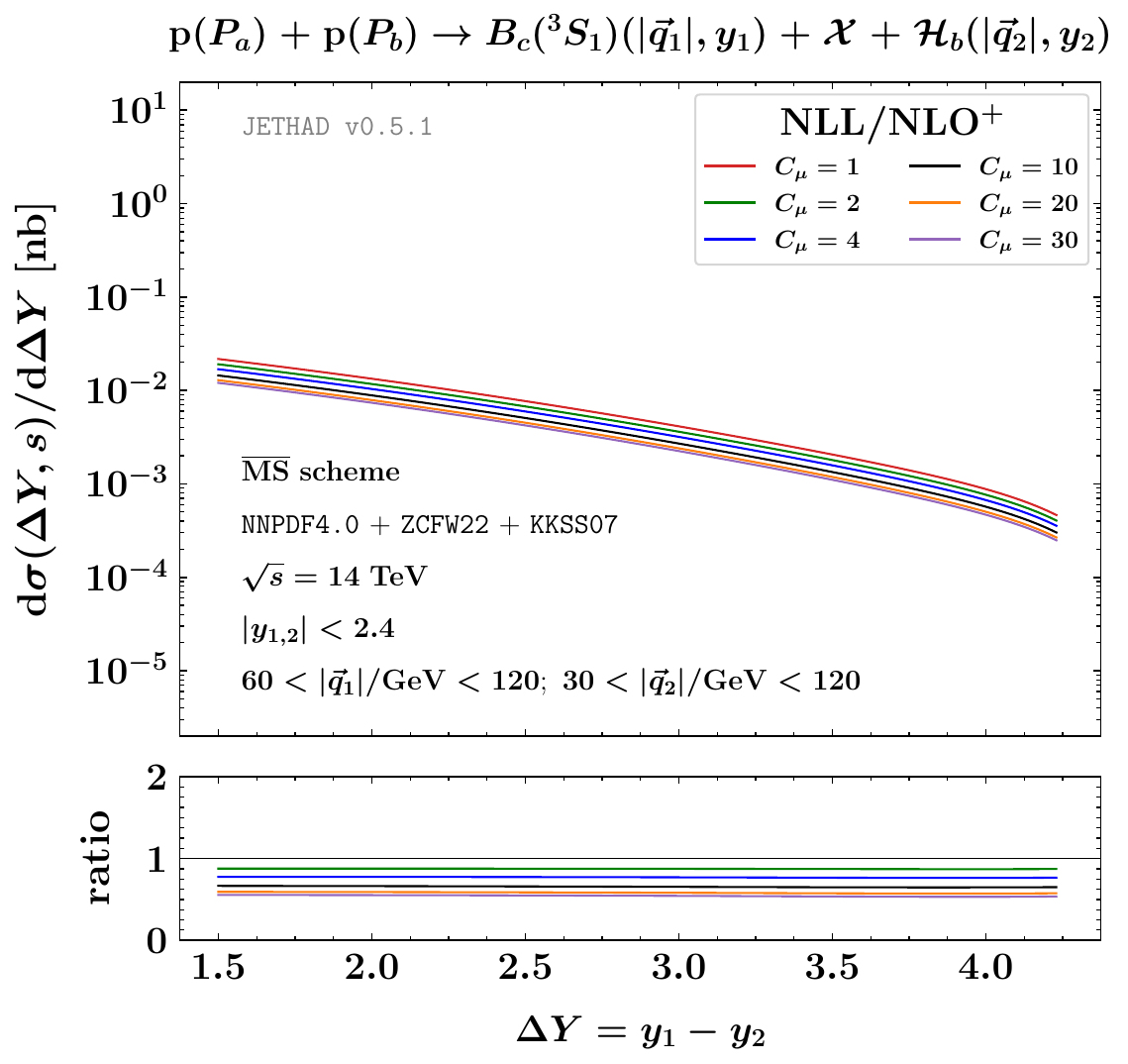}
   \hspace{0.50cm}
   \includegraphics[scale=0.36,clip]{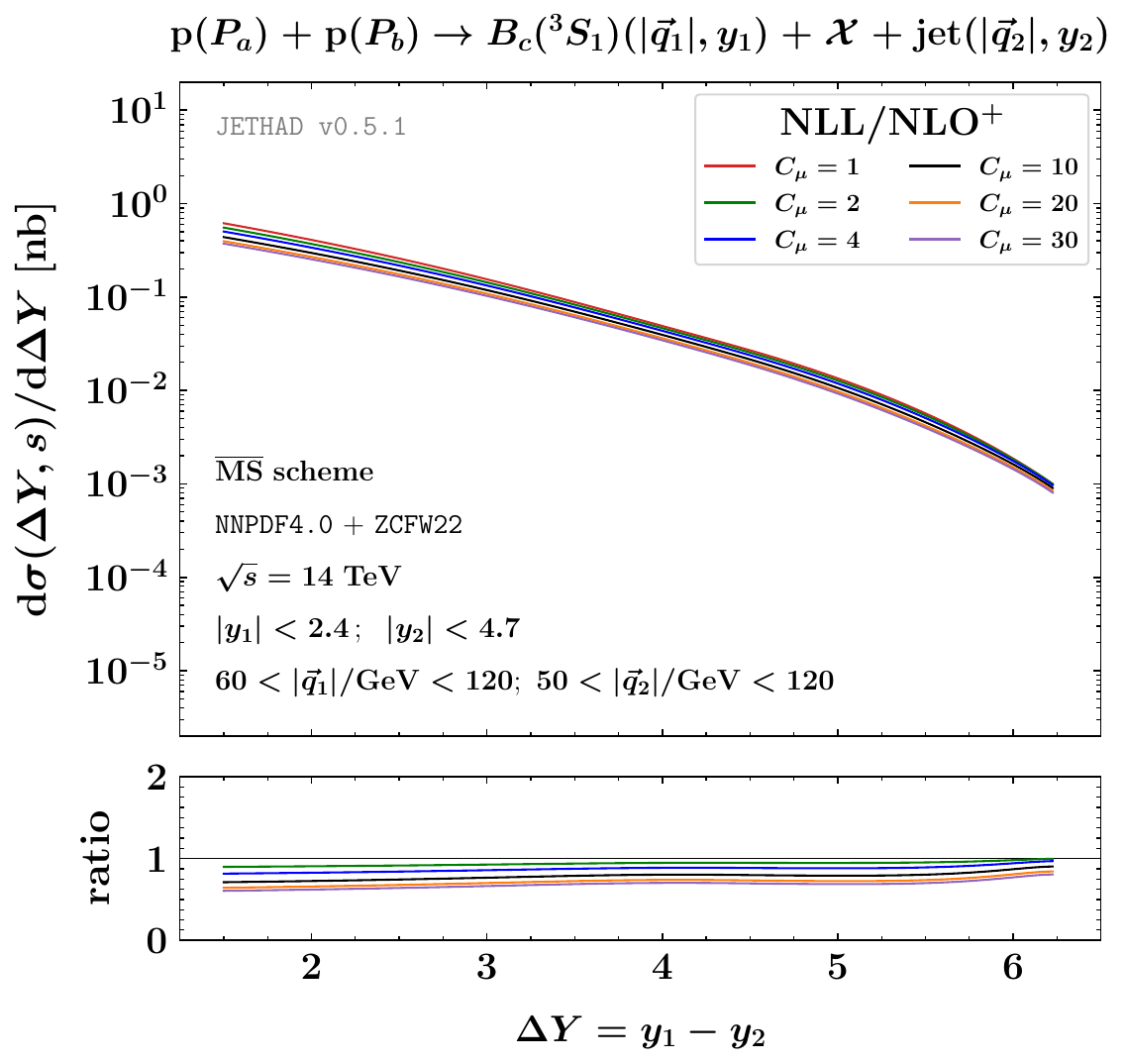}

\caption{$\DY$-rate of the double-hadron (left) and the hadron-plus-jet (right) channel at $\sqrt{s} = 14$ TeV and within the $\NLLp$ accuracy. Renormalization and factorization scales are progressively varied in the $1 < C_{\mu} < 30$ range.}
\label{fig:Y_psv}
\end{figure*}

We performed our numeric computations via the {\Jethad}~{\tt v0.5.1} interface~\cite{Celiberto:2020wpk,Celiberto:2022rfj,Celiberto:2023fzz}, a \textsc{Python}+\textsc{Fortran} multimodular working environment aimed at the calculation, management, and processing of high-energy distributions obtained from different theoretical approaches.
In particular, rapidity and transverse-momentum differential cross sections were calculated by means of {\Jethad} \textsc{Fortran 2008} core modules, whereas its \textsc{Python 3.0} analyzer was used to elaborate results.
Section~\ref{ssec:uncertainty} gives details on our procedure to gauge the size of uncertainties for our resummed observables. 
Final-state kinematic cuts are presented in Section~\ref{ssec:kinematics}.
Then, Section~\ref{ssec:rapidity_distributions} carries a discussion of our $\NLLp$ predictions.

\subsection{Uncertainty estimation}
\label{ssec:uncertainty}

\begin{figure*}[!t]
\centering

   \hspace{0.00cm}
   \includegraphics[scale=0.36,clip]{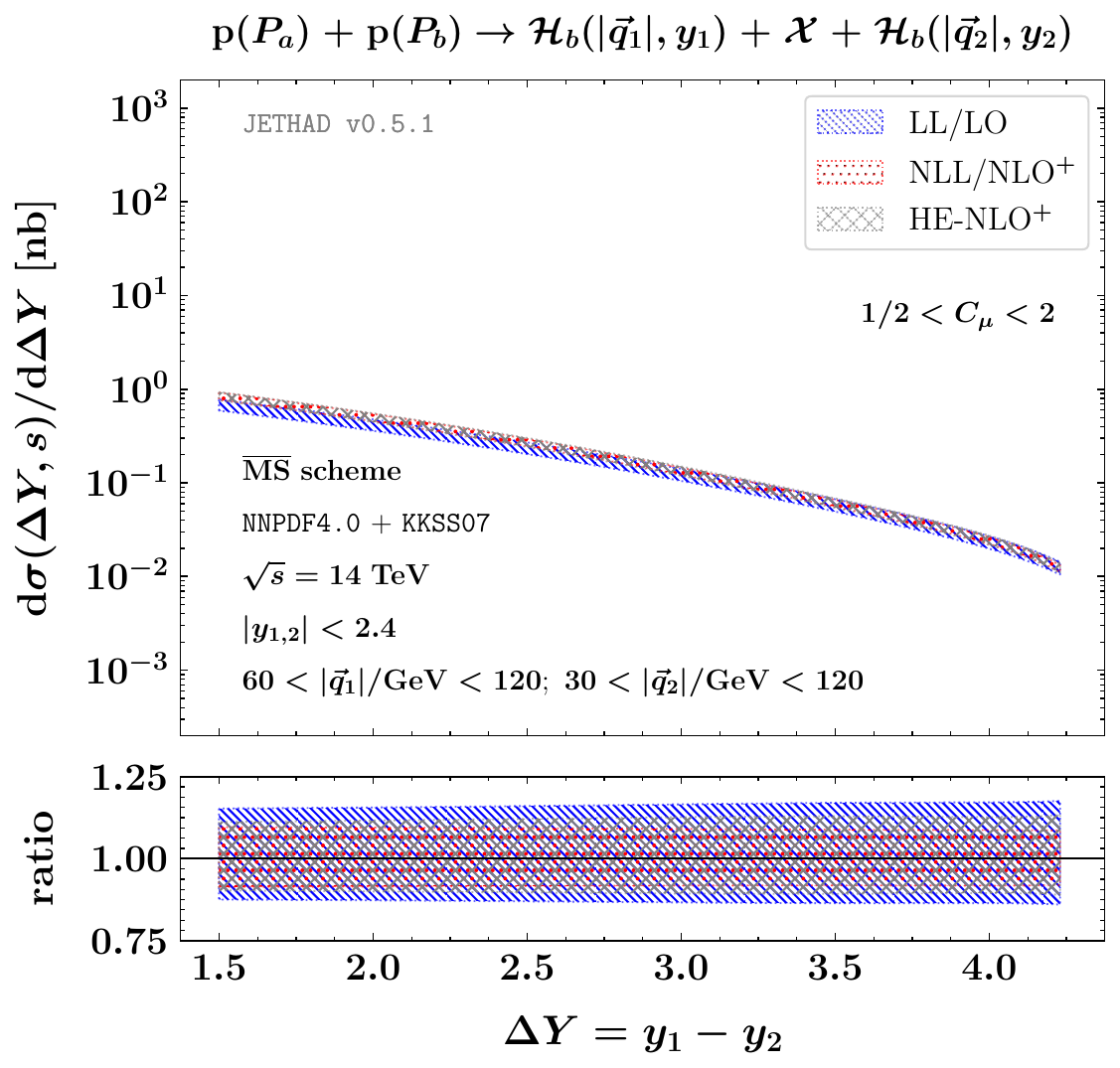}
   \hspace{0.50cm}
   \includegraphics[scale=0.36,clip]{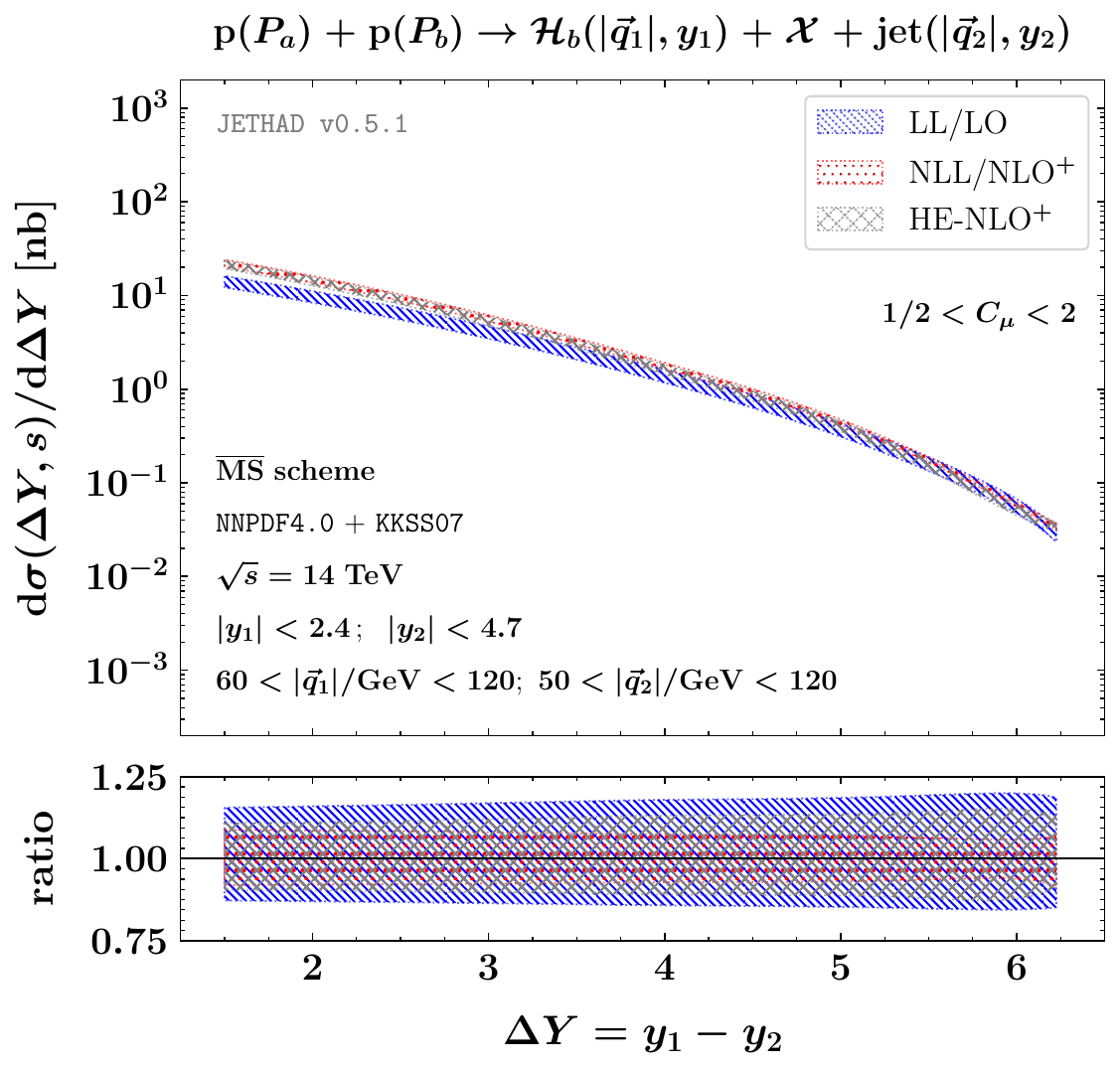}

   \includegraphics[scale=0.36,clip]{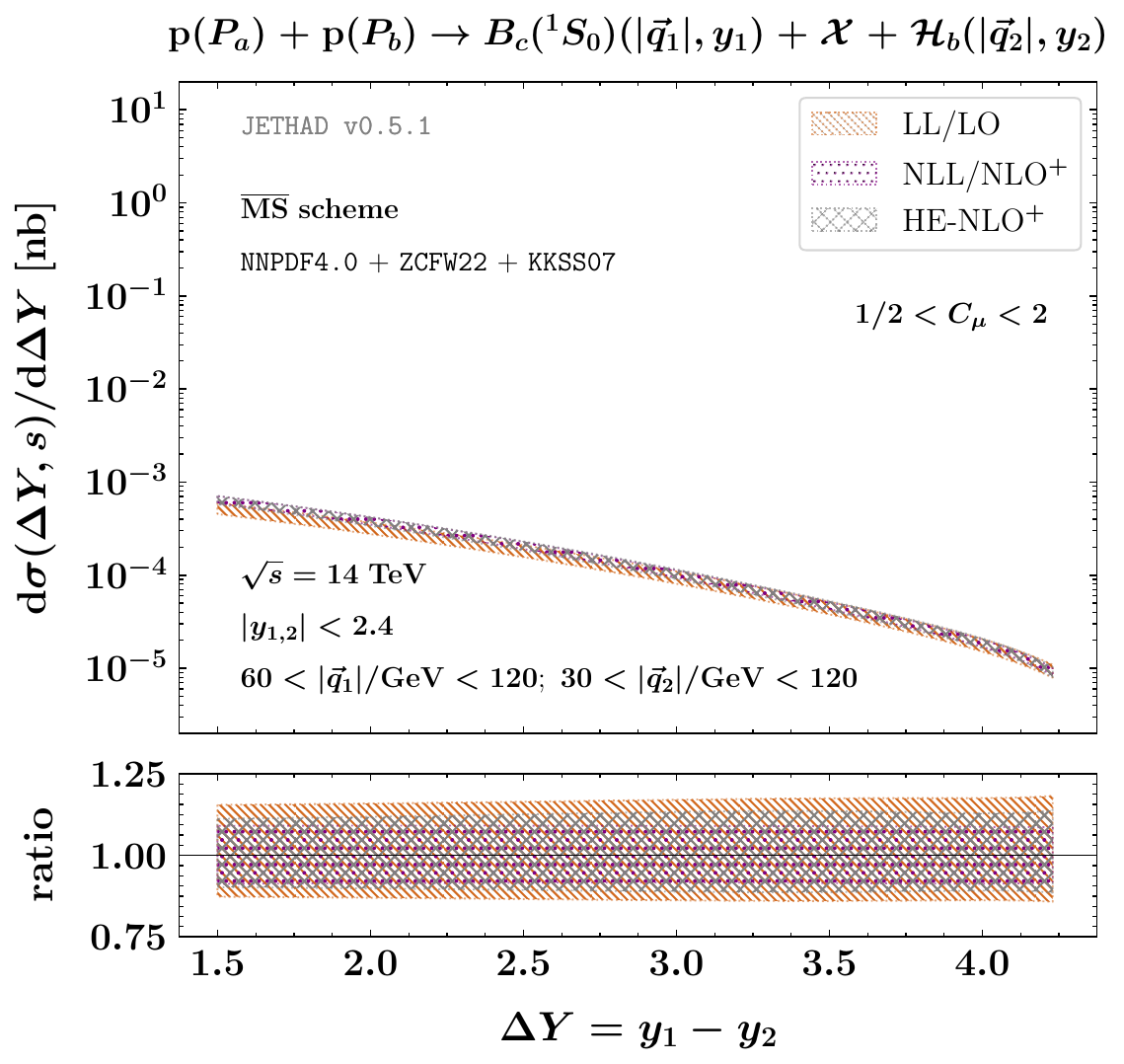}
   \hspace{0.50cm}
   \includegraphics[scale=0.36,clip]{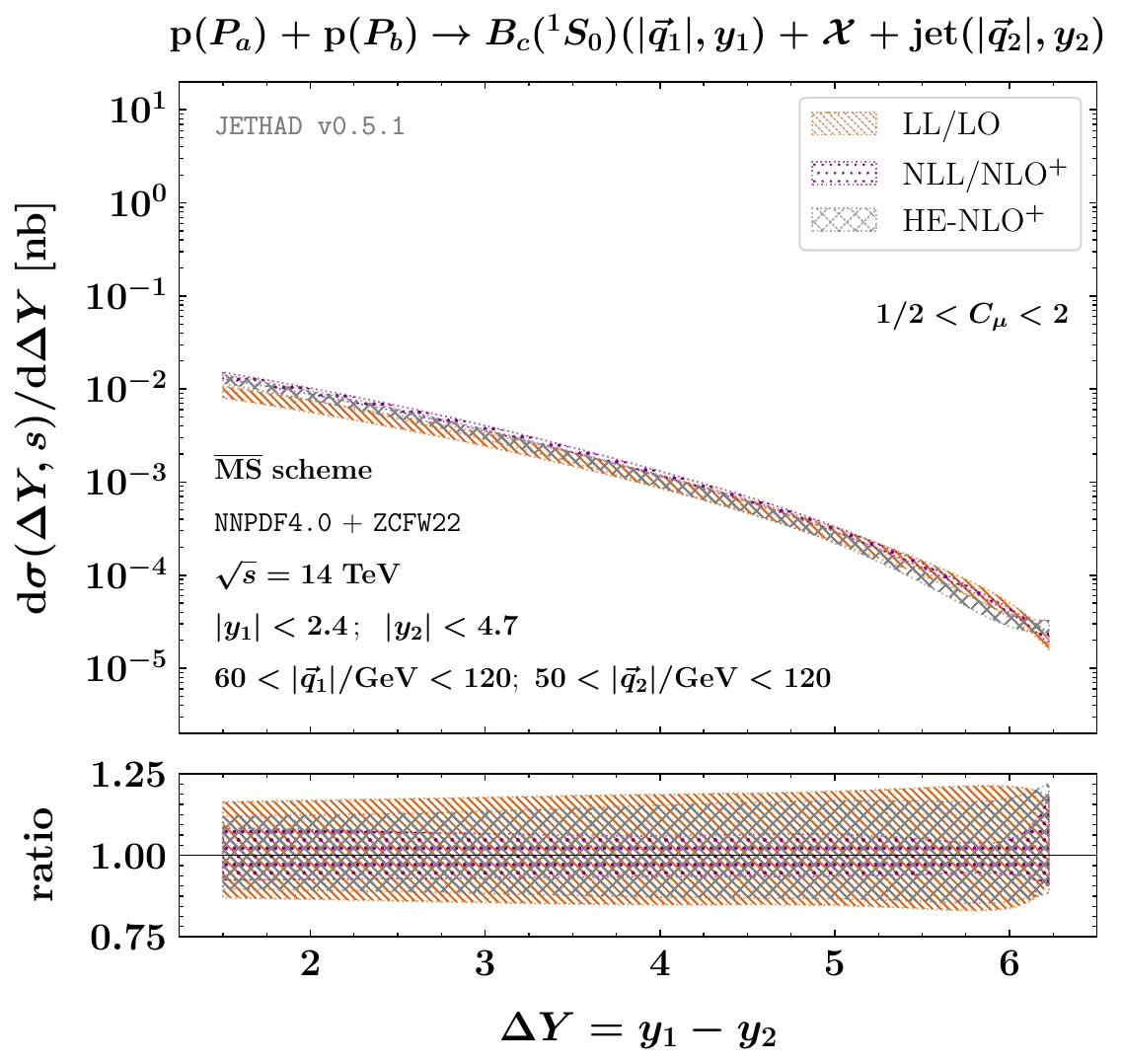}

   \includegraphics[scale=0.36,clip]{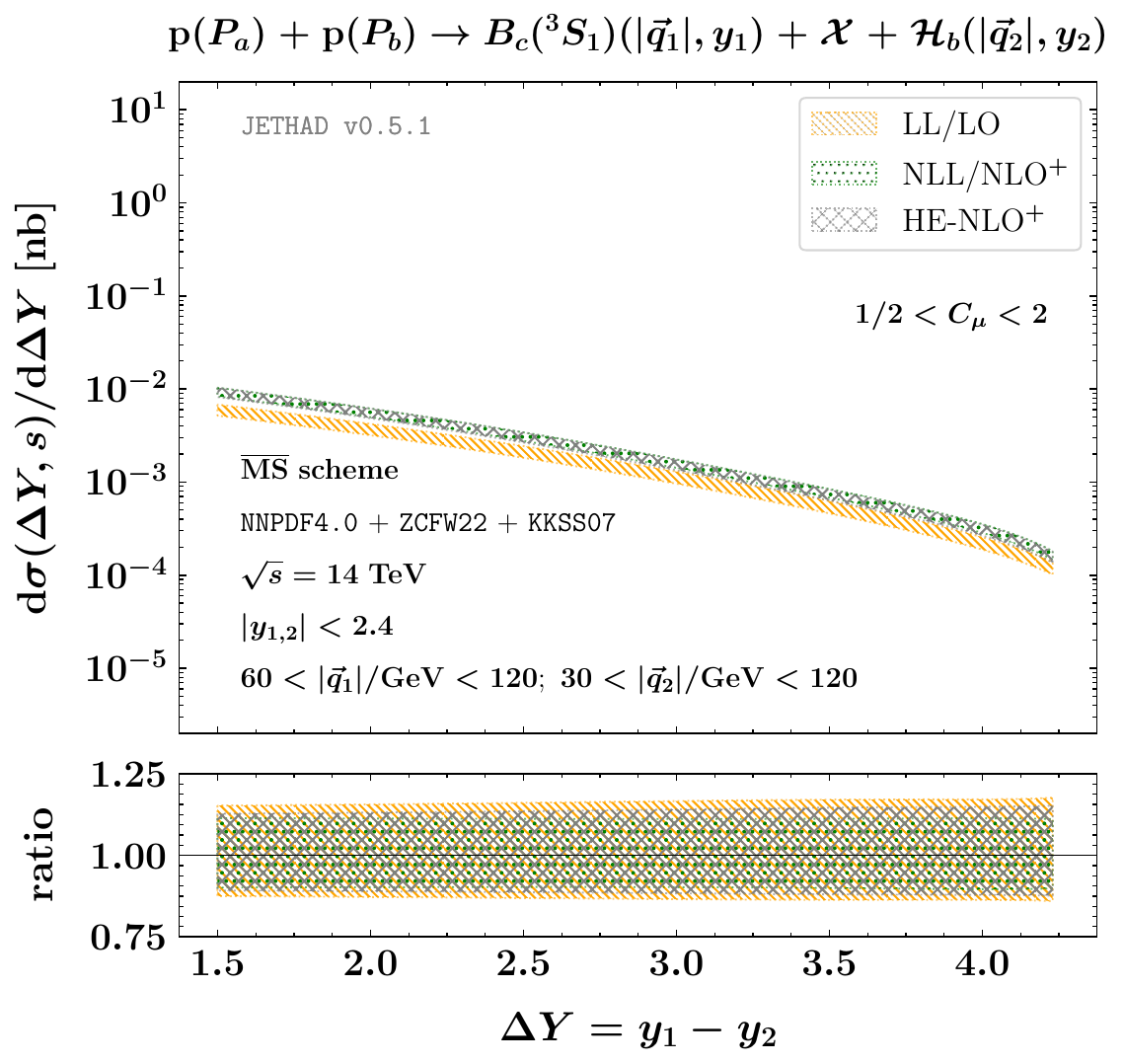}
   \hspace{0.50cm}
   \includegraphics[scale=0.36,clip]{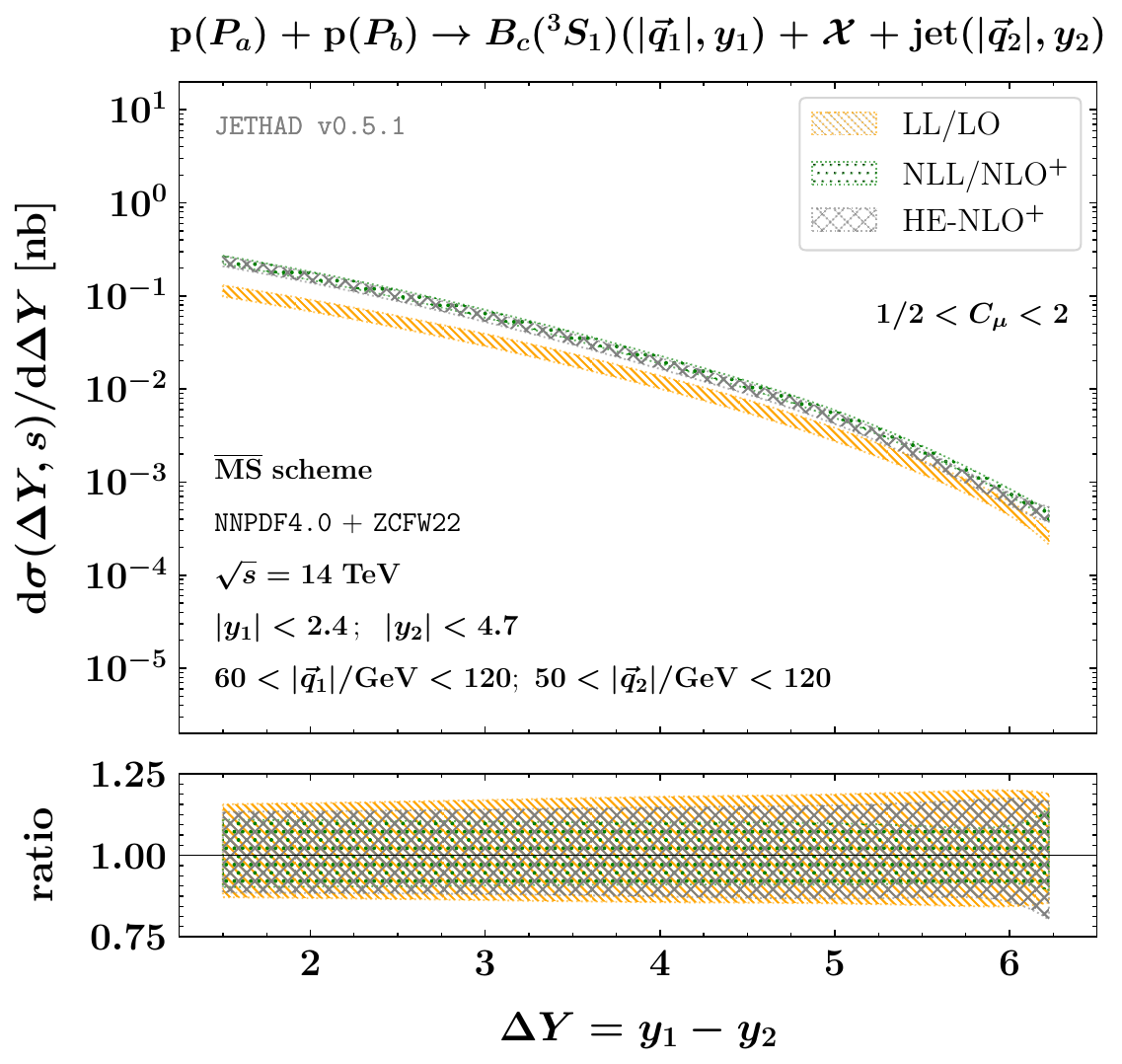}

\caption{$\DY$-rate of the double-hadron (left) and the hadron-plus-jet (right) channel at $\sqrt{s} = 14$ TeV and within the $\NLLp$ accuracy. 
Shaded bands exhibit the combined effect of renormalization- and factorization-scale variation in the$1/2 < C_{\mu} < 2$ range.}
\label{fig:Y_sc}
\end{figure*}

A widely-used methodology to quantify the weight of \ac{MHOUs} relies upon assessing how much our observables are sensitive to variations of factorization and renormalization scales around their natural values.
It is well-known that the effect MHOUs gives perhaps the major contribution to the overall uncertainty~\cite{Celiberto:2022rfj}.
For this reason, we will gauge the size of simultaneously varying $\mu_F$ and $\mu_R$ around $\mu_N/2$ and $2 \mu_N$.
The $C_{\mu}$ parameter in panels of figures of Sections~\ref{ssec:rapidity_distributions} and~\ref{ssec:TM_distributions} is just the $C_\mu \equiv \mu_{F}/\mu_N = \mu_{R}/\mu_N$ ratio.
This will represent our standard procedure to estimate the size of MHOUs. Then, for the sake of comparison with other approaches, where energy scales are allowed to assume values much larger than the natural ones (see, \emph{e.g.}, the BLM method~\cite{Brodsky:1996sg,Brodsky:2002ka,Caporale:2015uva}), we will also present an accessory, extended MHOU scan, with the $C_\mu$ parameter ranging from 1 to 30.
Another possibly relevant uncertainty is encoded in proton PDFs. 
Quite recent analyses on high-energy production rates have shown, however, that selecting distinct PDF parametrizations as well as distinct members inside the same set produces a very small effect~\cite{Bolognino:2018oth,Celiberto:2020wpk,Celiberto:2021fdp,Celiberto:2022rfj}. 
Therefore, we will calculate our observables by considering only the central member of the {\tt NNPDF4.0} parametrization.
Then, other potential uncertainties could come from a \emph{collinear improvement} of the NLO kernel (see Refs.~\cite{Salam:1998tj,Ciafaloni:2003rd,SabioVera:2005tiv} and reference therein), which consists in including renormalization-group (RG) terms to make the BFKL equation compatible with the DGLAP one in the collinear limit, or from changing the renormalization scheme.
The first case was deeply addressed in Ref.~\cite{Celiberto:2022rfj}. 
As a result, the impact of collinear-improvement techniques on rapidity-differential rates is fairly contained inside error bands produced by MHOUs, and it is even smaller in the case of when other differential observables.
The same article provides us with an overestimation of the effect of passing from $\MSb$~\cite{PhysRevD.18.3998} to MOM~\cite{Barbieri:1979be,PhysRevLett.42.1435} renormalization scheme.
MOM results for rapidity distributions are systematically higher than $\MSb$ ones, but still contained into MHOU bands. We stress, however, that a proper MOM analysis should be based on MOM-evolved PDFs and FFs, not available so far.

We will produce uncertainty bands for our distributions by combining MHOUs with the numeric error generated by the multidimensional integration (see Section~\ref{ssec:kinematics}). The latter will be constantly kept below $1\%$ by the {\Jethad} integration subinterfaces.

\subsection{Kinematic constraints on final-state ranges}
\label{ssec:kinematics}

We will consider two main observables: $(i)$ the rapidity-interval distribution, namely the cross section differential in the rapidity separation, $\DY = y_1 - y_2$, between the two produced particles, and $(ii)$ the doubly-differential transverse-momentum distribution.

The first observable genuinely corresponds the ${\cal C}_0$ azimuthal-angle coefficient, defined in Section~\ref{ssec:nll_cross_section}, integrated over final-state transverse momenta and rapidities and taken at fixed values of $\DY$
\begin{equation}
 \label{DY_distribution}
 \frac{\drv \sigma(\DY; s)}{\drv \DY} =
 \int_{|\vec q_2|^{\rm min}}^{|\vec q_2|^{\rm max}} 
 \!\!\drv |\vec q_1|
 \int_{|\vec q_2|^{\rm min}}^{|\vec q_2|^{\rm max}} 
 \!\!\drv |\vec q_2|
 \int_{\max \, (y_1^{\rm min}, \, \DY + y_2^{\rm min})}^{\min \, (y_1^{\rm max}, \, \DY + y_2^{\rm max})} \drv y_1
 \, \,
 {\cal C}_0^{\rm [order]}\left(|\vec q_1|, |\vec q_2|, y_1, y_2; s \right)
\Bigm \lvert_{y_2 \;=\; y_1 - \DY}
 \;,
\end{equation}
where the `${\rm [order]}$' label for ${\cal C}_0$ inclusively refers to: $\NLLp$, $\HENLOp$, or $\LL$.
Here we made use of a $\delta (\DY - y_1 + y_2)$ function to remove the integration over one of the two rapidities, say $y_2$.
Transverse momenta of the forward $b$-hadron and the backward one stay in the ranges $60 < |\vec q_1| /{\rm GeV} < 120$ and $30 < |\vec q_2| /{\rm GeV} < 120$, respectively. 
These cuts fairly meet the criteria of applicability of a VFNS-based fragmentation approach, in which energy scales must be sufficiently higher than thresholds for the DGLAP evolution of heavy-quark species.
Then, light jets are tagged in transverse-momentum ranges lightly different but compatible with current and forthcoming studies at the LHC~\cite{Khachatryan:2016udy}, say $50 < |\vec q_2| / \rm{GeV} < 60$.
The use of such \emph{asymmetric} transverse-momentum ranges eases disentangling the pure high-energy dynamics from the fixed-order signal~\cite{Celiberto:2015yba,Celiberto:2015mpa,Celiberto:2020wpk}. It also dampens large Sudakov logarithms due to almost back-to-back configurations which would require the employment of another resummation~\cite{Mueller:2013wwa,Marzani:2015oyb,Mueller:2015ael,Xiao:2018esv,Hatta:2020bgy,Hatta:2021jcd}.
Finally, it quenches instabilities connected to radiative corrections~\cite{Andersen:2001kta,Fontannaz:2001nq} as well as violations of the energy-momentum conservation relation~\cite{Ducloue:2014koa}.
As for the rapidity ranges, we select typical cuts of current LHC studies, where hadrons are detected only in the barrel calorimeter~\cite{Chatrchyan:2012xg}, say from $-2.4$ to $2.4$, whereas the jet can be also tagged in endcaps' regions~\cite{Khachatryan:2016udy}, say from $-4.7$ to $4.7$.

The second observable is given by the ${\cal C}_0$ azimuthal-angle coefficient, integrated over rapidities while $\DY$ is kept fixed, but this time being differential in the transverse momenta of the two outgoing objects
\begin{equation}
 \label{Yk12_distribution}
 \frac{\drv \sigma(|\vec q_1|, |\vec q_2|, \DY; s)}{\drv |\vec q_1| \drv |\vec q_2| \drv \DY} =
 \int_{\max \, (y_1^{\rm min}, \, \DY + y_2^{\rm min})}^{\min \, (y_1^{\rm max}, \, \DY + y_2^{\rm max})} \drv y_1
 \, \,
 {\cal C}_0^{\rm [order]}\left(|\vec q_1|, |\vec q_2|, y_1, y_2; s \right)
\Bigm \lvert_{y_2 \;=\; y_1 - \DY}
 \;. 
\end{equation}
Here we consider the previously mentioned kinematic cuts for rapidities, whereas the two transverse momenta can run from 10~to~100~GeV.
This observable was recently proposed, in the context of singly bottom-flavored hadrons~\cite{Celiberto:2021fdp} and cascade $\Xi$ baryons~\cite{Celiberto:2022kxx}, as a favorable testing ground whereby unraveling the interplay among distinct resummations.

\subsection{Rapidity-interval distributions}
\label{ssec:rapidity_distributions}

In Fig.~\ref{fig:Y_psv} we present rapidity-interval distributions for our reference processes, calculated within $\NLLp$ hybrid factorization and taken at $\sqrt{s} = 14$ TeV.
Left plots refer to double-hadron channels, while right plots are for the hadron-jet ones.
Final states with $\Hb$ hadrons, $\BCs$ mesons, or $\Bss$ resonances are considered in upper, central, and lower plots, respectively.
More in particular, any forward $b$-flavored hadron is always accompanied by a singly bottomed hadron $\Hb$ (left plots) or by a light-flavored jet (right plots).
This is to avoid the presence of two identified charmed $B$ mesons in the final state, which would lead to a substantial lowering in statistics.
We will adopt this choice also in the next Section~(\ref{ssec:TM_distributions}).
The minimum values of our distributions are the ones obtained at the maximum values of $\DY$. They are uniformly larger than $10^{-2}$~nb when only $\Hb$ hadrons (and light jets) are produced, and larger than $10^{-5}$~nb in the two $B_c^{(*)}$ cases. 
This represents a very promising statistics that can be easily accessed via dedicated experimental studies at the forthcoming Hi-Lumi upgrade of the LHC.
The decreasing trend with $\DY$ of our NLL-resummed distributions is a general feature shared by all the semi-hard hadroproduction final states investigated so far. 
It comes out as the interplay between two competing effects. 
Indeed, although the high-energy dynamics encoded in the partonic hard factor would make the cross section grow with energy, collinear PDFs and FFs fall off very fast with $\DY$, since this kinematically translates to longitudinal-momentum fractions very close to one. 
The net result is a dampening effect with $\DY$ of hadronic distributions.

The core analysis shown in Fig.~\ref{fig:Y_psv} is a study of our rapidity-interval differential distributions under a progressive variation of factorization and renormalization scales in an broad window, controlled by the $C_\mu$ parameter introduced in Section~\ref{ssec:uncertainty}, which ranges from $1$ to $30$.
This strategy actually represents an extended MHOU scan with respect to the traditional one, $1/2 < C_\mu < 2$.
Ancillary panels drawn right below primary ones show reduced distributions, namely the ones divided by their central values calculated at $C_\mu = 1$. 
Remarkably, $\NLLp$ results for all the six considered final states feature quite a weak sensitivity on $C_\mu$ variation in the whole spectrum of $\DY$ matter of our investigation. 
This supports the important message that a strong stabilization mechanism is encoded both in the {\tt KKSS07} functions extracted from global data on $B$ mesons and $\Lambda_b$ baryons, as well as in the novel {\tt ZCFW22} FFs built on the basis of NRQCD initial-scale inputs for $\BCs$ and $\Bss$ generalized-quarkonium states and then evolved via DGLAP in a VFNS scheme.
Results contained in plots of Fig.~\ref{fig:Y_psv} bring to light the emergence of the
\emph{natural stability}~\cite{Celiberto:2022grc} from semi-inclusive $b$-flavor emissions at high energies.

As a complement, in Fig.~\ref{fig:Y_sc} we show results for the same $\DY$-distributions, calculated within the $\NLLp$ accuracy and compared with $\HENLOp$ and $\LL$ corresponding ones.
Left (right) plots are for hadron-plus-hadron (hadron-plus-jet) observables.
Upper, central, and lower plots respectively correspond to $\Hb$, $B_c$, and $B_c^*$ sensitive final states.
This time, shaded bands are built by combining the uncertainty of the multidimensional integration, numerically performed by the {\Jethad} integrators, with the one coming from a traditional MHOU scan, namely with the $C_\mu$ parameter running from 1/2 to two.
As done before, lower ancillary panels exhibit reduced distributions, \emph{i.e.} divided by central values calculated when $C_\mu = 1$.
The global outcome is a strong stability under higher-order corrections.
This is confirmed by the fact that $\NLLp$ uncertainty bands are generally narrower and partially nested inside corresponding $\LL$ ones, in particular in the large-$\DY$ regime, where the impact of high-energy logarithms is expected to be large.
Their overlap is slightly weaker for hadron-plus-jet cases (right plots). 
This is not unexpected, since the emission of just one $b$-hadron rather then two reduces the stabilizing power by one half.

As discussed in previous works (see, \emph{e.g.} Section~3 of Ref.~\cite{Celiberto:2022keu}), shaded bands for fixed-order $\HENLOp$ results are almost completely nested in resummed $\NLLp$ ones.
Future analyses on more exclusive observables, such as \emph{angularity} distributions will shed light on a clearer disentanglement between our resummation and the fixed-order signal.
Finally, as a bonus result, we confirm the estimate, provided by the LHCb Collaboration~\cite{LHCb:2014iah,LHCb:2016qpe}, on the production-rate hierarchy between singly-bottomed hadrons and charmed $B$ mesons.
Production rates of the latters seems not to exceed 0.1\% of the $\Hb$ ones, at most.
To compare our predictions with these numbers, we can consider the hadron-plus-jet channel (right diagram of Fig.~\ref{fig:process}) at $\LL$, whose differential cross section is linear in the $b$-hadron FFs (see Eq.~\eqref{LOBIF}).
The inspection of right plots of Fig.~\ref{fig:Y_sc} in clearly in line with LHCb estimates for charmed $B$-meson sates.\footnote{This strictly holds for $B_c \equiv \BCs$ mesons, while statistics for $\Bss$ ones a bit larger.
We stress, however, that being the $\Bss$ particle a resonance of the $\BCs$ state, a more accurate description of its dynamical production mechanism might go beyond our treatment. 
Indeed, it could rely upon more intricate effects, such as spin correlations as well as higher-order,  relativistic corrections within the same NRQCD formalism.}
The hierarchy remains \emph{de facto} unaltered both at $\NLLp$ and at $\HENLOp$.
Remarkably, this fact not only represents a reliability test for our hybrid factorization, but it also definitely corroborates the validity of the single-parton fragmentation mechanism from NRQCD, at least for the $\BCs$ case.

\subsection{Doubly-differential transverse-momentum distributions}
\label{ssec:TM_distributions}

Rapidity-interval distributions represent golden channels to unveil the onset of high-energy dynamics at hadron colliders.
At the same time, pursuing the goal of exploring kinematic corners where other resummations are also at work, we need more differential observables in the transverse momentum.  
From one side, high transverse momenta or large imbalances between them heighten the size of DGLAP-type logarithms as well as soft, \emph{threshold} ones~\cite{Sterman:1986aj,Catani:1989ne,Catani:1996yz,Bonciani:2003nt,deFlorian:2005fzc,Ahrens:2009cxz,deFlorian:2012yg,Forte:2021wxe,Mukherjee:2006uu,Bolzoni:2006ky,Becher:2006nr,Becher:2007ty,Bonvini:2010tp,Ahmed:2014era,Muselli:2017bad,Banerjee:2018vvb,Duhr:2022cob,Shi:2021hwx,Wang:2022zdu,Bonvini:2023mfj}.
These logarithms, which are of a different nature with respect the high-energy ones, must also be resummed.

Combining threshold and high-energy logarithms is a cumbersome task. 
A first double resummation was obtained in the context of the inclusive Higgs hadroproduction from gluon-gluon fusion~\cite{Bonvini:2018ixe}.
The key ingredient there was the possibility of decoupling the two dynamics in the Mellin space, so that the small-$x$ (high-energy) resummation and the large-$x$ (threshold) one respectively ``control'' the small-$N$ and large-$N$ tail of the the Mellin projection of the cross section~\cite{Ball:2013bra,Bonvini:2014joa}.
In this way, the two kinds of logarithms can be separately resummed.
In our case, however, the semi-inclusive emission of two objects (see Fig.~\ref{fig:process}) genuinely leads to rapidity-differential observables, for which a theoretical framework to properly perform such a double resummation is still missing.

From the other side, low transverse momenta translate into large Sudakov logarithms which are also missed by our hybrid factorization. 
Furthermore, effects connected to the well-known \emph{diffusion pattern} rise~\cite{Bartels:1993du,Caporale:2013bva,Ross:2016zwl}.
The most natural way to catch these logarithms to all orders builds on the \ac{TM} resummation formalism~\cite{Catani:2000vq,Bozzi:2005wk,Bozzi:2008bb,Catani:2010pd,Catani:2011kr,Catani:2013tia,Catani:2015vma,Duhr:2022yyp}.

\begin{figure*}[!t]
\centering

   \includegraphics[scale=0.31,clip]{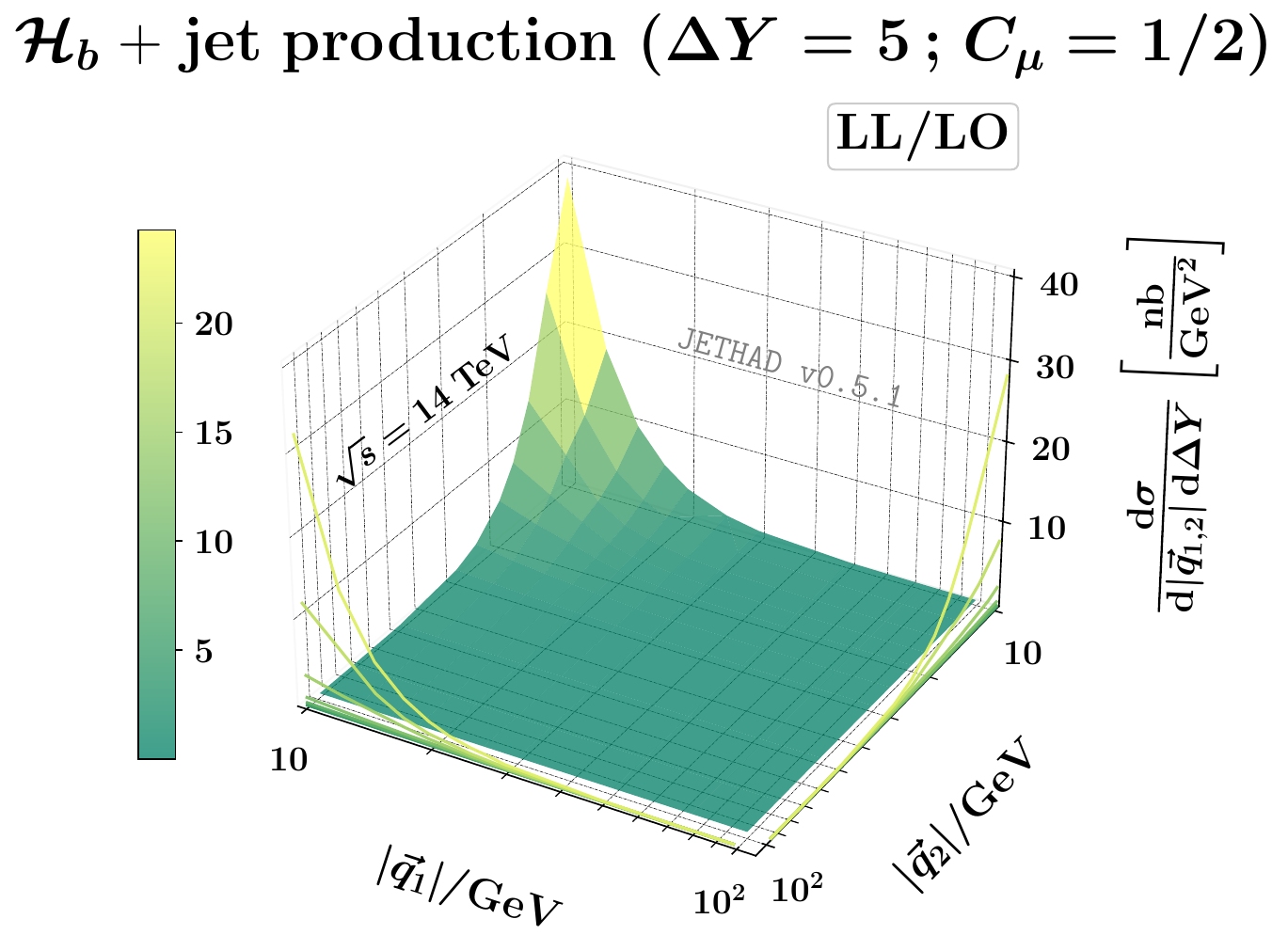}
   \hspace{0.25cm}
   \includegraphics[scale=0.31,clip]{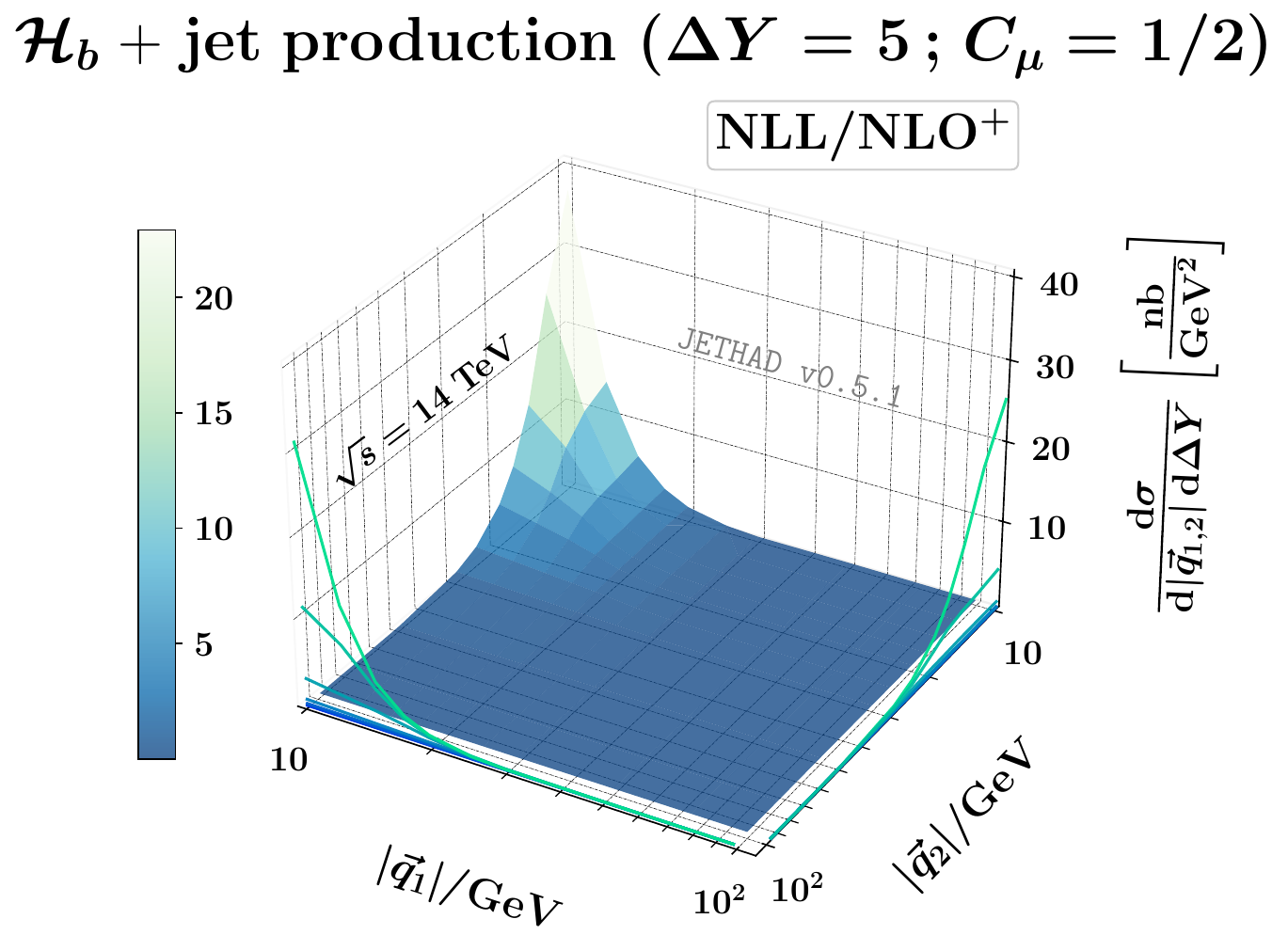}
   \vspace{0.45cm}

   \includegraphics[scale=0.31,clip]{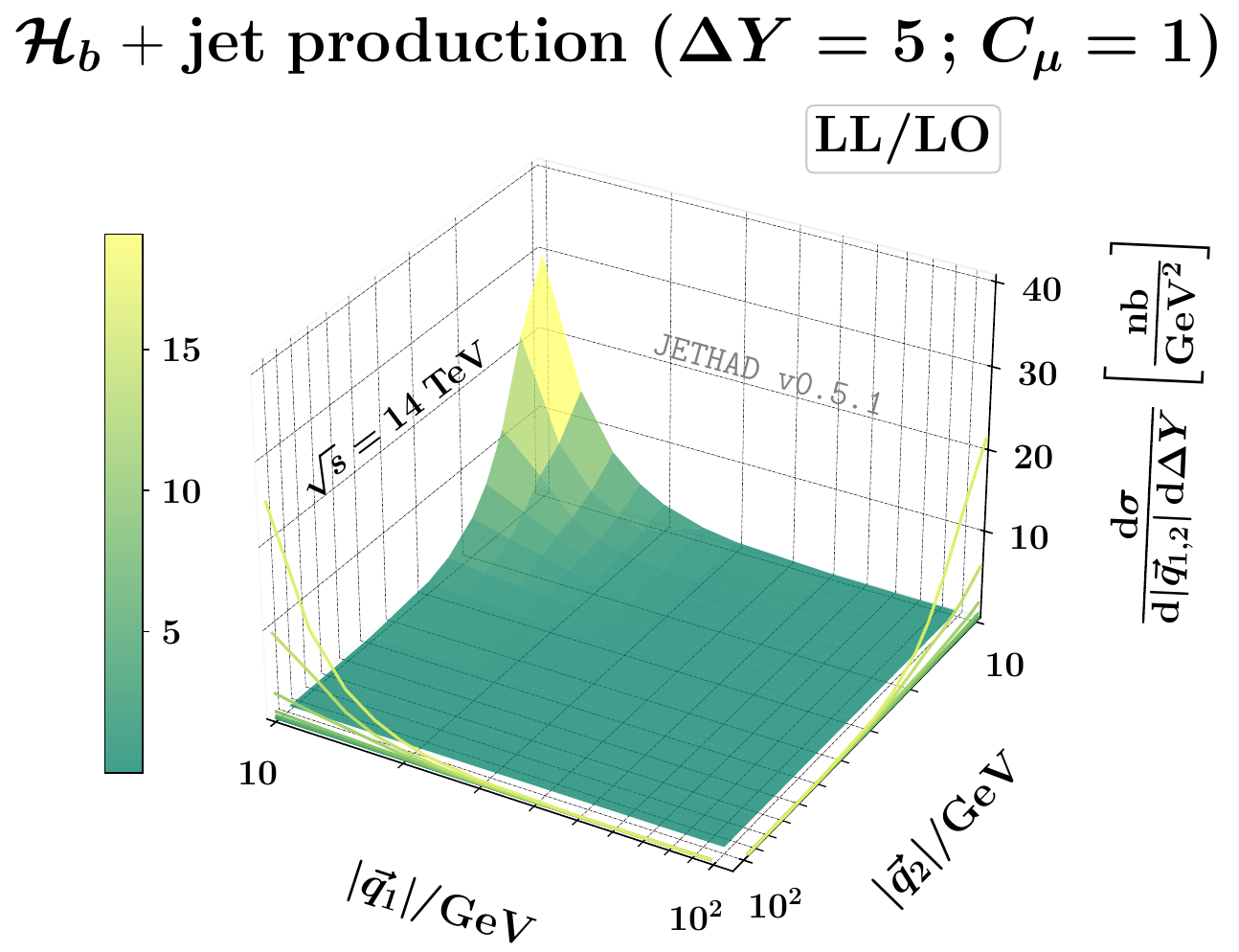}
   \hspace{0.25cm}
   \includegraphics[scale=0.31,clip]{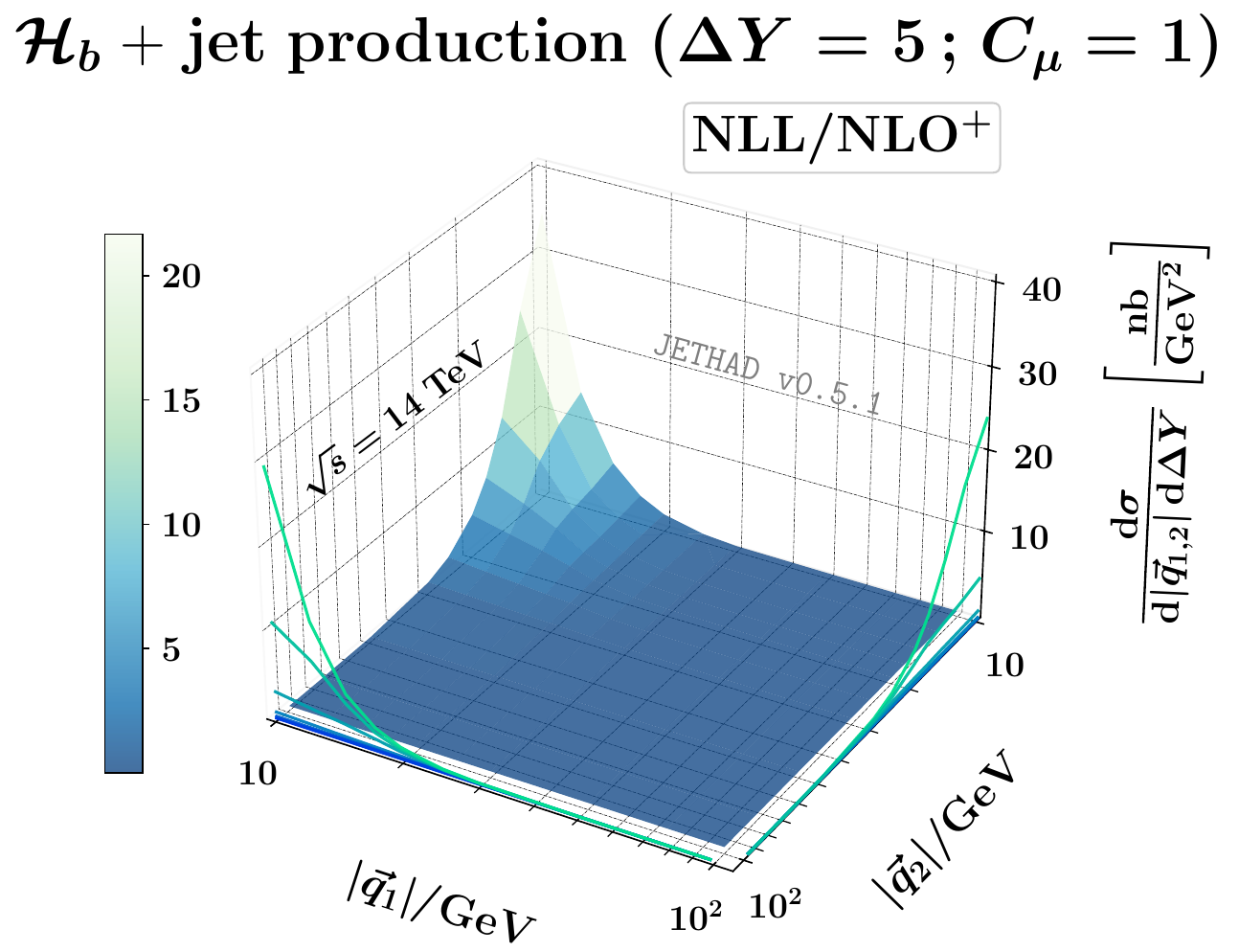}
   \vspace{0.45cm}

   \includegraphics[scale=0.31,clip]{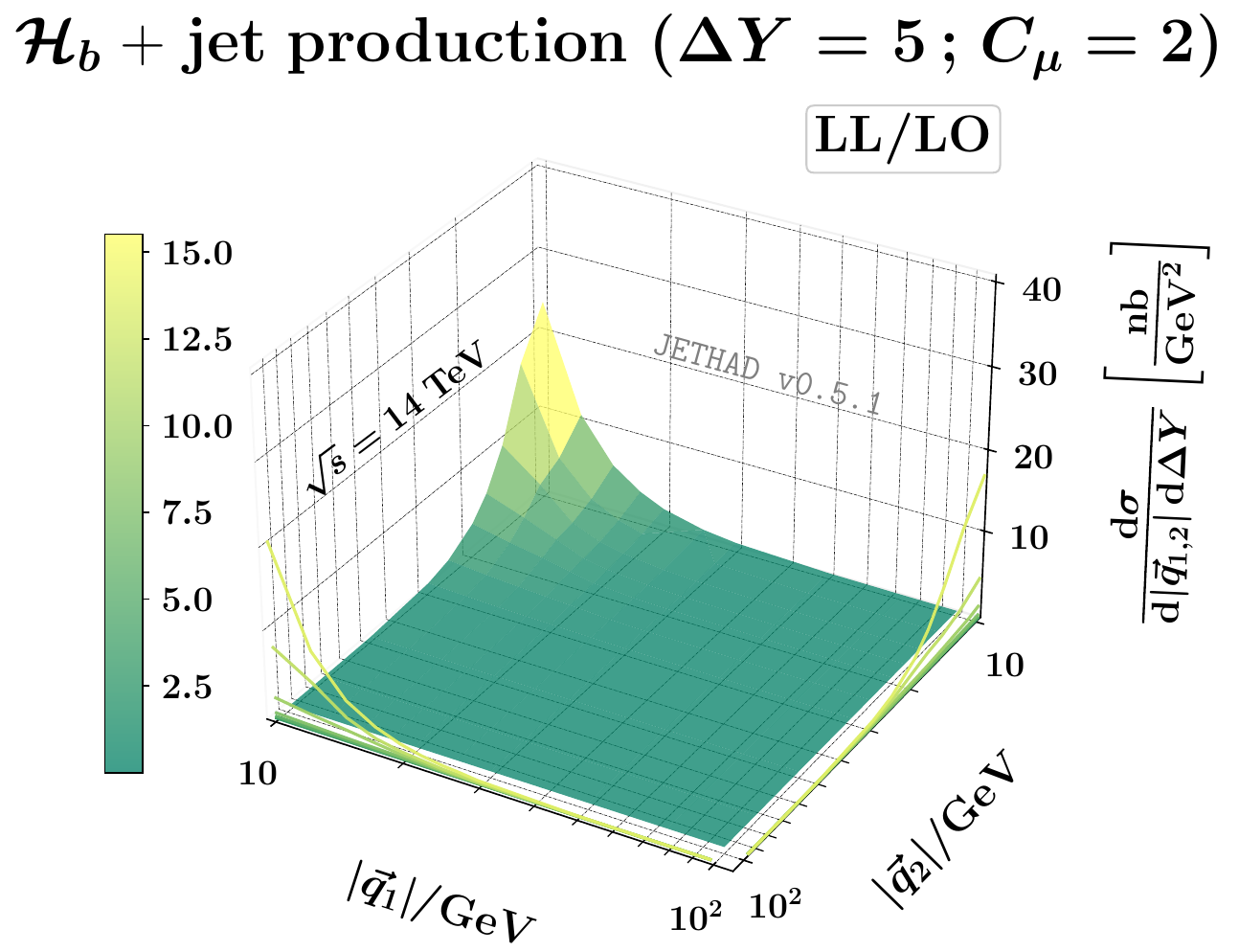}
   \hspace{0.25cm}
   \includegraphics[scale=0.31,clip]{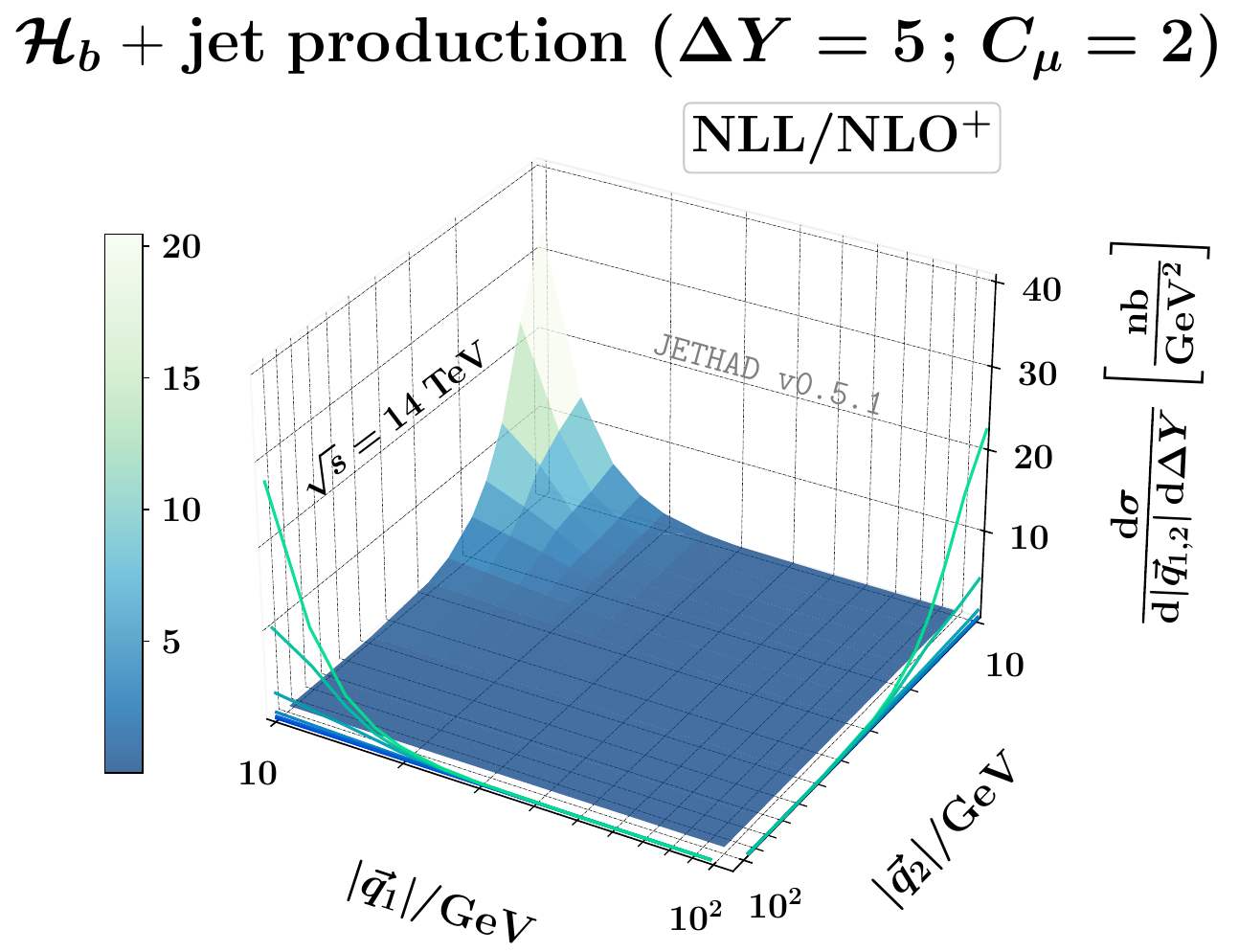}
   \vspace{0.45cm}

\caption{Doubly-differential transverse-momentum distribution for the $\Hb$-plus-jet production at $\DY=5$, $\sqrt{s} = 14$ TeV, and within the $\LL$ (left) and $\NLLp$ (right) accuracy. The $C_\mu$ scale parameter goes from 1/2 (top) to 2 (bottom).}
\label{fig:Y5q12_bJ}
\end{figure*}

\begin{figure*}[!t]
\centering

   \includegraphics[scale=0.31,clip]{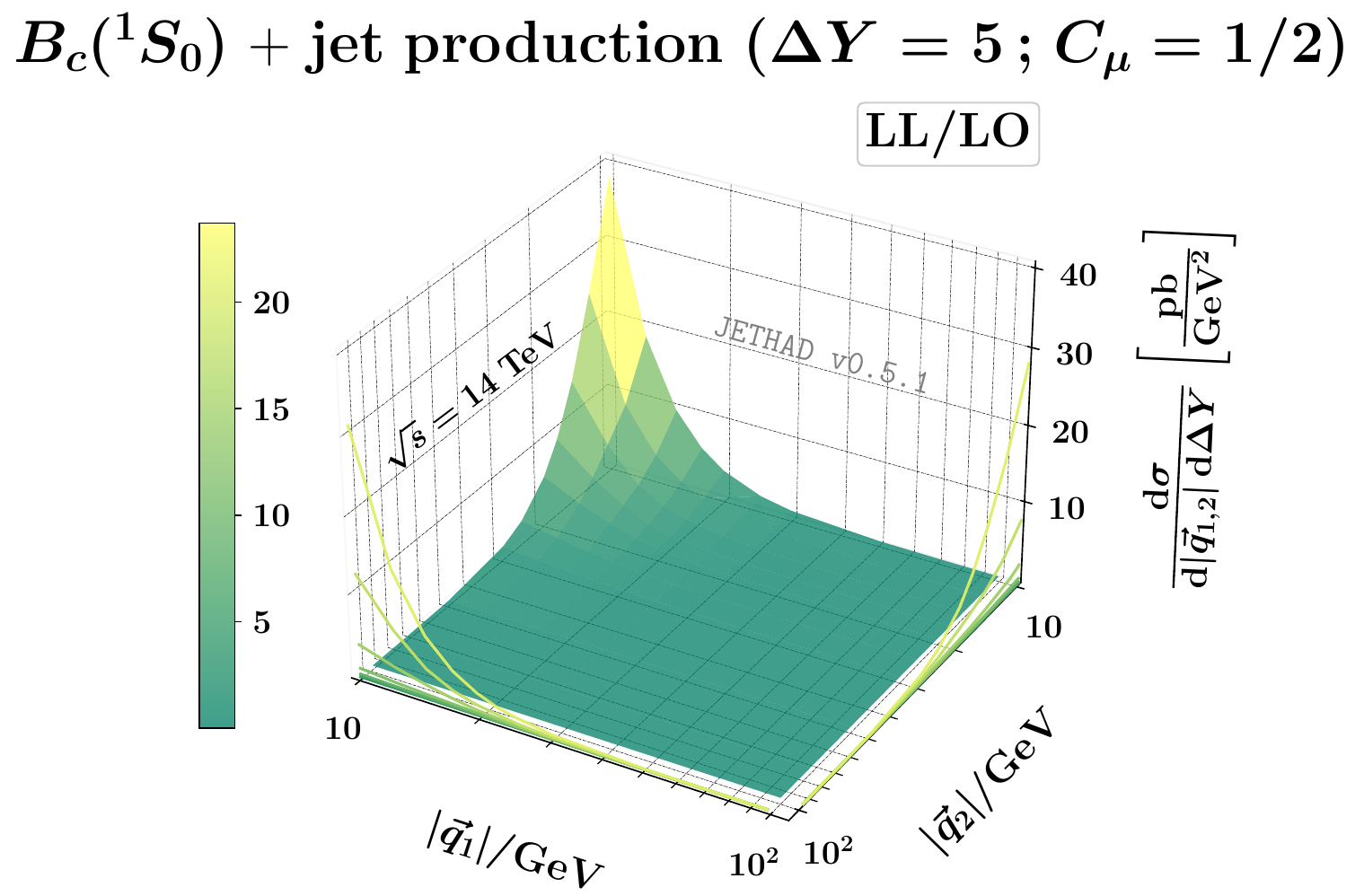}
   \hspace{0.25cm}
   \includegraphics[scale=0.31,clip]{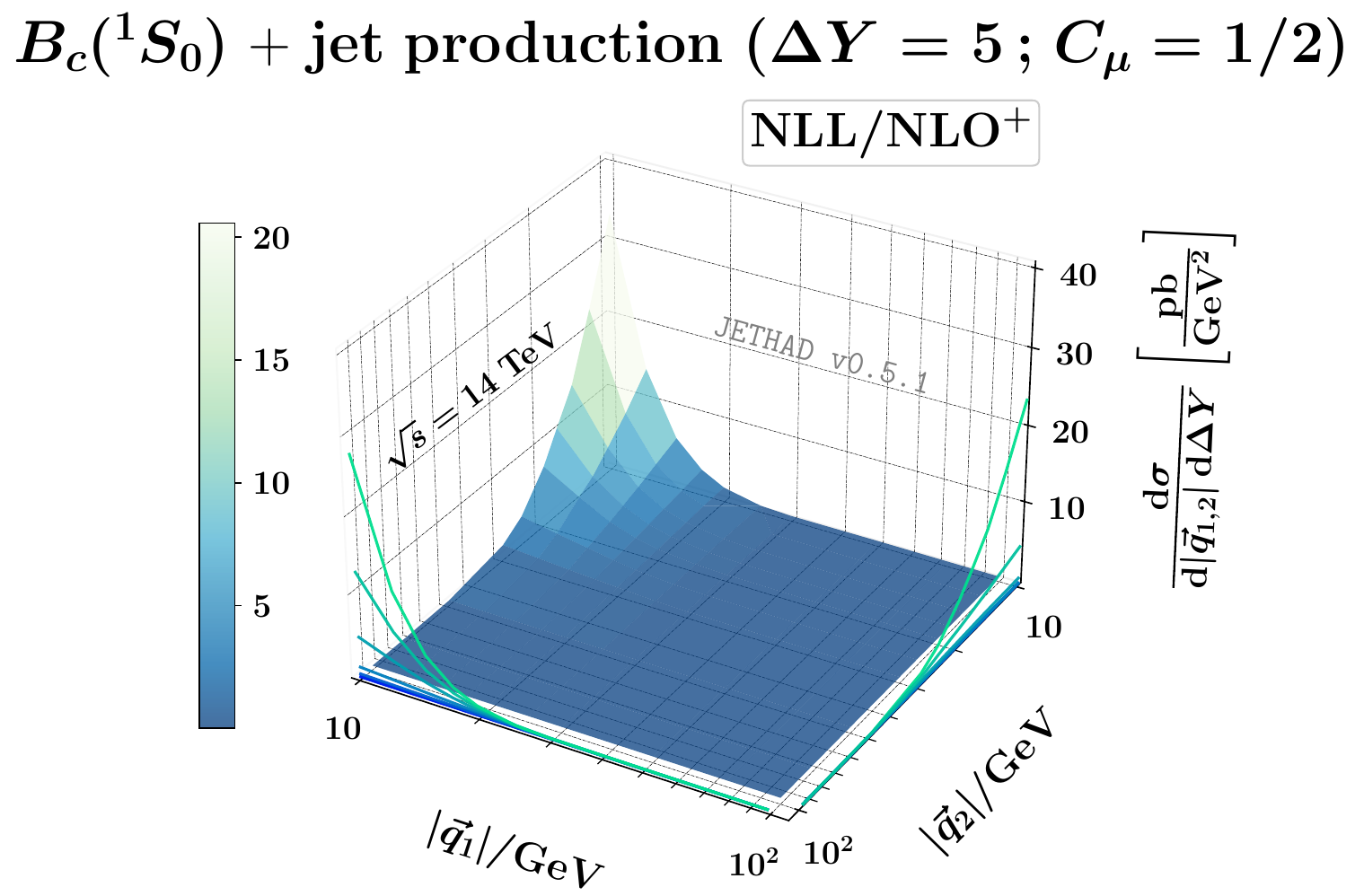}
   \vspace{0.45cm}

   \includegraphics[scale=0.31,clip]{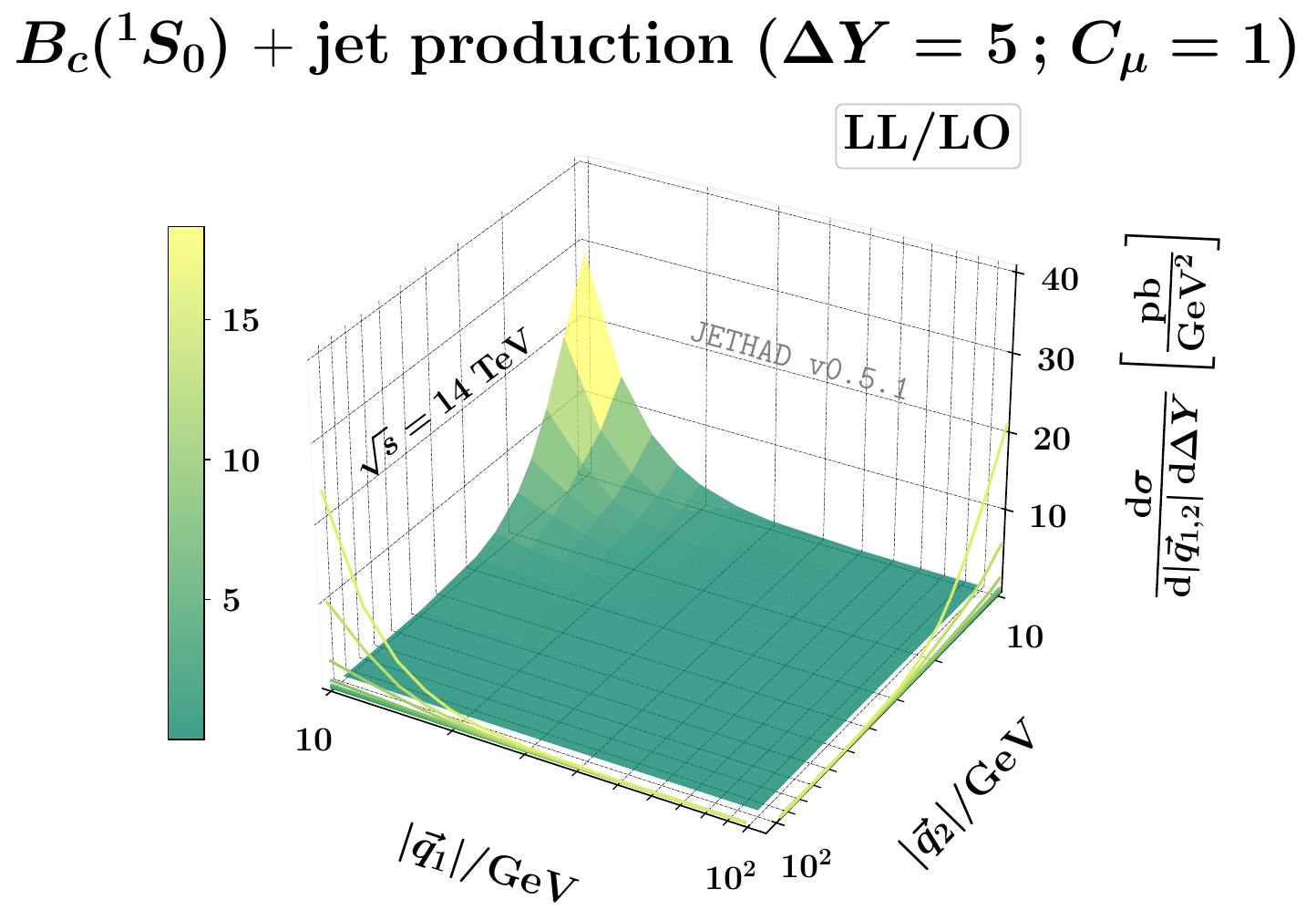}
   \hspace{0.25cm}
   \includegraphics[scale=0.31,clip]{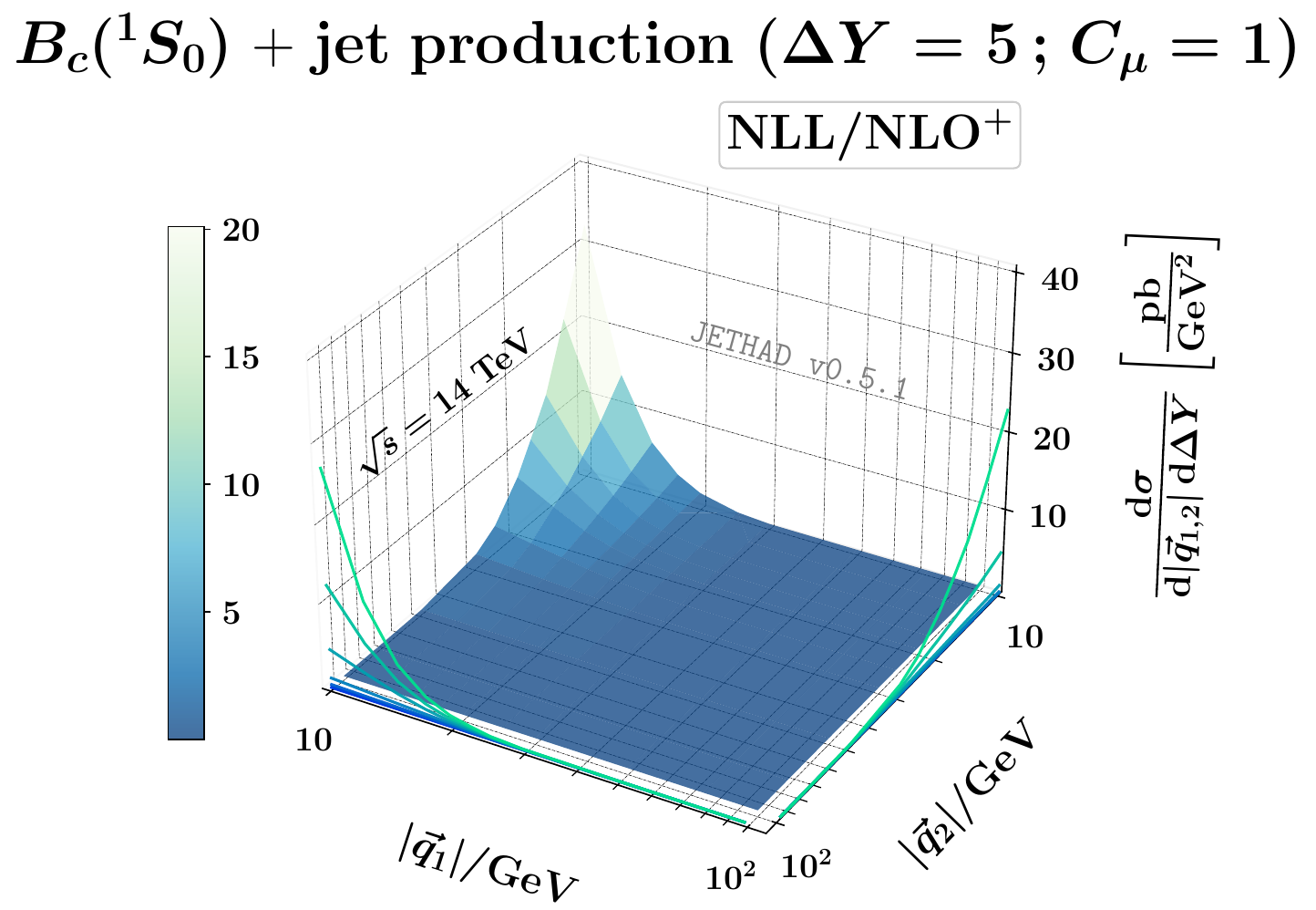}
   \vspace{0.45cm}

   \includegraphics[scale=0.31,clip]{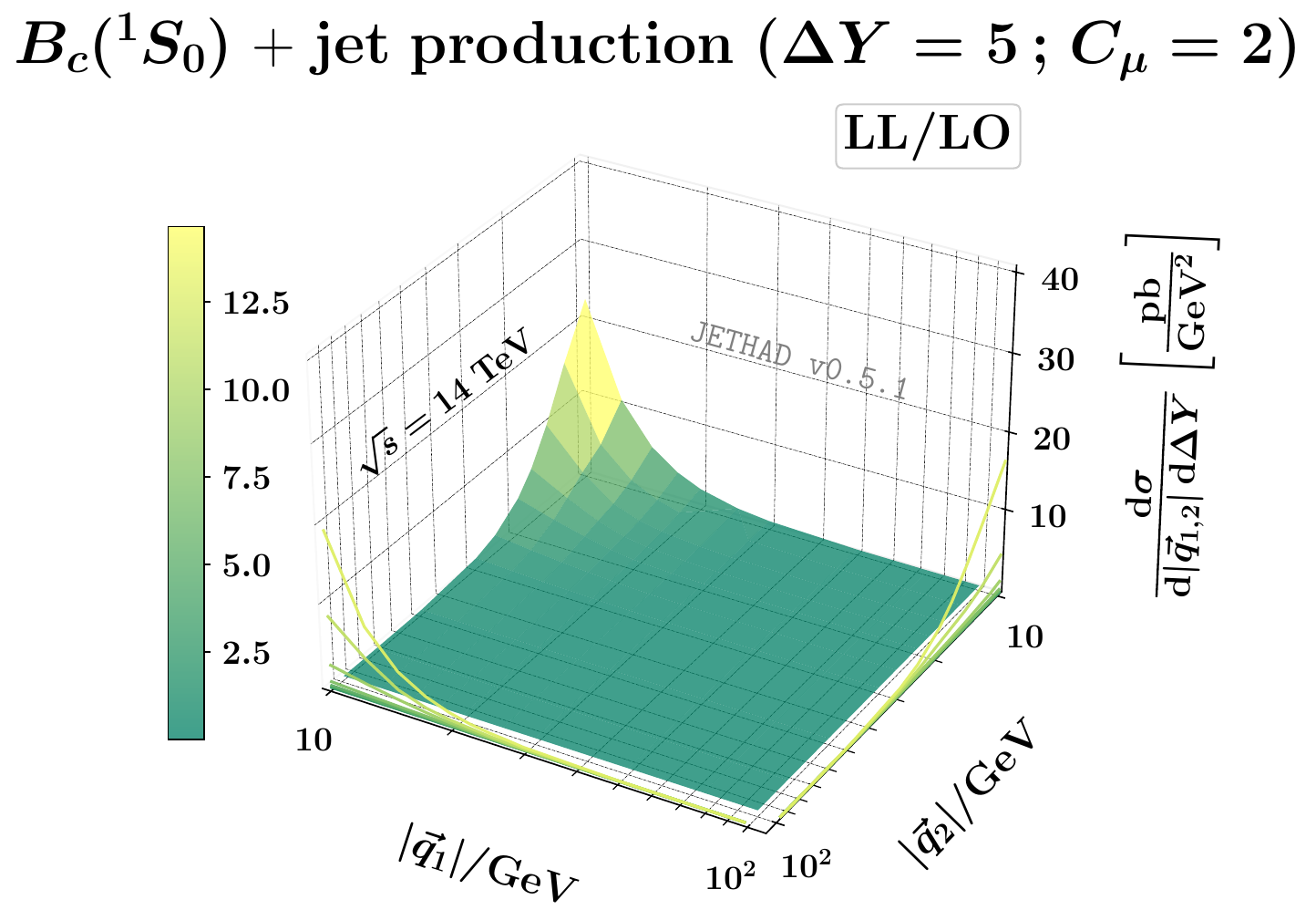}
   \hspace{0.25cm}
   \includegraphics[scale=0.31,clip]{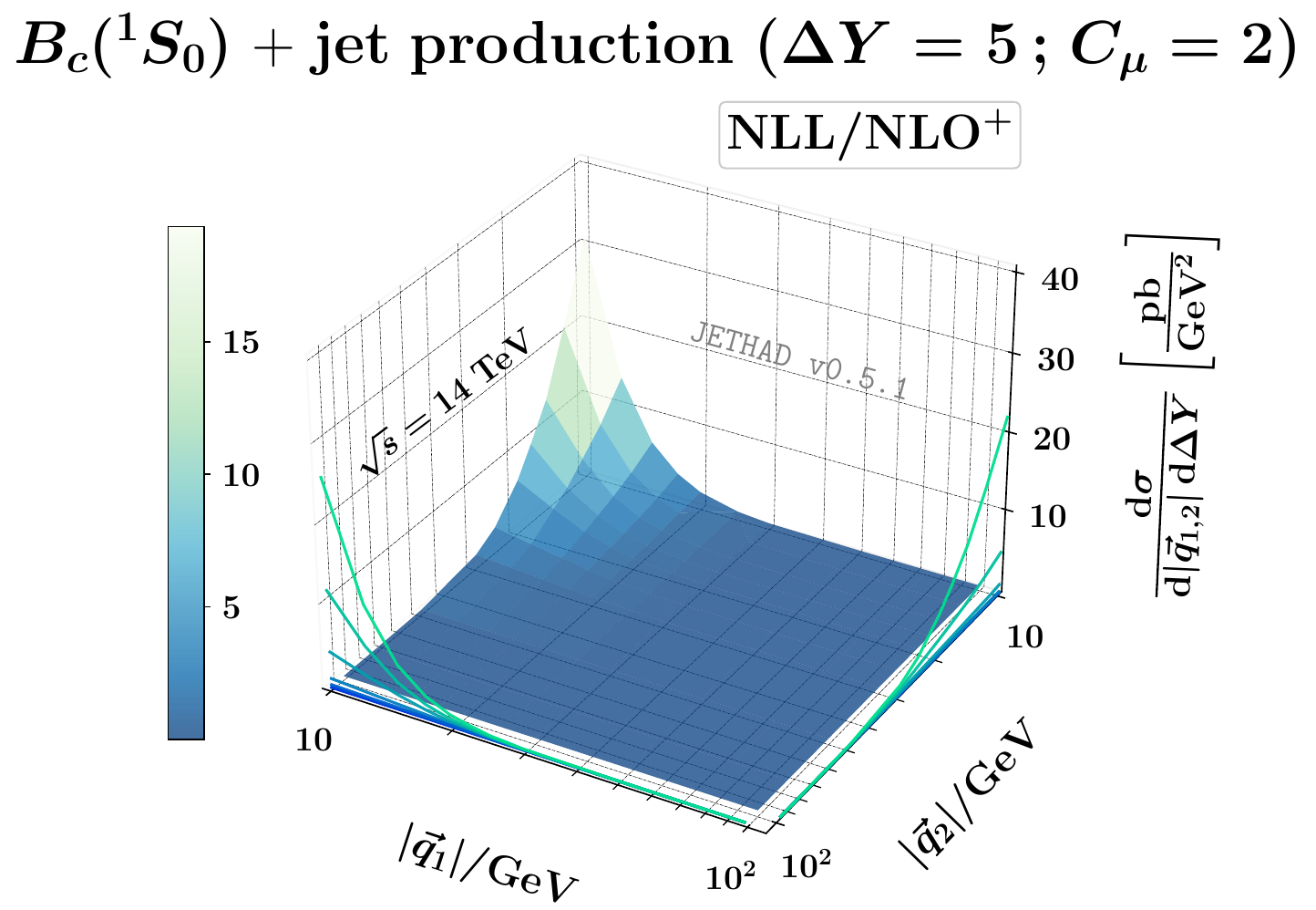}
   \vspace{0.45cm}

\caption{Doubly-differential transverse-momentum distribution for the $\BCs$-plus-jet production at $\DY=5$, $\sqrt{s} = 14$ TeV, and within the $\LL$ (left) and $\NLLp$ (right) accuracy. The $C_\mu$ scale parameter goes from 1/2 (top) to 2 (bottom).}
\label{fig:Y5q12_cJ}
\end{figure*}

\begin{figure*}[!t]
\centering

   \includegraphics[scale=0.31,clip]{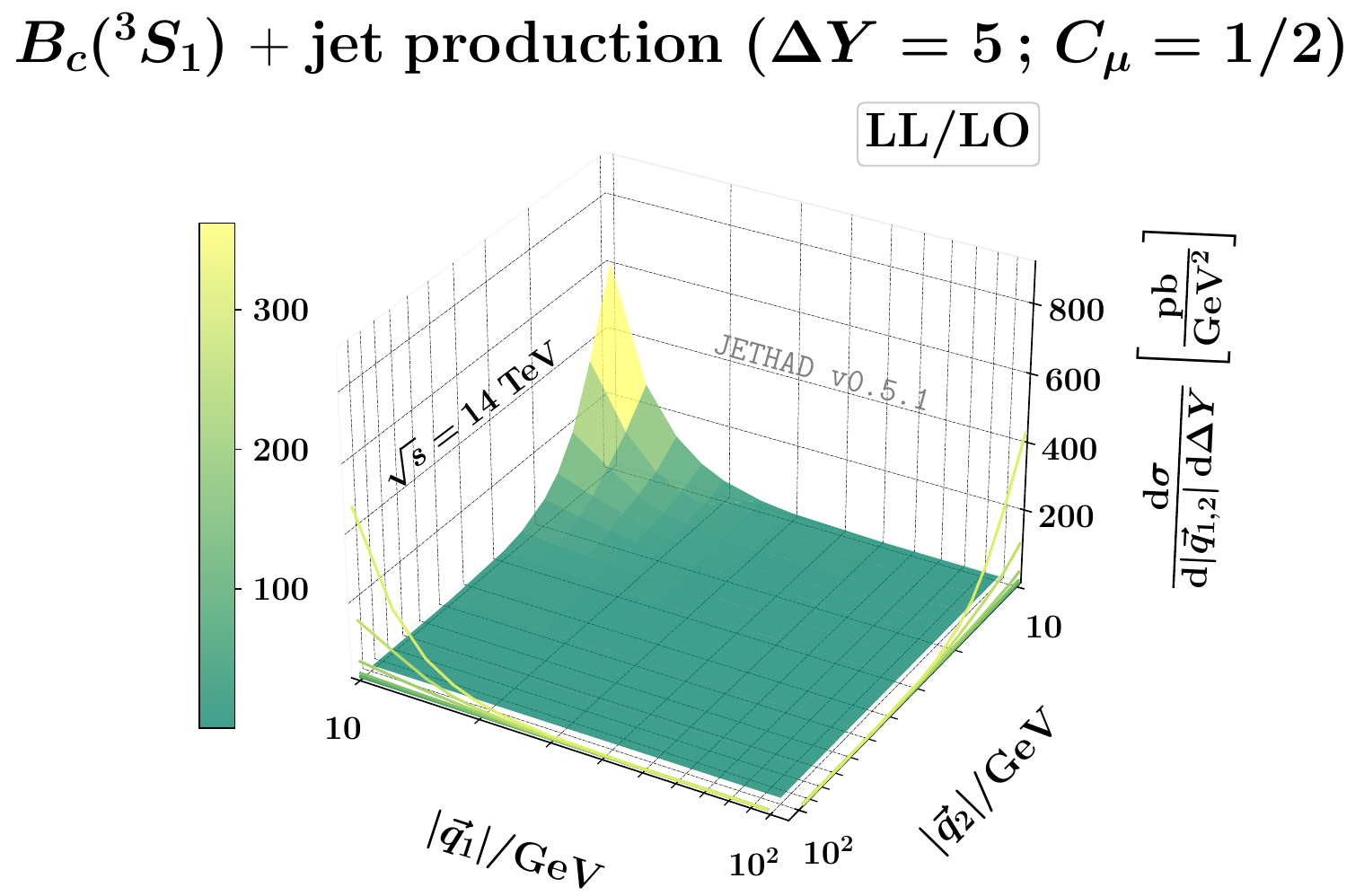}
   \hspace{0.25cm}
   \includegraphics[scale=0.31,clip]{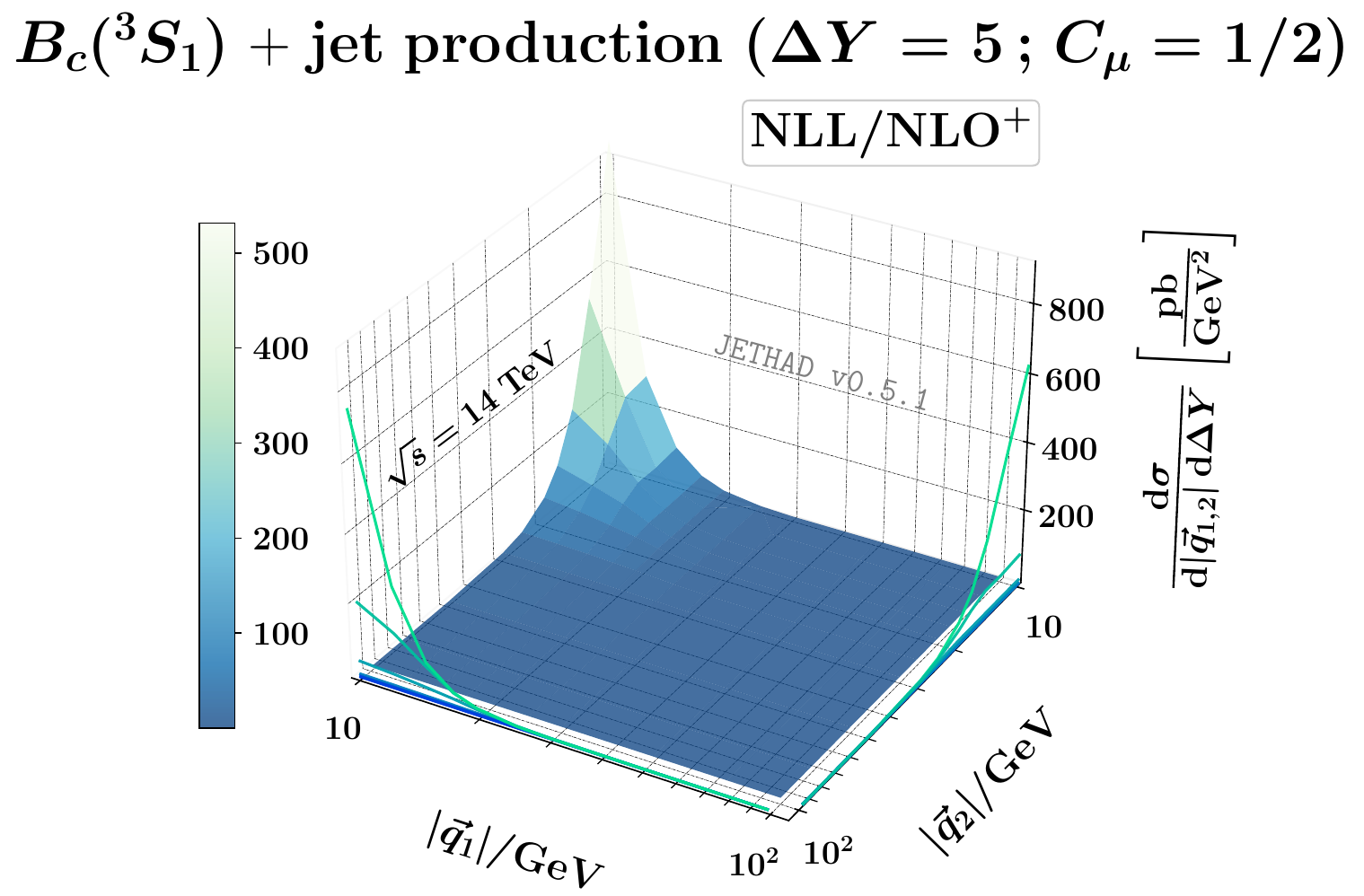}
   \vspace{0.45cm}

   \includegraphics[scale=0.31,clip]{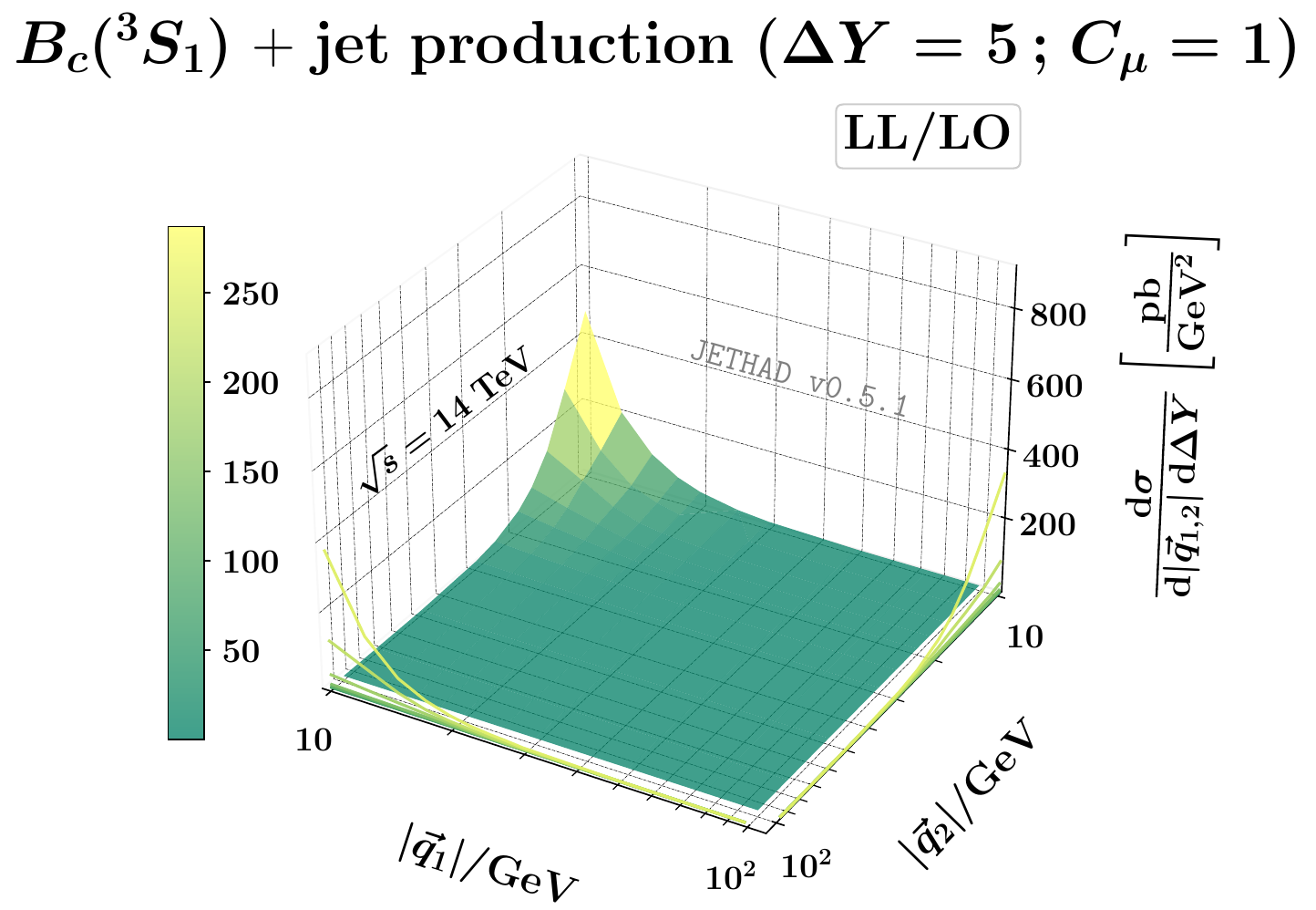}
   \hspace{0.25cm}
   \includegraphics[scale=0.31,clip]{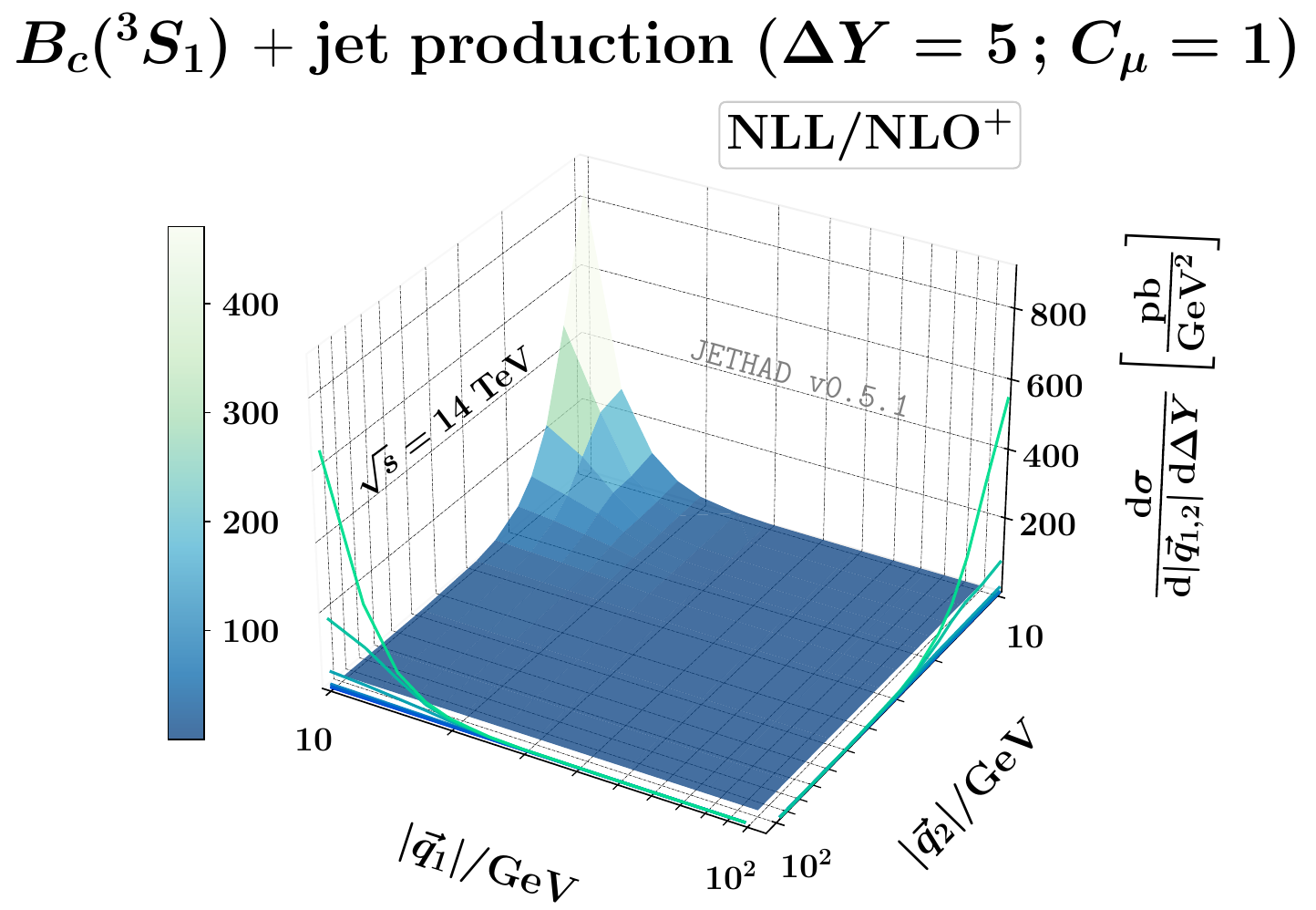}
   \vspace{0.45cm}

   \includegraphics[scale=0.31,clip]{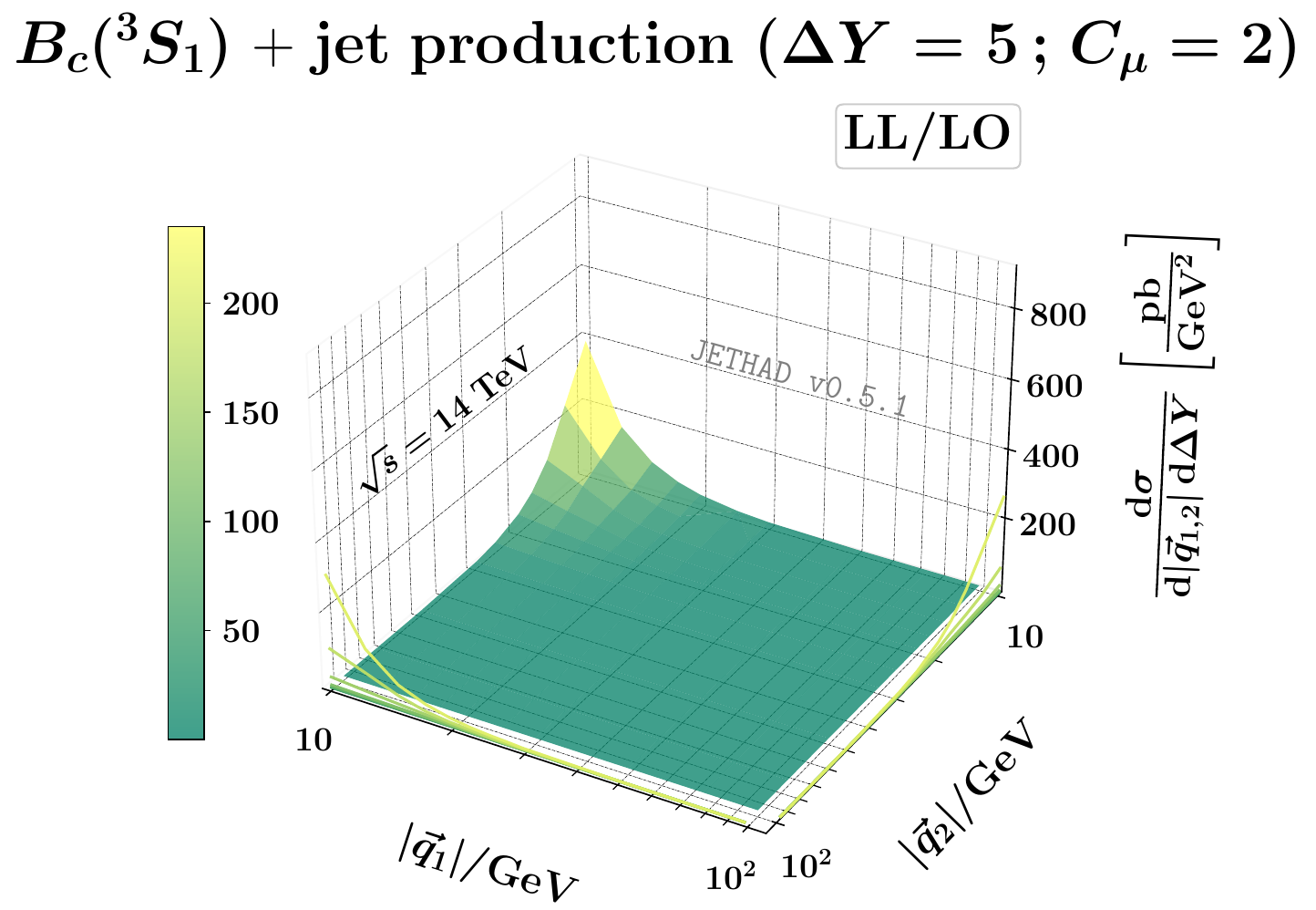}
   \hspace{0.25cm}
   \includegraphics[scale=0.31,clip]{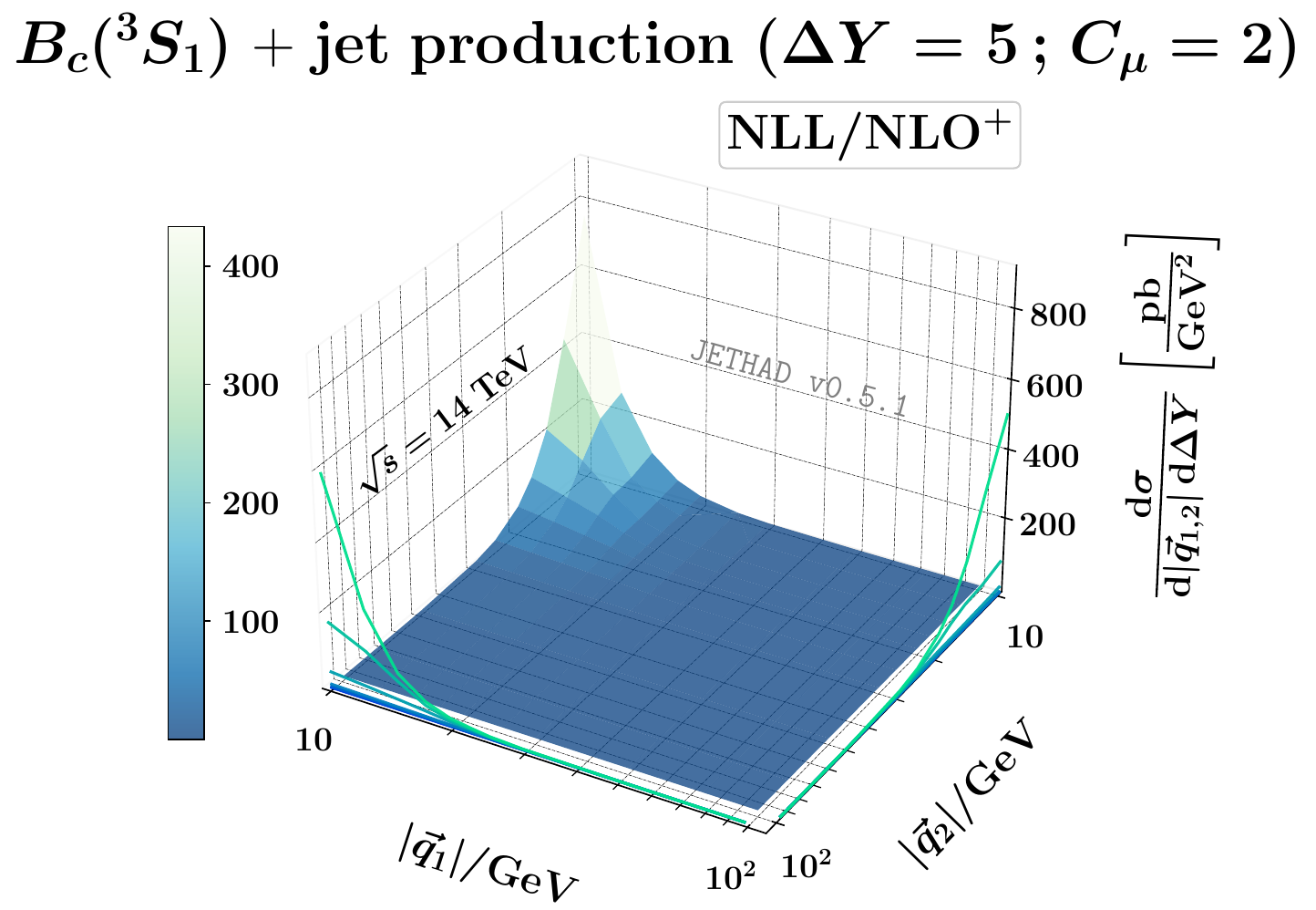}
   \vspace{0.45cm}

\caption{Doubly-differential transverse-momentum distribution for the $\BCs$-plus-jet production at $\DY=5$, $\sqrt{s} = 14$ TeV, and within the $\LL$ (left) and $\NLLp$ (right) accuracy. The $C_\mu$ scale parameter goes from 1/2 (top) to 2 (bottom).}
\label{fig:Y5q12_sJ}
\end{figure*}

Inclusive emissions of two-particle final states in proton collisions, such as photon~\cite{Cieri:2015rqa,Alioli:2020qrd,Becher:2020ugp,Neumann:2021zkb}, Higgs~\cite{Ferrera:2016prr} and $W^\pm$ boson~\cite{Ju:2021lah} pairs, as well as boson plus jet~\cite{Monni:2019yyr,Buonocore:2021akg} and $Z$ plus photon~\cite{Wiesemann:2020gbm} systems, were proposed as favorable channels whereby testing the TM resummation.
TM third-order fiducial rates for Drell--Yan and Higgs tags were respectively investigated in Refs.~\cite{Ebert:2020dfc,Re:2021con,Chen:2022cgv,Neumann:2022lft} and~\cite{Bizon:2017rah,Billis:2021ecs,Re:2021con,Caola:2022ayt}.
Ref.~\cite{Monni:2019yyr} presents a pioneering joint resummation of TM logarithms rising from the emission of two states.
There, the simultaneous measurement of transverse momenta of the Higgs boson and the leading jet was analyzed up to the NNLL accuracy by means of the {\RadISH} numeric technology~\cite{Bizon:2017rah}.
Ref.~\cite{Kallweit:2020gva} deals with a study of TM distributions for the fully leptonic $(W^+W^-)$ tag at the LHC.

Heavy-flavor distributions bring to a further complication. 
Indeed, when the transverse momentum of a heavy hadron diminishes, its transverse mass comes closer and closer to heavy-quark masses serving as DGLAP thresholds, even crossing then when energy scales are moved below their natural values (as an example, this happens when the $C_\mu$ parameter introduced in Section~\ref{ssec:uncertainty} is set to 1/2). 
In this case, the validity of a pure VFNS approach becomes questionable.

In this Section we consider cross sections at fixed values of $\DY$ and differential in the transverse momenta of both the final-state objects.
In the kinematic sectors matter of our investigations energy logarithms arising from the semi-hard scale hierarchy are large.
At the same time, however, also TM logarithms are potentially sizable. 
Therefore, we propose this study without claiming to catch the core features of these observables thanks to our hybrid-factorization framework, but rather to set the stage for futures analyses aimed at shedding light on the formal and phenomenological connection among distinct resummation mechanisms.
Figures presented here are for doubly differential transverse-momentum distributions at $\DY=5$. For the sake of simplicity, we present the hadron-plus-jet channel only (right diagram of Fig.~\ref{fig:process}).
Predictions for singly bottomed hadrons, $\BCs$ mesons, and their $\Bss$ resonances are shown in surface 3D plots of Figs.~\ref{fig:Y5q12_bJ}, \ref{fig:Y5q12_cJ}, and~\ref{fig:Y5q12_sJ}, respectively.

The common trend is a sharp falloff pattern when the two transverse momenta, $|\vec q_1|$ and $|\vec q_2|$, increase or when their mutual distance grows. 
$\LL$ cross sections (left plots) are much more sensitive to MHOU studies done through energy-scale variations than $\NLLp$ ones.
They generally lower with $C_\mu$ (from upper to lower plots).
$\NLLp$ distributions (right panels) instead generally oscillate around $C_\mu = 1$, which turns out to work as a critical point.
This is a clear signal that our $\NLLp$ doubly-differential cross sections are quite stable under MHOU analyses.
Notably, no distribution comes with a peak, which might however be present in the low-TM range (excluded from our study), where TM-resummation effects would dominate.
The information contained in the core of our 3D
figures is supplemented by 2D slice projections highlighting the behavior of our observables at $|\vec q_1| = 0$ and at $|\vec q_2| = 0$.
What emerges from the inspection of these 2D contour plots at low or intermediate transverse momentum is a smaller statistics in the $|\vec q_1| > |\vec q_2|$ range than in the opposite case.
This happens because cross sections are globally smaller when a bottomed hadron is produced rather than a light-flavored jet~\cite{Celiberto:2021fdp,Celiberto:2022keu}.

Finally, numeric checks on our transverse-momentum distributions have shown a noncharmed versus charmed $B$-meson production-rate hierarchy analogous to the one found in the case of rapidity-interval distributions (see Section~\ref{ssec:rapidity_distributions}).
$B_c$ production rates are almost always three orders of magnitude smaller than $\Hb$ ones, whereas rates of the $B_c^*$ resonance range from $10^{-2}$ to $1/60$ times the ones of $\Hb$ hadrons.

\section{Summary and Outlooks}
\label{sec:conclusions}

We considered the semi-inclusive production of a forward noncharmed or charmed $b$-hadron in association with a backward noncharmed $b$-hadron or a jet in high-energy regimes that can be accessed at current LHC analyses and its forthcoming Hi-Lumi upgrade.
We depicted the production of bottom-flavored hadrons by means of a VFNS collinear-fragmentation approach, whose validity is justified by the large transverse-momentum regime matter of our analysis.

Singly heavy-flavored particles, namely inclusive states consisting of $B$ mesons and $\Lambda_b$ baryons, were described via the {\tt KKSS07}~\cite{Kniehl:2008zza,Kramer:2018vde} NLO FFs based on a fit to global data and assuming a three-parameter power-like initial-scale parametrization for $b$ and $\bar b$ fragmentation channels~\cite{Kartvelishvili:1985ac}.
As for charmed $B$ mesons, \emph{i.e.} $B_c \equiv \BCs$ states and $B_c^* \equiv \Bss$ resonances, we adopted a more sophisticated strategy.
First, we modeled the initial-scale input for the fragmentation of heavy quarks (charm, bottom, and their antiparticles) and gluons to observed hadrons by means of corresponding NLO calculations within the NRQCD effective theory~\cite{Feng:2021qjm,Feng:2018ulg}.
Then, we plugged these inputs into the {\tt APFEL(++)} DGLAP evolution code~\cite{Bertone:2013vaa,Carrazza:2014gfa,Bertone:2017gds} to generate phenomenology-ready VFNS FF grids in {\tt LHAPDF} format~\cite{Buckley:2014ana}, which we named {\tt ZCFW22} NLO functions~\cite{Celiberto:2022keu}.
These novel FF sets will play a role in future phenomenological applications lying outside the semi-hard regime.
They also will serve as a support basis to shed light on the intersection corner between the collinear factorization and the NRQCD formalism.

By relying upon a traditional as well as an extended MHOU scan on our hybrid-factorization setup, we hunted for clues of stabilization of the high-energy resummation under next-to-leading logarithmic corrections.
We provided strong evidence that these effects are present and they allow for a fair description of high-energy distributions, differential in the rapidity interval and in the final-state transverse momenta.
As a bonus, we highlighted that the predicted production-rate hierarchy between noncharmed $b$-hadrons and charmed $\BCs$ mesons is in line with recent LHCb estimates.
In particular, the production rate of $B_c$ states is three orders of magnitude lower than the $\Hb$ one.
This served as a simultaneous benchmark for both the hybrid factorization and the NRQCD fragmentation applied to charmed $B$ mesons at our reference transverse masses (see Section~\ref{ssec:kinematics}).

Our analysis on bottom-flavor collinear fragmentation stands as a step forward in the study of heavy-flavored emissions within the $\NLLpp$ accuracy, originally started from the analytic calculation of singly off-shell emission functions depicting the production of heavy-quark pairs in the moderate-to-low transverse-momentum regime~\cite{Celiberto:2017nyx,Bolognino:2019ouc,Bolognino:2019yls}. 
It also represents a fair contact point with phenomenology of bottom-flavor production in the high-energy sector.
The extension of this study to wider kinematic ranges, like the ones typical of new-generation colliding machines~\cite{Chapon:2020heu,Anchordoqui:2021ghd,Feng:2022inv,Hentschinski:2022xnd,Accardi:2012qut,AbdulKhalek:2021gbh,Khalek:2022bzd,AlexanderAryshev:2022pkx,Arbuzov:2020cqg,Amoroso:2022eow,MuonCollider:2022xlm,Aime:2022flm,MuonCollider:2022ded,Accettura:2023ked,Black:2022cth,Dawson:2022zbb,Bose:2022obr,Begel:2022kwp,Abir:2023fpo,Accardi:2023chb}, as well as possible contact points with heavy-flavor fragmentation studies in the TMD formalism~\cite{Echevarria:2020qjk,Echevarria:2023dme,Copeland:2023wbu,Copeland:2023qed,vonKuk:2023jfd,Dai:2023rvd,Kang:2023elg}, is underway.

As a prospect, we plan to investigate bottom-flavor emissions in the moderate transverse-momentum sector, namely at the intersection line between the VFNS and the FFNS scheme, and possibly set the stage for a matching between the two descriptions.
Here, high-energy tags of charmed $B$ mesons will bring novelty to the plan.
In that case, indeed, the VFNS versus FFNS dichotomy translates into a short-distance versus fragmentation dichotomy.
This calls for a further enhancement of our {\tt ZCFW22}-based treatment, which currently relies upon NRQCD-modeled calculations for initial-scale inputs, but will serve in the medium-term future as a ``launch pad'' for extractions through fits to new data to be collected at the Hi-Lumi LHC.
In this respect, neural-network strategies already employed to extract collinear FFs for lighter hadron species will be an asset~\cite{Nocera:2017qgb,Bertone:2017xsf,Bertone:2017tyb,Bertone:2018ecm,Khalek:2021gxf,Khalek:2022vgy}.
As mentioned in Section~\ref{ssec:natural_stabilization}, a proper determination of FF uncertainties with possible inclusion of MHOU effects, not yet encoded in our {\tt ZCFW22} sets, is needed.
It might take inspiration from the {\tt MCscales} approach~\cite{Kassabov:2022orn} (for close-in-spirit studies on proton and pion PDFs, see Refs.~\cite{Harland-Lang:2018bxd,Ball:2021icz,McGowan:2022nag,NNPDF:2024dpb} and Ref.~\cite{Pasquini:2023aaf}).

The fair stability exhibited by our $\NLLp$ distributions motivates our interest in proposing the hybrid factorization as useful tool to enhance and extend the standard collinear description.
To reach the precision level, our $\NLLp$ hybrid approach needs to evolve into a \emph{multi-lateral} formalism where different resummation mechanisms are at work.
As a first steps, links with the soft-gluon~\cite{Hatta:2020bgy,Hatta:2021jcd,Caucal:2022ulg,Taels:2022tza} and the jet-radius~\cite{Dasgupta:2014yra,Dasgupta:2016bnd,Banfi:2012jm,Banfi:2015pju,Liu:2017pbb}\footnote{In Ref.~\cite{Celiberto:2022kxx} it was shown that high-energy angular multiplicities become more and more sensitive to jet-radius variations as the final-state azimuthal-angle distance grows.}
resummations should be found.
Possible common grounds with studies on jet angularities~\cite{Luisoni:2015xha,Caletti:2021oor,Reichelt:2021svh} also represent an intriguing perspective.

Then, the resummation of energy logarithms from BFKL could improve the description of DGLAP-evolved bottom-flavor FFs, its effect becoming sizable already at moderate values of final-state hadron longitudinal fraction~\cite{HF-HpT_2023_Conversations_UZC}.
Another engaging development would be exploring the validity of the use of a ZM-VFNS with higher HFMPs in the context of heavy-meson fragmentation~\cite{Bertone:2017djs}.
Finally, the striking evidence of intrinsic charm quarks in the proton~\cite{Ball:2022qks}, now corroborated by quite recent studies on its valence distribution~\cite{NNPDF:2023tyk}, traces the path towards searches for imprints of the possible existence of an intrinsic bottom component.
In this respect, any progress in widening the horizons of our knowledge of bottom physics is needed.
We believe that studies aimed at shedding light on the bottom-flavor collinear fragmentation go along these directions.

\section*{Acknowledgements}

The author would like to express his gratitude to colleagues of the \textbf{Heavy Flavours at High pT Workshop 2023} (Higgs Centre for Theoretical Physics, University of Edinburgh) and the \textbf{Milan Christmas Meeting 2023} (University of Milan) for the welcoming atmosphere and for insightful discussions.
The author is grateful to Alessandro Papa for a critical reading of the manuscript, useful suggestions, and encouragement.
This work is supported by the Atracci\'on de Talento Grant n. 2022-T1/TIC-24176 of the Comunidad Aut\'onoma de Madrid, Spain.

\appendix

\setcounter{appcnt}{0}
\hypertarget{app:NLOBIF}{
\section*{Appendix~A: NLO $b$-hadron emission function}}
\label{app:NLOBIF}

In this Appendix we provide the expression of the NLO correction for the emission function of a forward $b$-hadron tagged at large transverse momentum (see Ref.~\cite{Ivanov:2012iv} for technical details). We write

\begin{equation}
  \label{onium_IF_NLO}
  \hat \E_\B(n,\nu,|\vec q_\B|,x_\B)=
  \frac{1}{\pi}\sqrt{\frac{C_F}{C_A}}
  \left(|\vec q_\B|^2\right)^{i\nu-\frac{1}{2}}
  \int_{x_\B}^1\frac{\drv x}{x}
  \int_{\frac{x_\B}{x}}^1\frac{\drv \tau}{\tau}
  \left(\frac{x\tau}{x_\B}\right)^{2i\nu-1}
\end{equation}
  \[ \times \,
  \left[
  \frac{C_A}{C_F}f_g(x)D_g^\B\left(\frac{x_\B}{x\tau}\right)C_{gg}
  \left(x,\tau\right)+\sum_{i=q\bar q}f_i(x)D_i^\B
  \left(\frac{x_\B}{x\tau}
  \right)C_{qq}\left(x,\tau\right)
  \right.
  \]
  \[ \times \,
  \left.D_g^\B\left(\frac{x_\B}{x\tau}\right)
  \sum_{i=q\bar q}f_i(x)C_{qg}
  \left(x,\tau\right)+\frac{C_A}{C_F}f_g(x)\sum_{i=q\bar q}D_i^\B
  \left(\frac{x_\B}{x\tau}\right)C_{gq}\left(x,\tau\right)
  \right]\, ,
  \]
with
\begin{equation}
\stepcounter{appcnt}
\label{Cgg_hadron}
 C_{gg}\left(x,\tau\right) =  P_{gg}(\tau)\left(1+\tau^{-2\gamma}\right)
 \ln \left( \frac {|\vec q_\B|^2 x^2 \tau^2 }{\mu_F^2 x_\B^2}\right)
 -\frac{\beta_0}{2}\ln \left( \frac {|\vec q_\B|^2 x^2 \tau^2 }
 {\mu^2_R x_\B^2}\right)
\end{equation}
\[
 + \, \delta(1-\tau)\left[C_A \ln\left(\frac{s_0 \, x^2}{|\vec q_\B|^2 \,
 x_\B^2 }\right) \chi(n,\gamma)
 - C_A\left(\frac{67}{18}-\frac{\pi^2}{2}\right)+\frac{5}{9}n_f
 \right.
\]
\[
 \left.
 +\frac{C_A}{2}\left(\psi^\prime\left(1+\gamma+\frac{n}{2}\right)
 -\psi^\prime\left(\frac{n}{2}-\gamma\right)
 -\chi^2(n,\gamma)\right) \right]
 + \, C_A \left(\frac{1}{\tau}+\frac{1}{(1-\tau)_+}-2+\tau\bar\tau\right)
\]
\[
 \times \, \left(\chi(n,\gamma)(1+\tau^{-2\gamma})-2(1+2\tau^{-2\gamma})\ln\tau
 +\frac{\bar \tau^2}{\tau^2}{\cal I}_2\right)
\]
\[
 + \, 2 \, C_A (1+\tau^{-2\gamma})
 \left(\left(\frac{1}{\tau}-2+\tau\bar\tau\right) \ln\bar\tau
 +\left(\frac{\ln(1-\tau)}{1-\tau}\right)_+\right) \ ,
\]

\begin{equation}
\stepcounter{appcnt}
\label{Cgq_hadron}
 C_{gq}\left(x,\tau\right)=P_{qg}(\tau)\left(\frac{C_F}{C_A}+\tau^{-2\gamma}\right)\ln \left( \frac {|\vec q_\B|^2 x^2 \tau^2 }{\mu_F^2 x_\B^2}\right)
\end{equation}
\[
 + \, 2 \, \tau \bar\tau \, T_R \, \left(\frac{C_F}{C_A}+\tau^{-2\gamma}\right)+\, P_{qg}(\tau)\, \left(\frac{C_F}{C_A}\, \chi(n,\gamma)+2 \tau^{-2\gamma}\,\ln\frac{\bar\tau}{\tau} + \frac{\bar \tau}{\tau}{\cal I}_3\right) \ ,
\]

\begin{equation}
\stepcounter{appcnt}
\label{qg}
 C_{qg}\left(x,\tau\right) =  P_{gq}(\tau)\left(\frac{C_A}{C_F}+\tau^{-2\gamma}\right)\ln \left( \frac {|\vec q_\B|^2 x^2 \tau^2 }{\mu_F^2 x_\B^2}\right)
\end{equation}
\[
 + \tau\left(C_F\tau^{-2\gamma}+C_A\right) + \, \frac{1+\bar \tau^2}{\tau}\left[C_F\tau^{-2\gamma}(\chi(n,\gamma)-2\ln\tau)+2C_A\ln\frac{\bar \tau}{\tau} + \frac{\bar \tau}{\tau}{\cal I}_1\right] \ ,
\]
and
\begin{equation}
\stepcounter{appcnt}
\label{Cqq_hadron}
 C_{qq}\left(x,\tau\right)=P_{qq}(\tau)\left(1+\tau^{-2\gamma}\right)\ln \left( \frac {|\vec q_\B|^2 x^2 \tau^2 }{\mu_F^2 x_\B^2}\right)-\frac{\beta_0}{2}\ln \left( \frac {|\vec q_\B|^2 x^2 \tau^2 }{\mu^2_R x_\B^2}\right)
\end{equation}
\[
 + \, \delta(1-\tau)\left[C_A \ln\left(\frac{s_0 \, x_\B^2}{|\vec q_\B|^2 \, x^2 }\right) \chi(n,\gamma)+ C_A\left(\frac{85}{18}+\frac{\pi^2}{2}\right)-\frac{5}{9}n_f - 8\, C_F \right.
\]
\[
 \left. +\frac{C_A}{2}\left(\psi^\prime\left(1+\gamma+\frac{n}{2}\right)-\psi^\prime\left(\frac{n}{2}-\gamma\right)-\chi^2(n,\gamma)\right) \right] + \, C_F \,\bar \tau\,(1+\tau^{-2\gamma})
\]
\[
 +\left(1+\tau^2\right)\left[C_A (1+\tau^{-2\gamma})\frac{\chi(n,\gamma)}{2(1-\tau )_+}+\left(C_A-2\, C_F(1+\tau^{-2\gamma})\right)\frac{\ln \tau}{1-\tau}\right]
\]
\[
 +\, \left(C_F-\frac{C_A}{2}\right)\left(1+\tau^2\right)\left[2(1+\tau^{-2\gamma})\left(\frac{\ln (1-\tau)}{1-\tau}\right)_+ + \frac{\bar \tau}{\tau^2}{\cal I}_2\right] \; ,
\]

In our expressions $s_0$ is an energy scale that genuinely appears within the BFKL approach. We fix $s_0 \equiv \mu_C$.
Then we define $\bar \tau = 1 - \tau$ and $\gamma = i \nu - \frac{1}{2}$. The LO DGLAP $P_{i j}(\tau)$ splitting functions are
\begin{eqnarray}
\stepcounter{appcnt}
\label{DGLAP_kernels}
 P_{gq}(y)&=&C_F\frac{1+(1-y)^2}{y} \; , \\ \nonumber
 P_{qg}(y)&=&T_R\left[y^2+(1-y)^2\right]\; , \\ \nonumber
 P_{qq}(y)&=&C_F\left( \frac{1+y^2}{1-y} \right)_+= C_F\left[ \frac{1+y^2}{(1-y)_+} +{3\over 2}\delta(1-y)\right]\; , \\ \nonumber
 P_{gg}(y)&=&2C_A\left[\frac{1}{(1-y)_+} +\frac{1}{y} -2+y(1-y)\right]+\left({11\over 6}C_A-\frac{n_f}{3}\right)\delta(1-y) \; .
\end{eqnarray}
The ${\cal I}_{2,1,3}$ functions read
\begin{equation}
\stepcounter{appcnt}
\label{I2}
{\cal I}_2=
\frac{\tau^2}{\bar \tau^2}\left[
\tau\left(\frac{{}_2F_1(1,1+\gamma-\frac{n}{2},2+\gamma-\frac{n}{2},\tau)}
{\frac{n}{2}-\gamma-1}-
\frac{{}_2F_1(1,1+\gamma+\frac{n}{2},2+\gamma+\frac{n}{2},\tau)}{\frac{n}{2}+
\gamma+1}\right)\right.
\end{equation}
\[
 \stepcounter{appcnt}
 \left.
 +\tau^{-2\gamma}\left(\frac{{}_2F_1(1,-\gamma-\frac{n}{2},1-\gamma-\frac{n}{2},\tau)}{\frac{n}{2}+\gamma}-\frac{{}_2F_1(1,-\gamma+\frac{n}{2},1-\gamma+\frac{n}{2},\tau)}{\frac{n}{2} -\gamma}\right)
\right.
\]
\[
 \left.
 +\left(1+\tau^{-2\gamma}\right)\left(\chi(n,\gamma)-2\ln \bar \tau \right)+2\ln{\tau}\right] \; ,
\]
\begin{equation}
\stepcounter{appcnt}
\label{I1}
 {\cal I}_1=\frac{\bar \tau}{2\tau}{\cal I}_2+\frac{\tau}{\bar \tau}\left[\ln \tau+\frac{1-\tau^{-2\gamma}}{2}\left(\chi(n,\gamma)-2\ln \bar \tau\right)\right] \; ,
\end{equation}
and
\begin{equation}
\stepcounter{appcnt}
\label{I3}
 {\cal I}_3=\frac{\bar \tau}{2\tau}{\cal I}_2-\frac{\tau}{\bar \tau}\left[\ln \tau+\frac{1-\tau^{-2\gamma}}{2}\left(\chi(n,\gamma)-2\ln \bar \tau\right)\right] \; ,
\end{equation}
with ${}_2F_1$ being the standard hypergeometric function.

The \emph{plus~prescription} entering Eqs.~(\ref{Cgg_hadron}) and~(\ref{Cqq_hadron}) comes out as
\begin{equation}
\label{plus-prescription}
\stepcounter{appcnt}
\int_a^1 \drv x \frac{f(x)}{(1-x)_+}
=\int_a^1 \drv x \frac{f(x)-f(1)}{(1-x)}
-\int_0^a \drv x \frac{f(1)}{(1-x)}\; ,
\end{equation}
where $f(x)$ is a test function regularly behaved when $y$ approaches one.

\setcounter{appcnt}{0}
\hypertarget{app:NLOJIF}{
\section*{Appendix~B: NLO light-jet emission function}}
\label{app:NLOJIF}

We present here the NLO correction to the emission function for a forward light-flavored jet emission in the small-cone selection (see Refs.~\cite{Caporale:2012ih,Colferai:2015zfa} for technical details)
\hypertarget{jet_IF_NLO}{}
\begin{equation}
\stepcounter{appcnt}
\label{jet_IF_NLO}
 \hat \E_\J(n,\nu,|\vec q_\J|,x_\J)=
 \frac{1}{\pi}\sqrt{\frac{C_F}{C_A}}
 \left(|\vec q_\J|^2 \right)^{i\nu-1/2}
 \int^1_{x_\J}\frac{\drv \tau}{\tau}
 \tau^{-\bar\alpha_s(\mu_R)\chi(n,\nu)}
\end{equation}
\[
\times\;
\left\{\sum_{j=q,\bar q} f_j \left(\frac{x_\J}{ \tau}\right)\left[\left(P_{qq}(\tau)+\frac{C_A}{C_F}P_{gq}(\tau)\right)
\ln\frac{|\vec q_\J|^2}{\mu_F^2}\right.\right.
\]
\[
-\;2\tau^{-2\gamma} \ln \frac{r}{\max(\tau, \bar \tau)} \,
\left\{P_{qq}(\tau)+P_{gq}(\tau)\right\}-\frac{\beta_0}{2}
\ln\frac{|\vec q_\J|^2}{\mu_R^2}\delta(1-\tau)
\]
\[
+\;C_A\delta(1-\tau)\left[\chi(n,\gamma)\ln\frac{s_0}{|\vec q_\J|^2}
+\frac{85}{18}
\right.
\]
\[
\left.
+\;\frac{\pi^2}{2}+\frac{1}{2}\left(\psi^\prime
\left(1+\gamma+\frac{n}{2}\right)
-\psi^\prime\left(\frac{n}{2}-\gamma\right)-\chi^2(n,\gamma)\right)
\right]
\]
\[
+\;(1+\tau^2)\left\{C_A\left[\frac{(1+\tau^{-2\gamma})\,\chi(n,\gamma)}
{2(1-\tau)_+}-\tau^{-2\gamma}\left(\frac{\ln(1-\tau)}{1-\tau}\right)_+
\right]
\right.
\]
\[
\left.
+\;\left(C_F-\frac{C_A}{2}\right)\left[ \frac{\bar \tau}{\tau^2}{\cal I}_2
-\frac{2\ln\tau}{1-\tau}
+2\left(\frac{\ln(1-\tau)}{1-\tau}\right)_+ \right]\right\}
\]
\[
+\;\delta(1-\tau)\left(C_F\left(3\ln 2-\frac{\pi^2}{3}-\frac{9}{2}\right)
-\frac{5n_f}{9}\right)
+C_A\tau+C_F\bar \tau
\]
\[
\left.
+\;\frac{1+\bar \tau^2}{\tau}
\left(C_A\frac{\bar \tau}{\tau}{\cal I}_1+2C_A\ln\frac{\bar\tau}{\tau}
+C_F\tau^{-2\gamma}(\chi(n,\gamma)-2\ln \bar \tau)\right)\right]
\]
\[
+\;f_{g}\left(\frac{x_\J}{\tau}\right)\frac{C_A}{C_F}
\left[
\left(P_{gg}(\tau)+2 \,n_f \frac{C_F}{C_A}P_{qg}(\tau)\right)
\ln\frac{|\vec q_\J|^2}{\mu_F^2}
\right.
\]
\[
\left.
-\;2\tau^{-2\gamma} \ln \frac{r}{\max(\tau, \bar \tau)} \left(P_{gg}(\tau)+2 \,n_f P_{qg}(\tau)\right)
-\frac{\beta_0}{2}\ln\frac{|\vec q_\J|^2}{4\mu_R^2}\delta(1-\tau)
\right.
\]
\[
\left.
+\; C_A\delta(1-\tau)
\left(
\chi(n,\gamma)\ln\frac{s_0}{|\vec q_\J|^2}+\frac{1}{12}+\frac{\pi^2}{6}
\right.\right.
\]
\[
\left.
+\;\frac{1}{2}\left[\psi^\prime\left(1+\gamma+\frac{n}{2}\right)
-\psi^\prime\left(\frac{n}{2}-\gamma\right)-\chi^2(n,\gamma)\right]
\right)
\]
\[
+\,2C_A (1-\tau^{-2\gamma})\left(\left(\frac{1}{\tau}-2
+\tau\bar\tau\right)\ln \bar \tau + \frac{\ln (1-\tau)}{1-\tau}\right)
\]
\[
+\,C_A\, \left[\frac{1}{\tau}+\frac{1}{(1- \tau)_+}-2+\tau\bar\tau\right]
\left((1+\tau^{-2\gamma})\chi(n,\gamma)-2\ln\tau+\frac{\bar \tau^2}
{\tau^2}{\cal I}_2\right)
\]
\[
\left.\left.
+\,n_f\left[\, 2\tau\bar \tau \, \frac{C_F}{C_A} +(\tau^2+\bar \tau^2)
\left(\frac{C_F}{C_A}\chi(n,\gamma)+\frac{\bar \tau}{\tau}{\cal I}_3\right)
-\frac{1}{12}\delta(1-\tau)\right]\right]\right\} \; ,
\]
with $r$ being the radius of the jet cone.

\bibliographystyle{elsarticle-num}

\bibliography{bibliography}

\end{document}